\documentclass[twocolumn,twocolappendix]{aastex63}

\graphicspath{{./}{figures/}}
\usepackage{color}
\usepackage{soul}
\usepackage{natbib}
\usepackage{hyperref}
\usepackage{amsmath}
\usepackage{xspace}
\usepackage{graphicx}
\usepackage{enumitem}
\usepackage{amssymb}
\usepackage{xifthen}
\usepackage{hyperref}
\usepackage[normalem]{ulem}
\usepackage{comment}
\usepackage{float}
\usepackage{CJKutf8}
\usepackage{bbm}
\hypersetup{    
  colorlinks      = {true},
  linkcolor       = {blue},
  citecolor       = {blue},
  urlcolor        = {blue},
}

\newcommand{\code}[1]{\texttt{#1}\xspace}

\newcommand{\teff}{\ensuremath{T_\mathrm{eff}}\xspace}

\usepackage{bm}
\usepackage{graphicx,pifont}
\let\oldding\ding
\renewcommand{\ding}[2][1]{\scalebox{#1}{\oldding{#2}}}

\usepackage{array}
\newcolumntype{H}{>{\setbox0=\hbox\bgroup}c<{\egroup}@{}}

\newcommand{\gm}{\textbf}

\shorttitle{DESI DR1 Milky Way Mass}
\shortauthors{Medina, Li, Eadie, et al.}

\begin{document}

\title{The mass of the Milky Way from outer halo stars measured by DESI DR1}

\author[0000-0003-0105-9576]{Gustavo E. Medina}
\affiliation{David A. Dunlap Department of Astronomy \& Astrophysics, University of Toronto, 50 St George Street, Toronto ON M5S 3H4, Canada}
\affiliation{Dunlap Institute for Astronomy \& Astrophysics, University of Toronto, 50 St George Street, Toronto, ON M5S 3H4, Canada}

\author[0000-0002-9110-6163]{Ting S. Li}
\affiliation{David A. Dunlap Department of Astronomy \& Astrophysics, University of Toronto, 50 St George Street, Toronto ON M5S 3H4, Canada}
\affiliation{Dunlap Institute for Astronomy \& Astrophysics, University of Toronto, 50 St George Street, Toronto, ON M5S 3H4, Canada}

\author[0000-0003-3734-8177]{Gwendolyn M. Eadie}
\affiliation{David A. Dunlap Department of Astronomy \& Astrophysics, University of Toronto, 50 St George Street, Toronto ON M5S 3H4, Canada}
\affiliation{Department of Statistical Sciences, University of Toronto, 9th Floor, Ontario Power Building, 700 University Ave, Toronto, ON M5G 1Z5, Canada}
\affiliation{Data Sciences Institute, University of Toronto, 17th Floor, Ontario Power Building, 700 University Ave, Toronto, ON M5G 1Z5, Canada}

\author[0000-0001-5805-5766]{Alexander H. Riley}
\affiliation{Institute for Computational Cosmology, Department of Physics, Durham University, South Road, Durham DH1 3LE, UK}
\affiliation{Lund Observatory, Division of Astrophysics, Department of Physics, Lund University, SE-221 00 Lund, Sweden}

\author[0000-0002-6257-2341]{Monica Valluri}
\affiliation{Department of Astronomy, University of Michigan, Ann Arbor, MI 48109, USA}
\affiliation{University of Michigan, Ann Arbor, MI 48109, USA}

\author[0000-0002-5683-2389]{Nabeel Rehemtulla}
\affiliation{Department of Physics and Astronomy, Northwestern University, 2145 Sheridan Road, Evanston, IL 60208, USA}
\affiliation{Center for Interdisciplinary Exploration and Research in Astrophysics (CIERA), 1800 Sherman Ave., Evanston, IL 60201, USA}
\affiliation{NSF-Simons AI Institute for the Sky (SkAI), 172 E. Chestnut St., Chicago, IL 60611, USA}

\author[0000-0002-8010-6715]{Jiaxin Han}
\affiliation{Department of Astronomy, School of Physics and Astronomy, and Shanghai Key Laboratory for Particle Physics and Cosmology, Shanghai Jiao Tong University, Shanghai 200240, People's Republic of China}

\author[0000-0002-5762-7571]{Wenting Wang}
\affiliation{Department of Astronomy, School of Physics and Astronomy, and Shanghai Key Laboratory for Particle Physics and Cosmology, Shanghai Jiao Tong University, Shanghai 200240, People's Republic of China}
\affiliation{State Key Laboratory of Dark Matter Physics, School of Physics and Astronomy, Shanghai Jiao Tong University, Shanghai 200240, People's Republic of China}

\author[0000-0002-5689-8791]{Amanda Bystr\"om}
\affiliation{Institute for Astronomy, University of Edinburgh, Royal Observatory, Blackford Hill, Edinburgh EH9 3HJ, UK}

\author[0000-0002-0740-1507]{Leandro {Beraldo e Silva}}
\affiliation{Steward Observatory, University of Arizona, 933 N, Cherry Ave, Tucson, AZ 85721, USA}
\affiliation{Observat\'orio Nacional, Rio de Janeiro - RJ, 20921-400, Brasil}

\author[0000-0003-2644-135X]{S.~E.~Koposov}
\affiliation{Institute for Astronomy, University of Edinburgh, Royal Observatory, Blackford Hill, Edinburgh EH9 3HJ, UK}

\author[0000-0002-7393-3595]{N.~R.~Sandford}
\affiliation{Department of Astronomy and Astrophysics, University of Toronto, 50 St. George Street, Toronto, ON M5P 0A2, Canada}

\author[0000-0002-7667-0081]{R.~G.~Carlberg}
\affiliation{Department of Astronomy \& Astrophysics, University of Toronto, Toronto, ON M5S 3H4, Canada}

\author[0000-0002-2527-8899]{M. Lambert}
\affiliation{Department of Astronomy \& Astrophysics, University of California, Santa Cruz, Santa Cruz, CA 95064, USA}

\author[0000-0001-9852-9954]{O.~Y.~Gnedin}
\affiliation{University of Michigan, Ann Arbor, MI 48109, USA}

\author[0000-0001-8274-158X]{A.~P.~Cooper}
\affiliation{Institute of Astronomy and Department of Physics, National Tsing Hua University, 101 Kuang-Fu Rd. Sec. 2, Hsinchu 30013, Taiwan}

\author[0000-0002-9370-8360]{J.~Garc\'ia-Bellido}
\affiliation{Instituto de F\'isica Te\'orica UAM/CSIC, Universidad Aut\'onoma de Madrid, Nicol\'as Cabrera 13, 28049 Madrid, Spain}

\author[0000-0003-0853-8887]{N.~Kizhuprakkat}
\affiliation{Institute of Astronomy and Department of Physics, National Tsing Hua University, 101 Kuang-Fu Rd. Sec. 2, Hsinchu 30013, Taiwan}

\author{B.~A.~Weaver}
\affiliation{NSF NOIRLab, 950 N. Cherry Ave., Tucson, AZ 85719, USA}

\author{J.~Aguilar}
\affiliation{Lawrence Berkeley National Laboratory, 1 Cyclotron Road, Berkeley, CA 94720, USA}
\author[0000-0001-6098-7247]{S.~Ahlen}
\affiliation{Department of Physics, Boston University, 590 Commonwealth Avenue, Boston, MA 02215 USA}
\author[0000-0003-2923-1585]{A.~Anand}
\affiliation{Lawrence Berkeley National Laboratory, 1 Cyclotron Road, Berkeley, CA 94720, USA}
\author[0000-0001-9712-0006]{D.~Bianchi}
\affiliation{Dipartimento di Fisica ``Aldo Pontremoli'', Universit\`a degli Studi di Milano, Via Celoria 16, I-20133 Milano, Italy}
\affiliation{INAF-Osservatorio Astronomico di Brera, Via Brera 28, 20122 Milano, Italy}
\author{D.~Brooks}
\affiliation{Department of Physics \& Astronomy, University College London, Gower Street, London, WC1E 6BT, UK}
\author{T.~Claybaugh}
\affiliation{Lawrence Berkeley National Laboratory, 1 Cyclotron Road, Berkeley, CA 94720, USA}
\author[0000-0002-2169-0595]{A.~Cuceu}
\affiliation{Lawrence Berkeley National Laboratory, 1 Cyclotron Road, Berkeley, CA 94720, USA}
\author[0000-0002-1769-1640]{A.~de la Macorra}
\affiliation{Instituto de F\'{\i}sica, Universidad Nacional Aut\'{o}noma de M\'{e}xico,  Circuito de la Investigaci\'{o}n Cient\'{\i}fica, Ciudad Universitaria, Cd. de M\'{e}xico  C.~P.~04510,  M\'{e}xico}
\author{P.~Doel}
\affiliation{Department of Physics \& Astronomy, University College London, Gower Street, London, WC1E 6BT, UK}
\author[0000-0003-4992-7854]{S.~Ferraro}
\affiliation{Lawrence Berkeley National Laboratory, 1 Cyclotron Road, Berkeley, CA 94720, USA}
\affiliation{University of California, Berkeley, 110 Sproul Hall \#5800 Berkeley, CA 94720, USA}
\author[0000-0002-2890-3725]{J.~E.~Forero-Romero}
\affiliation{Departamento de F\'isica, Universidad de los Andes, Cra. 1 No. 18A-10, Edificio Ip, CP 111711, Bogot\'a, Colombia}
\affiliation{Observatorio Astron\'omico, Universidad de los Andes, Cra. 1 No. 18A-10, Edificio H, CP 111711 Bogot\'a, Colombia}
\author[0000-0001-9632-0815]{E.~Gaztañaga}
\affiliation{Institut d'Estudis Espacials de Catalunya (IEEC), c/ Esteve Terradas 1, Edifici RDIT, Campus PMT-UPC, 08860 Castelldefels, Spain}
\affiliation{Institute of Cosmology and Gravitation, University of Portsmouth, Dennis Sciama Building, Portsmouth, PO1 3FX, UK}
\affiliation{Institute of Space Sciences, ICE-CSIC, Campus UAB, Carrer de Can Magrans s/n, 08913 Bellaterra, Barcelona, Spain}
\author[0000-0001-9822-6793]{J.~Guy}
\affiliation{Lawrence Berkeley National Laboratory, 1 Cyclotron Road, Berkeley, CA 94720, USA}
\author{R.~Kehoe}
\affiliation{Department of Physics, Southern Methodist University, 3215 Daniel Avenue, Dallas, TX 75275, USA}
\author[0000-0003-4207-7420]{S.~Kent}
\affiliation{Department of Astronomy and Astrophysics, University of Chicago, 5640 South Ellis Avenue, Chicago, IL 60637, USA}
\affiliation{Fermi National Accelerator Laboratory, PO Box 500, Batavia, IL 60510, USA}
\author[0000-0003-3510-7134]{T.~Kisner}
\affiliation{Lawrence Berkeley National Laboratory, 1 Cyclotron Road, Berkeley, CA 94720, USA}
\author[0000-0003-1838-8528]{M.~Landriau}
\affiliation{Lawrence Berkeley National Laboratory, 1 Cyclotron Road, Berkeley, CA 94720, USA}
\author[0000-0001-7178-8868]{L.~Le~Guillou}
\affiliation{Sorbonne Universit\'{e}, CNRS/IN2P3, Laboratoire de Physique Nucl\'{e}aire et de Hautes Energies (LPNHE), FR-75005 Paris, France}
\author[0000-0003-1887-1018]{M.~E.~Levi}
\affiliation{Lawrence Berkeley National Laboratory, 1 Cyclotron Road, Berkeley, CA 94720, USA}
\author[0000-0003-4962-8934]{M.~Manera}
\affiliation{Departament de F\'{i}sica, Serra H\'{u}nter, Universitat Aut\`{o}noma de Barcelona, 08193 Bellaterra (Barcelona), Spain}
\affiliation{Institut de F\'{i}sica d’Altes Energies (IFAE), The Barcelona Institute of Science and Technology, Edifici Cn, Campus UAB, 08193, Bellaterra (Barcelona), Spain}
\author[0000-0002-1125-7384]{A.~Meisner}
\affiliation{NSF NOIRLab, 950 N. Cherry Ave., Tucson, AZ 85719, USA}
\author{R.~Miquel}
\affiliation{Instituci\'{o} Catalana de Recerca i Estudis Avan\c{c}ats, Passeig de Llu\'{\i}s Companys, 23, 08010 Barcelona, Spain}
\affiliation{Institut de F\'{i}sica d’Altes Energies (IFAE), The Barcelona Institute of Science and Technology, Edifici Cn, Campus UAB, 08193, Bellaterra (Barcelona), Spain}
\author[0000-0003-3188-784X]{N.~Palanque-Delabrouille}
\affiliation{IRFU, CEA, Universit\'{e} Paris-Saclay, F-91191 Gif-sur-Yvette, France}
\affiliation{Lawrence Berkeley National Laboratory, 1 Cyclotron Road, Berkeley, CA 94720, USA}
\author[0000-0002-0644-5727]{W.~J.~Percival}
\affiliation{Department of Physics and Astronomy, University of Waterloo, 200 University Ave W, Waterloo, ON N2L 3G1, Canada}
\affiliation{Perimeter Institute for Theoretical Physics, 31 Caroline St. North, Waterloo, ON N2L 2Y5, Canada}
\affiliation{Waterloo Centre for Astrophysics, University of Waterloo, 200 University Ave W, Waterloo, ON N2L 3G1, Canada}
\author[0000-0001-7145-8674]{F.~Prada}
\affiliation{Instituto de Astrof\'{i}sica de Andaluc\'{i}a (CSIC), Glorieta de la Astronom\'{i}a, s/n, E-18008 Granada, Spain}
\author[0000-0001-6979-0125]{I.~P\'erez-R\`afols}
\affiliation{Departament de F\'isica, EEBE, Universitat Polit\`ecnica de Catalunya, c/Eduard Maristany 10, 08930 Barcelona, Spain}
\author{G.~Rossi}
\affiliation{Department of Physics and Astronomy, Sejong University, 209 Neungdong-ro, Gwangjin-gu, Seoul 05006, Republic of Korea}
\author[0000-0002-9646-8198]{E.~Sanchez}
\affiliation{CIEMAT, Avenida Complutense 40, E-28040 Madrid, Spain}
\author{D.~Schlegel}
\affiliation{Lawrence Berkeley National Laboratory, 1 Cyclotron Road, Berkeley, CA 94720, USA}
\author{M.~Schubnell}
\affiliation{Department of Physics, University of Michigan, 450 Church Street, Ann Arbor, MI 48109, USA}
\affiliation{University of Michigan, 500 S. State Street, Ann Arbor, MI 48109, USA}
\author[0000-0002-3461-0320]{J.~Silber}
\affiliation{Lawrence Berkeley National Laboratory, 1 Cyclotron Road, Berkeley, CA 94720, USA}
\author{D.~Sprayberry}
\affiliation{NSF NOIRLab, 950 N. Cherry Ave., Tucson, AZ 85719, USA}
\author[0000-0003-1704-0781]{G.~Tarl\'{e}}
\affiliation{University of Michigan, 500 S. State Street, Ann Arbor, MI 48109, USA}
\author[0000-0002-6684-3997]{H.~Zou}
\affiliation{National Astronomical Observatories, Chinese Academy of Sciences, A20 Datun Road, Chaoyang District, Beijing, 100101, P.~R.~China}

\correspondingauthor{Gustavo E. Medina}
\email{gustavo.medina@utoronto.ca}


\begin{abstract}
As a benchmark for galaxy evolution and dark matter studies, the total mass of the Milky Way is a parameter of cosmological significance, and its value at large radii from the Galactic center remains highly uncertain. 
Following a hierarchical Bayesian inference approach, we measure the cumulative mass of the Milky Way using full 6D phase-space information of stars from the first data release of the Dark Energy Spectroscopic Instrument (DESI). 
We employ 330 blue horizontal-branch stars (BHBs) and 110 RR Lyrae stars (RRLs) in DESI covering Galactocentric distances in the range $\sim$50--100\,kpc. 
Within 100\,kpc from the Galactic center, we report an enclosed mass of $M(<100\ {\rm kpc}) = 0.57^{+0.08}_{-0.07}\times10^{12}$\,M$_\odot$  and $M(<100\ {\rm kpc}) = 0.55^{+0.12}_{-0.10}\times10^{12}$\,M$_\odot$  when using BHBs and RRLs, respectively.
Extrapolating our mass profiles beyond the extent of our data, we find the virial mass of the Galaxy to be $M_{200}=0.85^{+0.16}_{-0.14}\times10^{12}$\,M$_\odot$ and $M_{200}=0.78^{+0.19}_{-0.15}\times10^{12}$\,M$_\odot$, respectively. 
We validate the effectiveness and limitations of our method using mock BHBs and RRLs from two AuriDESI halos.
These tests show that the code recovers the enclosed mass of the mock galaxy with high precision and accuracy between 50 and 200\,kpc, independent of the stellar tracer used and their spatial distribution. 
The tests also suggest an underestimation of the galaxy's cumulative mass at a level of up to $\sim20$\% if stars close to the Galactic center are used in the models. 
Our mass estimates lay the groundwork for future inference of the Galactic mass with upcoming DESI data releases and spectroscopic surveys mapping the halo.
\end{abstract}





\section{Introduction} \label{sec:intro}

The total mass of the Milky Way (MW) is a key parameter to place our Galaxy into a broad cosmological context.
Indeed, since the MW is arguably the best studied galaxy in the Universe \citep[][]{Bland-Hawthorn2016}, it has become customary to compare its properties, such as its mass, shape, and assembly history, to those of MW-analog galaxies in extragalactic surveys and cosmological simulations in the $\Lambda$ Cold Dark Matter framework \citep[e.g.,][]{Tissera2013,Han2016,Xia2025,Apfel2025}. 
Therefore, a reliable determination of the total enclosed mass of the MW out to large radii is essential. 
Unfortunately, to this day, determinations of the total mass of the MW within 150\,kpc suffer from large scatter \citep[see e.g.,][]{Fritz2020,Wang2020,Shen2022,Bobylev2023,Hunt2025}, with results that are inconsistent, sensitive to choices of methodology and tracers of the potential, and limited by the challenges of measuring distances in the outer halo.

Estimating the total mass of our Galaxy or the mass enclosed within a given radius typically relies on kinematical information from different dynamical tracers: globular clusters, dwarf galaxies, stellar streams, and even field stars. 
A method that has been used extensively is to study the escape velocity from a sample of mass tracers, measuring the enclosed mass from the velocity distribution of high velocity stars and identifying the limits at which such tracers become gravitationally unbound to the MW \citep[see e.g.,][]{Piffl2014,Monari2018,Deason2019,Prudil2022,Necib2022b}. 
Another common method to estimate the MW mass is to measure the circular velocity of tracers in the MW disk and to characterize their rotation given a certain model and Galactic mass \citep[e.g.,][]{McMillan2017,Cautun2020,Karukes2020}. 
Other approaches rely on the timing argument, which measures the relative velocity and position between the Galaxy and some tracers to infer the conditions under which the MW potential halted the initial expansion of the universe \citep[see e.g.,][]{Zaritsky1989,Li2008}. 
Additionally, studying the Jeans equations, obtained from the steady-state collisionless Boltzmann equation and under the assumption of dynamical equilibrium, provides tools to measure the enclosed mass of the Galaxy at any radius \citep[see e.g.,][]{Jeans1915,Xue2008,Gnedin2010,Rehemtulla+22}.
For a review of methods for estimating the Galactic mass and a compilation of recent results, we refer the reader to \citet{Bobylev2023}. 

One of the most commonly used methods to derive the mass of the MW halo relies on phase-space distribution functions \citep[see e.g.,][]{Little1987,Eadie2017,Li2020,Deason2021,Hattori2021}. 
Phase-space distribution functions are used to describe the dynamical state of the Galaxy by combining positions (right ascensions, declinations, and distances) and velocities (tangential velocities from proper motions and line-of-sight velocities), in what is typically referred to as a 6D analysis. 
Measuring the MW mass profile using 6D information can, however, be a challenging task, as incomplete datasets are the norm and missing components of the phase space (e.g., line-of-sight velocities, distances, and/or proper motions) can substantially affect the analysis.

Relying on the expression of the distribution function discussed by \citet{Evans1997} (also used by \citealt{Deason2011a, Deason2012b, Deason2012a}), \citet{Eadie2015} developed a hierarchical Bayesian method to measure the MW mass under the assumption of a steady state halo with a fixed radial density distribution. 
This method is able to deal with limited or incomplete data effectively, treating the unknown velocity components of a given tracer population as nuisance parameters in the model while taking full advantage of the complete data that are available.

Recent works have exploited and expanded the formalism developed by \citet{Eadie2015} to derive the cumulative mass profile (CMP) of the Galaxy based on specific datasets. 
\citet{Eadie2019}, for instance, used this Bayesian approach with different spatial distribution models to model {\it Gaia} DR2-based positions and proper motions of globular clusters as tracers of the Galactic potential.
These authors inferred the MW mass within 125 kpc to be $0.52\times10^{12}$\,M$_\odot$, and a virial mass $M_{200}$\footnote{$M_{200}$ is defined as the mass enclosed within the radius at which the average density of the halo is equal to 200 times the critical density of the universe.} of $0.68\times10^{12}$\,M$_\odot$.
Similarly, \citet{Slizewski2022} used dwarf galaxies beyond 45 kpc as kinematic tracers and found a halo mass within 100 kpc of $0.78\times10^{12}$\,M$_\odot$ and a virial mass of $1.19\times10^{12}$\,M$_\odot$, though they stress that the resulting cumulative mass is sensitive to the choice of tracer and is particularly influenced by disturbances in the potential (e.g., by massive satellites) and assumptions on the bound/unbound state of distant tracers.
More recently, \citet{Shen2022} incorporated all available 6D phase-space measurements (including measurement correlations) of halo stars with data from the H3 survey and {\it Gaia} EDR3. 
These authors found the MW mass within 100\,kpc to be $0.69\times10^{12}$\,M$_\odot$ and a virial mass of $1.08\times10^{12}$\,M$_\odot$, highlighting that their results are sensitive to a number of substructures in the halo, which they claim limit the precision of their mass estimates to $\sim15\%$.

\begin{figure*}
     \centering    
    \includegraphics[width=0.999\textwidth]{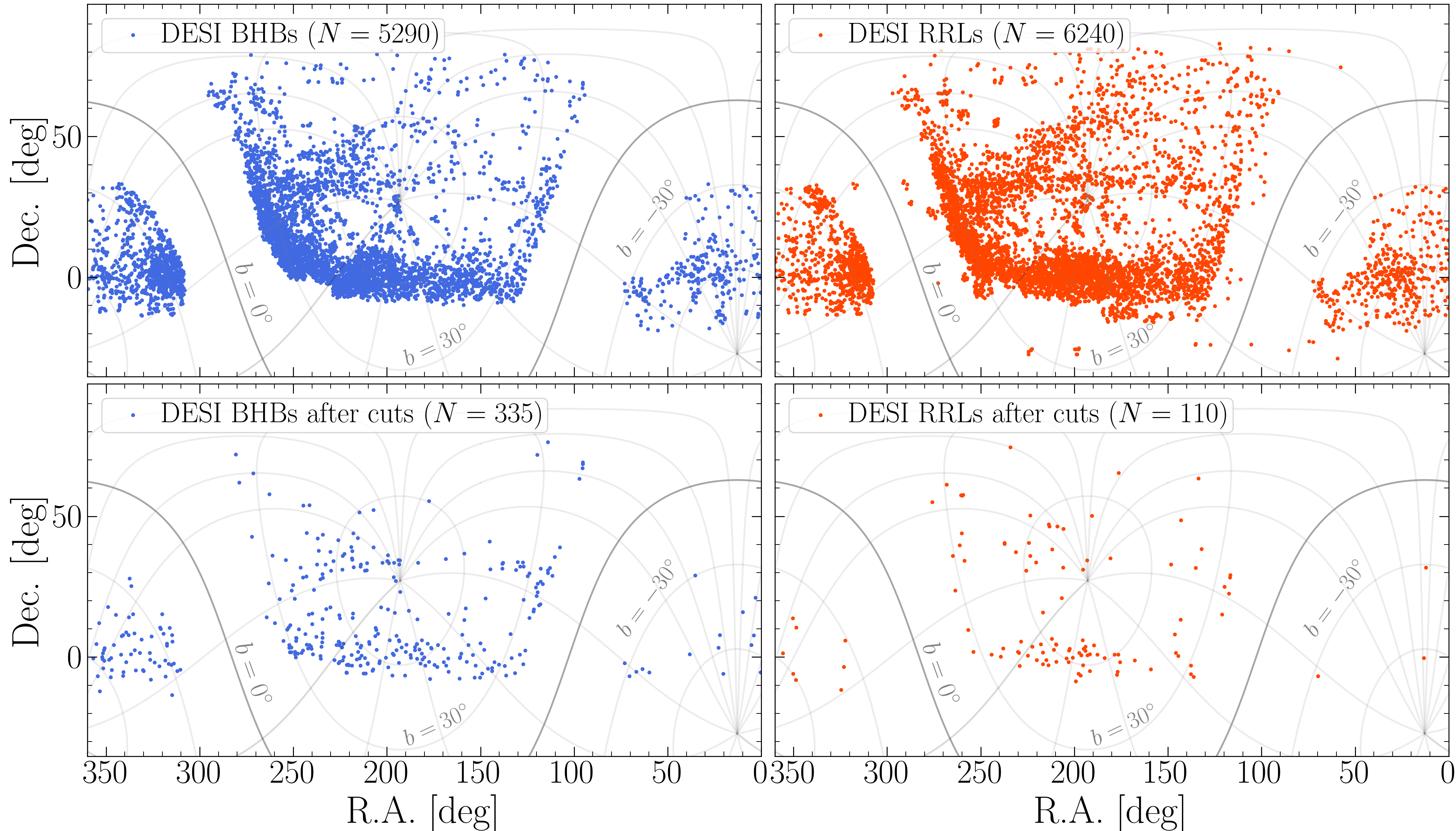}
    \caption {Spatial distribution in Equatorial coordinates of the DESI BHBs and RRLs (left and right panels, respectively). 
    The upper panels show the full catalogs of BHBs and RRLs used in this work, whereas the bottom panels depict the samples used for our GME analysis, resulting from employing the cuts described in Section~\ref{sec:cuts}. 
    }
    \label{fig:beforeAfter}
\end{figure*}

To take full advantage of the aforementioned methods, the availability of line-of-sight velocities  and precise distances of a large number of tracers is often required, in particular in the outer regions of the MW. 
The Dark Energy Spectroscopic Instrument (DESI) survey is a large-scale spectroscopic campaign designed to map the Northern hemisphere by obtaining medium-resolution (R$\sim$2,000-5,000) spectra of distant galaxies, quasars, and stars \citep[][]{DESI2016a,DESI2016b,DESI2022,DESI_DR1_cosmology_2024,DESI_DR1_2025,desi2025ii_bao}. 
Indeed, the DESI survey has measured spectra of millions of stars in the Galaxy as part of its Milky Way Survey \citep[MWS;][]{Cooper2023}, yielding high-precision line-of-sight velocities and chemical abundances \citep[][]{Koposov2024,Koposov2025}. 
More specifically, DESI-MWS observed blue horizontal-branch stars (hereafter BHBs, or BHB for single stars) and RR Lyrae stars (hereafter RRLs, or RRL for single stars) which are particularly useful as dynamical tracers due to their nature as precise distance estimators 
(see e.g., \citealt{Deason2011b,Fermani2013,Barbosa2022} for BHBs, and \citealt{Catelan2015,Bhardwaj2023,Narloch2024} for RRLs). 
In fact, the DESI DR1 catalog contains over 5,000 BHBs and over 6,000 RRLs 
with typical distance uncertainties at a $\sim6$\% and $5$\% level \citep[respectively;][]{Bystroem2024,Medina2025a}.
These catalogs include a large number of stars beyond 50\,kpc, and even stars beyond 100\,kpc. 
A natural caveat of the use of these datasets for the inference of the Galactic potential is the lack of (reliable) proper motions at large distances, where halo mass estimations are the most sensitive. 
This, however, makes them suitable testbeds for the hierarchical Bayesian methodology developed by \citet{Eadie2015} and \citet{Eadie2019}, and later expanded by \citet{Shen2022}. 
The combination of these catalogs represents a tantalizing opportunity to test the sensitivity of current MW mass estimation models to substructures, different tracers, incomplete data, and the effects of the Large Magellanic Cloud (LMC), as well as to reduce the uncertainties in the MW mass measurements out to its virial radius.

In this work, we employ the sample of BHBs and RRLs 
observed in DESI's first year of operations to measure the cumulative mass of the MW out to its virial radius. 
In Section~\ref{sec:data}, we describe the sample used, including the steps used to build the BHB and RRL catalogs, the determination of distances for these stars, and the selection cuts implemented for their use in our study.
Section~\ref{sec:model} provides an overview of the hierarchical Bayesian methodology developed by \citet{Shen2022} and used in this work. 
A validation of our methodology is presented in  Section~\ref{sec:mocks}, for which we use mock BHB and RRL catalogs that resemble the DESI data used as input in our inference model.  
In Section~\ref{sec:ResultsAndDiscussion}, we discuss the results of our analysis, 
strengths and limitations of our methodology, put the outcome of our analysis into a broader context, and compare them with other results from the literature. 
Finally, a summary of our work and the impact of upcoming surveys in the measurement of the MW mass is presented in Section~\ref{sec:conclusions}.

\section{Data}
\label{sec:data}

\subsection{The DESI survey}

DESI, mounted on the Mayall 4-m telescope at Kitt Peak National Observatory (KPNO), is a massively multiplexed fiber-fed spectrograph designed to study the effects of dark energy on the formation and expansion history of the universe. 
The instrument is composed of 5,000 robotic fibers that feed 10 thermally-controlled 3-channel spectrographs (DESI's blue, red, and near-infrared arms), covering a wavelength range of $\sim$3,600-9,800\,\AA\ with a resolution varying between $\sim$2,500 and 5,000.
The data acquisition for the DESI survey is structured in surveys (dark and bright time surveys, depending on the observing conditions) and programs (commissioning, science verification, and main program), and the wavelength and flux calibration of the survey's data are carried out using DESI's processing and reduction pipeline \citep[see e.g.,][]{Guy2023,DESI_Survey_Operation_2023, DESI_Corrector_2024,DESI_Fiber_System_2024}.
Moreover, although the DESI survey is primarily focused on studying the large-scale structure of the universe, and thus its main targets are galaxies, there is a MW survey component to it (DESI-MWS), which has observed millions of MW stars with Galactic latitudes $|b|>20$\,deg (i.e., focusing mostly on the halo). 
A comprehensive description of the design and goals of DESI-MWS is provided by \citet{Cooper2023} and \citet{Koposov2024}.

The spectroscopic properties of the stars observed by DESI are derived using two main processing pipelines, RVS  and the SP, both of which rely on the interpolation of stellar templates. 
On the one hand, the RVS pipeline employs an adapted version of the Python package RVSpecfit \citep[][]{Koposov:2019}
to derive radial velocities and atmospheric parameters of stars based on PHOENIX stellar models \citep[][]{Husser2013}. 
On the other hand, the SP pipeline uses the FORTRAN code FERRE \citep[][]{AllendePrieto2006} to measure atmospheric parameters and chemical abundances, combining PHOENIX and Kurucz ATLAS9 models \citep[][]{Kurucz1979,Kurucz1993} for the spectral fitting optimization. 
In this work, we use the spectroscopic properties derived with DESI's RVS pipeline. 

For determining the mass of the MW, we take advantage of the large number of precise stellar distance indicators that have been observed by the DESI survey and {\it Gaia} DR3, most of which possess 6D phase-space information. 
More specifically, we use line-of-sight velocities from DESI DR1 \citep[][]{Koposov2025}  and proper motions from the {\it Gaia} survey. 
We focus our attention on RRLs and BHBs, both of which are considered high-priority targets in DESI and, as of year 1, were observed in bright and dark time.

\subsection{The RRL and BHB samples}
\label{sec:RRL-BHB-samples}

The RRL catalog used in this work is based on DESI DR1 \citep[][]{DESI_DR1_2025} and was presented by \citet{Medina2025a}. 
The DESI DR1 RRL sample consists of stars observed as part of DESI's main program and its science verification (SV) program, crossmatched with the {\it Gaia} DR3 RRL catalog \citep[][]{Clementini2023} using a radius of 0.5\,arcsec.
These stars are selected as DESI targets using the {\it Gaia} DR2 RRL catalog \citep[][]{Clementini2019} and the Pan-STARRS1 RRL catalog \citep[][]{Sesar2017}.
The RRL catalog is constructed from a total of 12,301 single-epoch spectra corresponding to 6,240 stars with unique {\it Gaia} identification numbers ({\sc source\_ids}). 
Most of these stars (4,980 RRLs, $\sim$82\%) have less than three DESI epochs, whereas 11 stars have spectra for over 20 epochs. 
This catalog contains information that is relevant for the use of RRLs for MW-related studies, obtained from the photometric and astrometric characterization of {\it Gaia}'s RRLs, as well as the spectroscopic parameters determined by the RVSpec pipeline. 
Notably, it includes estimates of the RRLs' iron abundances (relevant for the determination of distances; Section~\ref{sec:distances}) and systemic velocities, corrected for the pulsating component of the (observed) line-of-sight velocities. 
Indeed, radial velocity and atmospheric parameter determinations are challenging to conduct in RRLs, owing to their variability on short time-scales ($<1$\,d).
Because RRLs are variable stars, their systemic (or center-of-mass) velocities differ from their observed radial velocities, and thus, a correction must be applied to account for their pulsating nature. 
We further describe these corrections in Section~\ref{sec:RVs}.

In DESI DR1, BHB targets are selected using photometric and astrometric cuts, which result in a BHB catalog that is later cleaned using spectroscopic information determined from co-added spectra. 
Here, we present a brief summary of these cuts \citep[explained in detail by][]{Cooper2023}, and refer the reader to \citet{Bystroem2024} for a full description of the methodology followed to assemble the catalog.
The BHB targets in the DESI main survey were selected using the following criteria: 

\begin{equation}
\begin{aligned}
   G &> 10   \\
    \varpi \ {\rm [mas]}&  \leq 0.1 + 3 \sigma_\varpi\\
    -0.35 & \leq (g-r) \leq -0.02\\
    -1.5 &> r - 2.3(g-r) - W1_\text{faint} .
\end{aligned}
\end{equation}
\label{eq:BHB_selection1}

\noindent In the previous equations, $G$, $\varpi$, and $\sigma_\varpi$ represent the mean $G$ magnitude, parallax, and parallax error from the {\it Gaia} DR3 catalog.  
The color cuts used above use $g$ and $r$ mean magnitudes from the DESI Legacy Imaging Survey, and the term $W1_\text{faint} = 22.5 - 2.5\log_{10}{(W1 - 3\sigma_{W1})}$ computed from the  WISE 3.4 $\mu$m flux ($W1$) and its error ($\sigma_{W1}$). 
Moreover, a boundary in color-color space,  defined as

\begin{equation}
\begin{aligned}
    X_\text{BHB} &= (r-z) - [1.07163(g-r)^5 - 1.42272(g-r)^4 \\
    & + 0.69476(g-r)^3 -0.12911(g-r)^2 \\
    & + 0.66993(g-r) - 0.11368]
\end{aligned}
\end{equation}
\label{eq:BHB_boundary}

\noindent \citep[as defined by][]{Li2019}, 
is used to distinguish between BHBs from blue stragglers, such that

\begin{equation}
    -0.05 \leq X_\text{BHB} \leq 0.05.
\end{equation}
\label{eq:BHB_boundary2}

Secondary BHB targets are selected to be observed in dark time, so that $19<r<21$ and 

\begin{equation}
\begin{aligned}
   -0.3 &<(g-r)<0.0  \\
    -0.7&<(r-z)<-0.05\\
    -0.3 \cdot (g-r)^2 &< \\& ((r-z) - 0.75 \cdot ((g-r) + 0.15) + 0.27 - 0.025) \\&\ \ \ \ \ \ < 0.06 - 0.6 \cdot ((g-r) + 0.15)^2) \\
    W1_\text{ratio} &< 0.3 \cdot (g-r) + 0.15 + 3W1_\text{ratio,error} + 0.3, 
\end{aligned}
\end{equation}
\label{eq:BHB_selection2}

\noindent where $W1_\text{ratio} = \dfrac{W1}{g}$ and $W1_\text{ratio,error} = \dfrac{1}{\sqrt{W1_{\text{ivar}}} \cdot g}$, and $W1_{\text{ivar}}$ represents the inverse variance of W1.
Adding the secondary targets to the main targets described above results in a total of 10,695 BHB targets in DESI DR1, after removing duplicates. 

\citet{Bystroem2024} performed additional cuts to reduce contamination in the catalog of observed BHB stars, 
including the removal of quasar-like sources flagged as such by the template-fitting code Redrock (Bailey et al., in prep.; \citealt{Anand2024}) and stars classified as RRLs in the {\it Gaia} DR3 and the PS-1 catalogs. 
Line-of-sight velocity quality flags from the RVS pipeline are then applied to ensure that only high-quality velocities are used in the analysis (as defined by the pipeline's success flags). 
Furthermore, based on their position in the Kiel diagram (surface gravity vs. effective temperature, $\log g$ vs. $T_{\rm eff}$), only stars meeting the conditions

\begin{equation}
\begin{aligned}
7,000 &< T_\text{eff} < 14,500 \text{ K} \\
    1.50 &< \text{log(}g) - k \cdot T_\text{eff} < 2.65
\end{aligned}
\end{equation}
\label{eq:BHB_kiel-cut}

\noindent (where $k = 0.00014$) are kept as reliable 
spectroscopically confirmed BHBs. 
Lastly, only BHB candidates with $-0.3 < g - r < 0.0$ are considered for the analysis, since this represents the $g-r$ color range in which reliable distances can be computed. The resulting catalog contains 5,290 unique BHBs, which are used as a starting point for the analysis (concerning BHBs) presented in this work.

Figure~\ref{fig:beforeAfter} (top row) shows the spatial distribution, in Equatorial coordinates, of the stars in the DESI DR1 RRL and BHB catalogs.  
Figure~\ref{fig:beforeAfter} (lower row) also shows the spatial distribution of the stars in these catalogs after applying selection cuts to the data (as described in Section~\ref{sec:cuts}).
This is an important step for us to use these stars as inputs for our mass measurement methodology.

\begin{figure}
    \centering    
    \includegraphics[width=0.45\textwidth]{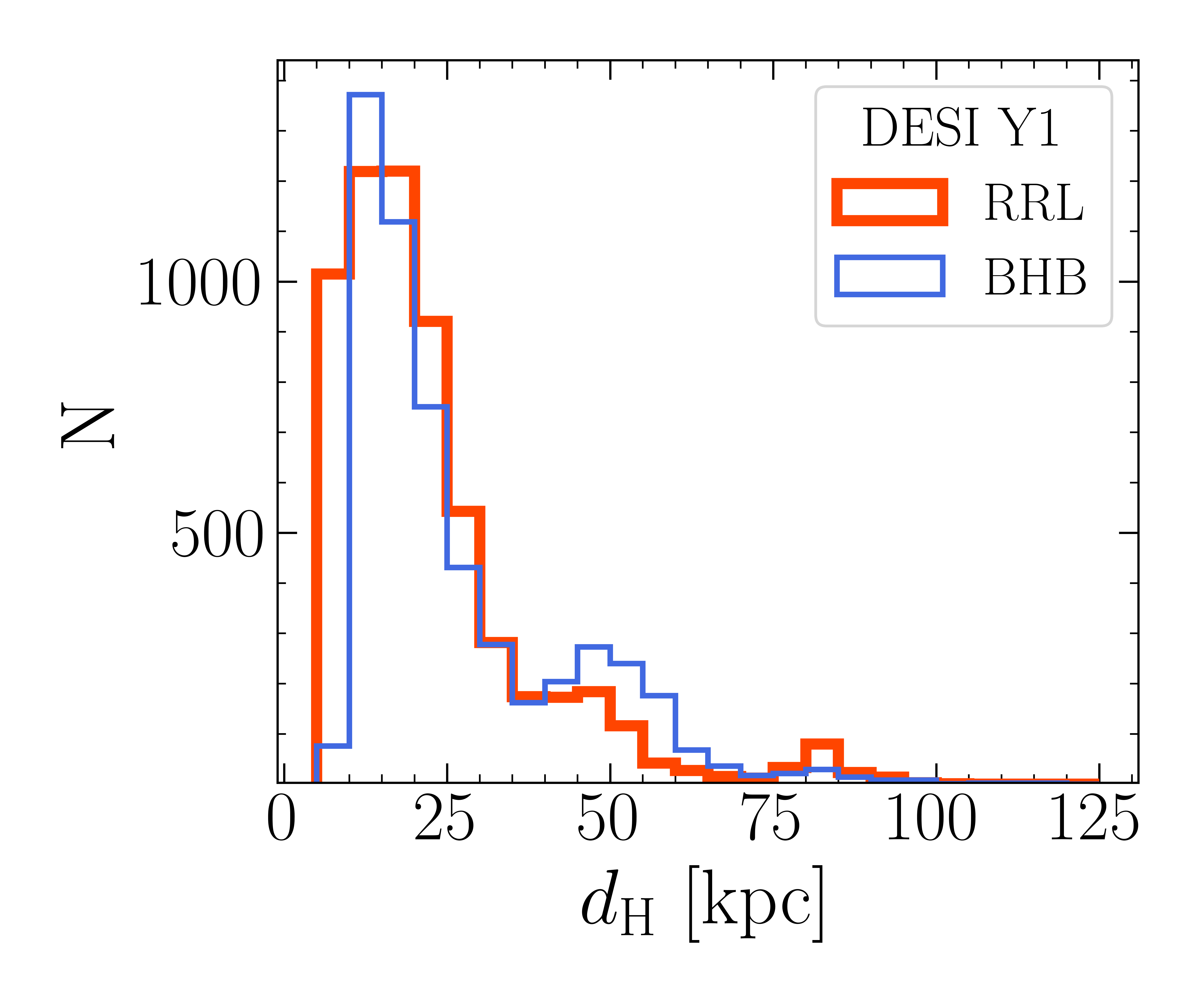}
    \caption{Heliocentric distance distribution of the BHBs and RRLs in DESI DR1. 
    The significant overdensities of stars at $d_{\rm H}>30$\,kpc correspond to various substructures in the halo (intact satellites, stellar streams, and remnants from accretion events). 
    }
    \label{fig:dh_dist}
\end{figure}

\subsection{Distance determination}
\label{sec:distances}

To estimate the intrinsic brightness of the stars in our BHB and RRL samples, we adopt absolute magnitude-color and -metallicity relations available in the literature. 
For the former, absolute magnitudes are computed from the absolute magnitude-color relations adopted by \citet{Bystroem2024}, that is:

\begin{equation}
\begin{array}{lc}
M_g = 0.566 - 0.392\ \left(g - r\right) + 2.729\ \left(g - r\right)^2 \\ \ \ \ \ \ \ \ \ \ +\ 29.1128\ \left(g - r\right)^3 + 113.569\ \left(g - r\right)^4.  
\end{array}
\label{eq:magcolor_relation_bhbs}
\end{equation}

\noindent for BHBs with $-0.30 < \left(g-r\right) <-0.05$, and 

\begin{equation}
\begin{array}{lc}
M_g = 0.564 - 0.500\ \left(g-r\right)
\end{array}
\label{eq:magcolor_relation_bhbs2}
\end{equation}

\noindent for BHBs with $-0.05 < \left(g-r\right) < 0.00$. In these equation, $g$ and $r$ are extinction-corrected magnitudes in the Dark Energy Camera (DECam) photometric system from the DESI Legacy Imaging Survey and $M_g$ is the absolute magnitude in the $g$. 

For RRLs, we employ the 
absolute magnitude-metallicity relation inferred by \citet{Garofalo2022} in the {\it Gaia} $G$ band, using the intensity-averaged $G$ magnitude provided in the {\it Gaia} DR3 RRL catalog ({\sc int\_average\_g}).
Thus, we use the following relation:

\begin{equation}
\begin{array}{lc}
M_G = (0.28^{+0.36}_{-0.36}) {\rm [Fe/H]} + (0.97^{+0.49}_{-0.52}),  
\end{array}
\label{eq:magfeh_relation_rrls}
\end{equation}

\noindent where $M_G$ represents the absolute magnitude in $G$. 
For each star in the RRL catalog, similar to \citet{Medina2025a} and \citet{Medina2025b}, 
we use the average of the [Fe/H] values determined by the DESI RVS pipeline from single-epoch spectra as the iron abundances required in Equation~\ref{eq:magfeh_relation_rrls}. 
For stars with more than one available epoch, we adopt the mean [Fe/H] as a sensible representation of their true [Fe/H].

Heliocentric distances ($d_{\rm H}$) are computed via distance modulus using the absolute magnitudes from Equations~\ref{eq:magcolor_relation_bhbs} and \ref{eq:magfeh_relation_rrls} and extinction-corrected apparent magnitudes in the corresponding filters. 
For the extinction correction, we use the dust maps from \citet{Schlafly2011}, adopting the standard value of relative visibility ($R_V$) in the diffuse interstellar medium, $R_V = 3.1$ \citep{Schultz1975,Cardelli1989}. 
For the BHBs, we use the extinction ratios $A_g/A_V = 1.04419$ and $A_r/A_V = 0.70194$. 
For the RRLs, we adopt $A_G/A_V = 0.85926$\footnote{http://stev.oapd.inaf.it/cgi-bin/cmd\_3.3}.

Figure~\ref{fig:dh_dist} depicts the heliocentric distance ($d_{\rm H}$) distribution of BHBs and RRLs. 
The figure shows the wide range of $d_{\rm H}$ that BHBs and RRLs cover, both of which start at $<10$\,kpc and extend out to $>100$\,kpc. 
A non-negligible fraction of the stars in these samples are associated with intact satellites, satellites undergoing tidal disruption, or already accreted satellites,  
most notably from the Draco dwarf galaxy, the Sagittarius (Sgr) stream, and the Gaia-Sausage-Enceladus (GSE) merger event.
More details about stars in these substructures are provided in Section~\ref{sec:cuts}. 
When converted to Galactocentric distances ($R_{\rm GC}$) assuming a spherical halo and a distance of 8.2\,kpc \citep{Gravity2021} of the Sun to the Galactic center, 
the range of distances covered by our BHB and RRL samples are $\sim$2-115 and 1-120\,kpc, respectively. 

\subsection{Radial velocities}
\label{sec:RVs}

In general, kinematic studies involving RRLs are difficult to conduct, owing to their short variation timescales.
Thus, for RRLs neither coadded spectra nor those from single exposures can provide an unbiased estimation of their center-of-mass velocities necessarily, due to their pulsating nature.  
In fact, the amplitude of a radial velocity curve can easily reach 60\,km\,s$^{-1}$ peak to peak. 
Moreover, the amplitude of the observed velocity variation throughout the star's cycle depends on the lines used to compute the wavelength shift, as they are produced at different depths in the stellar atmosphere (e.g., metallic lines are formed deeper in the atmosphere than Balmer lines). 

By adopting a radial velocity curve (RVC) model, it is possible to estimate the systemic velocity if the phase of a given observation (or a set of observations) is available. 
As described by \citet{Medina2025a}, the DESI DR1 catalog contains two independent estimates of the systemic velocity of its RRLs: one determined using the radial velocity curves of \citet{Braga2021} and one obtained following a Bayesian inference approach. 
For this work, we use the latter, which is based on single-epoch spectra with velocities derived using the RVS pipeline. 
The outcome of this methodology provides RVC models with which it is possible to estimate the systemic velocities of RRLs even when a single observation is available, provided that the phase of the pulsation in which that observation was taken is known.
This, in turn, is an interesting outcome that could be used for a wide range of applications.

For the BHBs in our sample, we directly use the velocities computed from the RVS pipeline applied to their coadded spectra as line-of-sight velocities.

\subsection{Selection cuts}
\label{sec:cuts}

Similar to \citet{Shen2022}, we perform selection cuts to remove stars from our sample with parameters that normally cause issues for our methodology, while preserving a meaningful sample for a MW cumulative mass estimation. 
These cuts are based on the mean signal-to-noise ratio of the stellar spectra, the $R_{\rm GC}$ of the stars, their azimuthal velocity $v_{\phi}$, their velocities in the galactic standard of rest  $v_{\rm GSR}$, 
and their membership to satellite systems. 
More specifically, we adopt the following criteria to select our sample:

\begin{itemize}
 \item A mean signal-to-noise ratio $>3$, to remove stars with unreliable spectroscopic parameters derived with the RVS pipeline (the median signal-to-noise of the BHB and RRL samples are 28 and 32, respectively);

\item $R_{\rm GC} > 50$\,kpc, to remove tracers associated with substructures in the inner halo that could 
bias our mass estimations (as they can introduce changes in the radial density slopes and velocity anisotropy not covered by our model;  see Section~\ref{sec:aurigamass})\footnote{As reported by \citet{Medina2025a}, $\sim35$\% of the RRLs in DESI DR1 are likely associated with GSE and most of them lie within 50\,kpc from the Galactic center. };

\item $|v_{\phi}| < 500$\,km\,s$^{-1}$ and $|v_{\rm GSR}| < 400$\,km\,s$^{-1}$ to remove potentially unbound stars or stars with unreliably large velocities (e.g., with large proper motion errors), which are predominantly located in the Galactic outskirts ($R_{\rm GC}>80$\,kpc);

\item Stars listed as dwarf galaxy or globular cluster members by \citet{Medina2025a} are removed from our sample, limiting our sample to field stars only.   

\end{itemize}

We note that, as part of DESI DR1, several dwarf galaxies and globular clusters were observed, including systems like Draco, Ursa~Major~II, Pal~5, and NGC~5024.
Membership to these systems is based on a 0.5\,arcsec crossmatch with high probability members in the catalogs of stars in globular clusters and dwarf galaxies prepared by \citet{vasiliev_GCs} and \citet{Pace2022}, respectively. 
In addition, we remove stars likely associated with the Sgr stream, as this substructure can be seen as a clear overdensity in our sample (see Figures~\ref{fig:beforeAfter} and ~\ref{fig:dh_dist}). 
Moreover, similar to \citet{Bystroem2024}, we label stars as likely Sgr stream members based on their position relative to the stream and their heliocentric distance. 
In addition, we consider the radial velocity of the stars to assess their association with the Sgr stream, similar to \citet{Erkal2021}. 
For the positions, we transform the stars' equatorial coordinates into a Sgr stream coordinate system $\tilde{\Lambda}_{\odot}$, $\tilde{B}_{\odot}$. 
These coordinates represent stream longitude and latitude, respectively. 
We follow the convention of \citet{Belokurov2014} and adopt a stream longitude $\tilde{\Lambda}_{\odot}$ that increases in the direction of Sgr's motion.
Therefore, $\tilde{\Lambda}_{\odot} = 360-{\Lambda}_{\odot}$, where ${\Lambda}_{\odot}$ is the Sgr stream longitude defined by \citet{Majewski2003}.  
We only consider stars with $|\tilde{B}_{\odot}|<15$\,deg for the distance- and velocity-based assessment of association with Sgr. 
For the distances, we use the values defined in Section~\ref{sec:distances} and label stars within three standard deviations from the distance splines of \citet{Hernitschek2017} (which reach distances out to $\sim100$\,kpc). 
For the radial velocities, we use the values defined in Section~\ref{sec:RVs} and select stars within 3.5 standard deviations from the values reported by \citet{Vasiliev2021} as likely Sgr stream stars.
The limits in distance and velocity standard deviations allow us to remove obvious Sgr-like overdensities from our sample, as shown in Figure~\ref{fig:beforeAfter}. 
Based on these selection limits, a total of 867 BHBs and 600 RRLs meet both the spatial and radial velocity requirements to be considered likely Sgr stars.

\begin{figure}
    \centering    
    \includegraphics[width=0.49\textwidth]{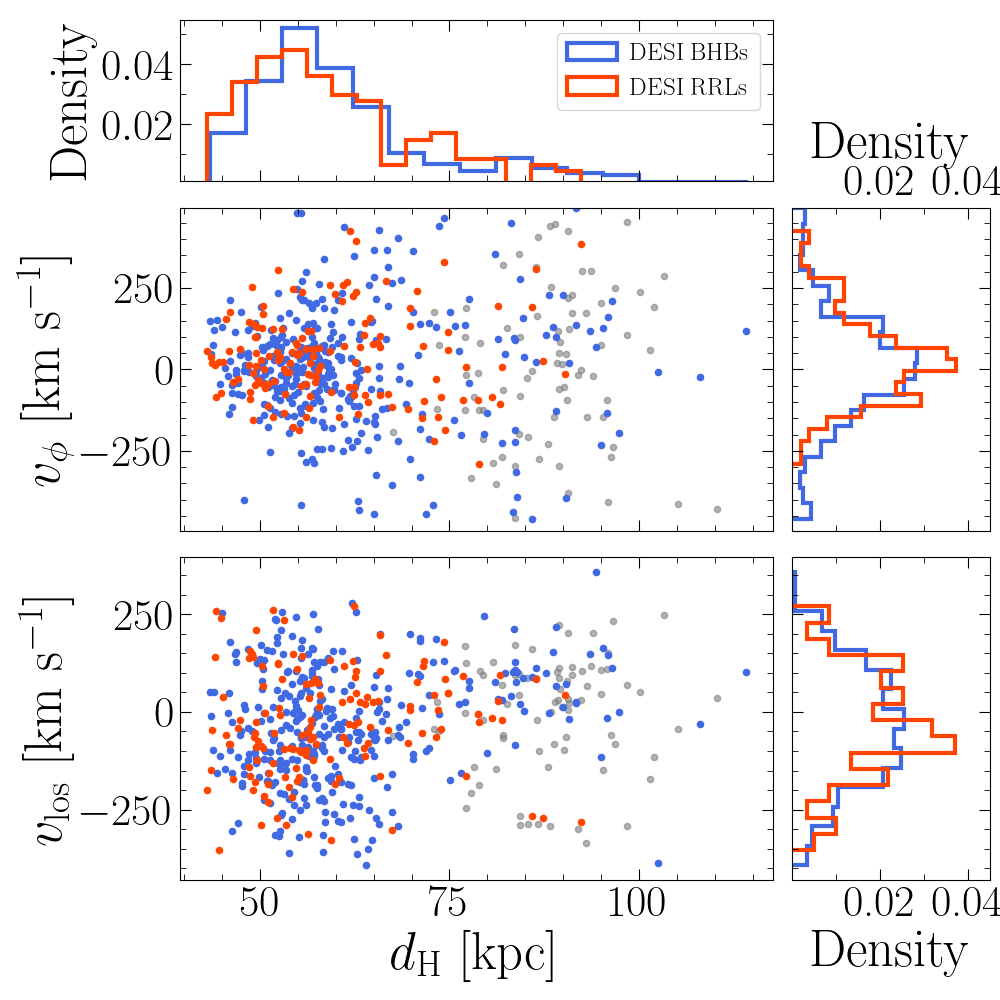}
    \caption {Distribution of heliocentric distances ($d_{\rm H}$), 
    azimuthal velocities ($v_{\phi}$), 
    and line-of-sight velocities ($v_{\rm los}$) of the DESI BHBs and RRLs used in this work for the MW mass determination (i.e., after the selection cuts described in Section~\ref{sec:cuts}). 
    Stars in the RRL sample beyond the limit at which we consider the catalog to be complete ({\sc int\_average\_g} $>20$) are represented by grey markers. 
    }
    \label{fig:dh_vel_distribution_aftercuts}
\end{figure}

The aforementioned cuts result in a clean sample of
335 BHBs with $R_{\rm GC}$ out to 115\,kpc and 
143 RRLs with distances out to 110\,kpc from the Galactic center. 
The distribution of these stars in heliocentric distance, azimuthal velocity ($v_{\phi}$), and line-of-sight velocity ($v_{\rm los}$) is shown in  Figure~\ref{fig:dh_vel_distribution_aftercuts}. 
The figure shows that, for stars at large radii, the selection cuts successfully remove the bulk of the overdensities associated with known substructures in the halo and shown in Figure~\ref{fig:dh_dist}.
In addition, we observe a moderately different velocity distribution between BHBs and RRLs, with BHBs displaying higher $v_{\phi}$ than RRLs at a given distance. 
This difference is likely caused by the BHB sample being intrinsically fainter (and consequently with larger proper motion errors) than the RRLs at a given distance.

The effects of these selection cuts in our mass measurements are described in more detail in Section~\ref{sec:aurigamass}.

\begin{figure*}
    \centering    
    \includegraphics[width=0.43\textwidth]{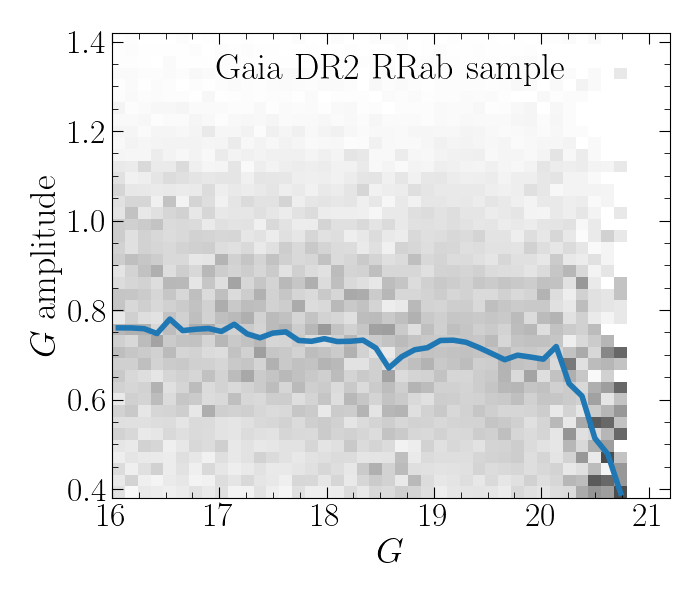}
    \includegraphics[width=0.43\textwidth]{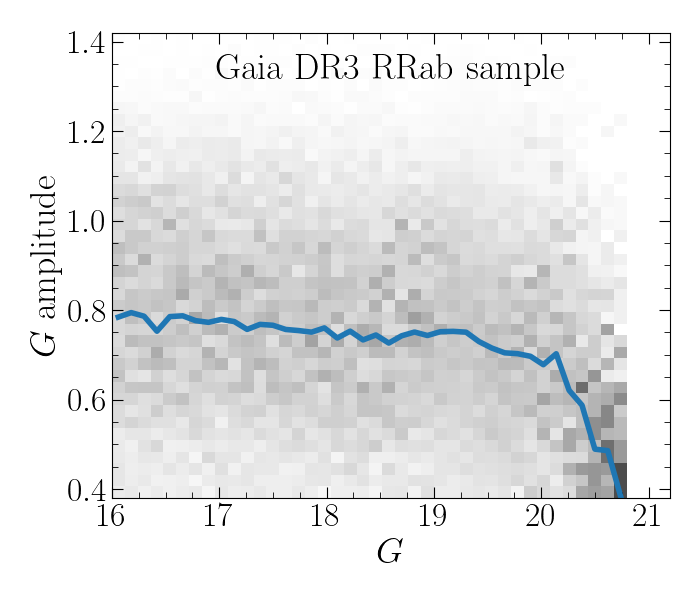}
    \includegraphics[width=0.43\textwidth]{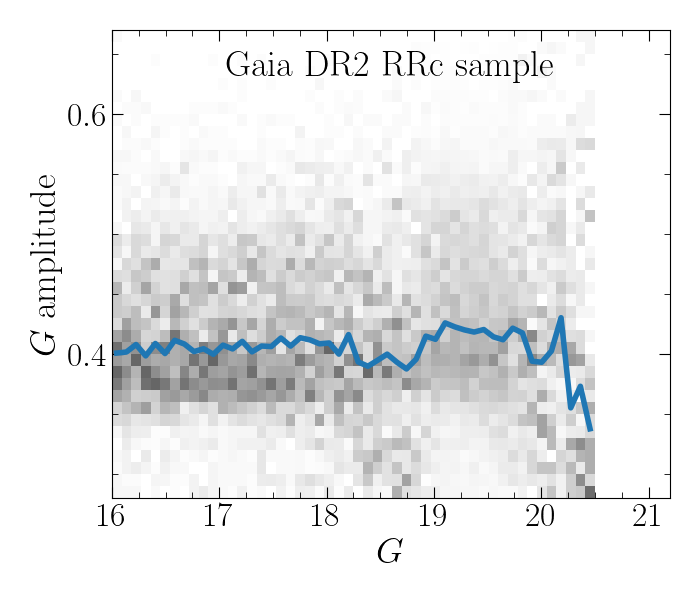}
    \includegraphics[width=0.43\textwidth]{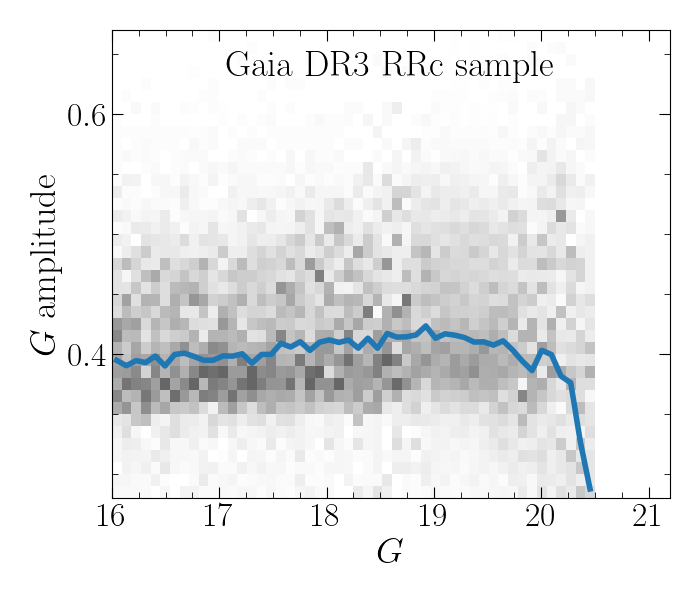}
    \caption{Column-normalized 2D histograms showing the (peak-to-peak) $G-$band amplitude as a function of $G-$band intensity-averaged magnitude for the RRab and RRc stars in the {\it Gaia} DR2 and DR3 RRL catalogs. 
    A blue solid line is used to depict the variation of the median $G-$band amplitude per mean $G$.  
    These panels show a relatively constant amplitude down to a limiting magnitude, after which both RRab and RRc exhibit a marked decrease. 
    This transition occurs at $G\sim20$ and $G\sim20.25$ for RRab and RRc stars, respectively. 
    }
    \label{fig:gaia-completeness}
\end{figure*}

\subsubsection{Completeness}
\label{sec:completeness_rrl}

As explained later in Section~\ref{sec:hierarchicalbayesianmodel}, this model does not explicitly account for the completeness of the catalogs used to infer the mass of the Galaxy. 
Therefore, 
understanding the decrease in completeness of our sample of outer halo tracers with distance is key to assess the limitations of our analysis and to compare our results with previous Galactic mass estimations.  
The completeness of each of our tracer samples can be understood as the product of two selection effects -- the completeness of the catalogs used as inputs for target selection in DESI and the fraction of those targets that were observed as part of DR1.

For BHBs, we assume that the completeness of the input catalog does not vary significantly as a function of magnitude or color. 
Moreover, we find that, for a given target set and program in DESI \citep[stars with primary or secondary priority by DESI targeting strategy, observed in bright or dark conditions; see e.g.,][]{Cooper2023,Koposov2025}, the fraction of BHBs of the input catalog observed in DR1 does not change as a function of magnitude or color, although it can vary between different programs and target sets. 
In Section~\ref{sec:selection_cuts_biases}, we test the effect of using stars from different programs and target sets on our mass estimates. 

For RRL variables, the selection of RRL as observable targets in DESI is based on the {\it Gaia} DR2 \citep[][]{Clementini2019} and the RRL catalog in the PS-1 survey \citet{Sesar2017}. 
Moreover, the DESI DR1 RRL catalog is constructed by crossmatching the (single-epoch) sources observed by DESI with the {\it Gaia} DR3 catalog \citep[][]{Clementini2023}. 

For the {\it Gaia}-based catalogs, the completeness is estimated from the comparison with external RRL catalogs. 
\citet{Clementini2019} recognized that the completeness of the {\it Gaia} DR2 catalog highly depends on the number of epochs available at the time it was compiled (hence, this quantity is a strong function of the source position on the sky).
These authors report recovery percentages that vary between 60 and 82\% when excluding the bulge region from their analysis. 
Similarly, the completeness of the {\it Gaia} DR3 RRL catalog varies between 74 and 94\%, depending on the database it is compared against \citep{Clementini2023}.

Although the aforementioned completeness estimates recognize some of the limitations of the {\it Gaia} RRL catalogs, they use these catalogs in their entirety and additional information is required to address the decrease in completeness as a function of distance.

We estimate the completeness by inspecting the distribution of RRLs light curve amplitudes as a function of mean apparent magnitude (as an indicator of heliocentric distance). 
Under the assumption that there is no inherent strong correlation between the light-curve amplitude of RRLs and their distance (i.e., the light-curve amplitude of halo RRLs is not a function of apparent magnitude), this distribution can be interpreted as a proxy of the survey's completeness. 
In other words, the declination of the observed light curve amplitudes as a function of magnitude could indicate where the RRL detection completeness starts to drop.
We would expect high-amplitude RRab to be observed less frequently than short-amplitude RRab at the catalog's fainter end, 
as single-epoch observations at the faint end of the RRLs light curves (required to reliably identify RRLs) would be affected by large photometric errors or would not be observable at all for {\it Gaia}. 
A similar effect would be expected for RRc, although the large photometric uncertainties at the faint end would also affect the characterization of their lower $G-$band amplitudes, making their magnitude variation harder to distinguish.

Figure~\ref{fig:gaia-completeness} depicts the $G-$band pulsation amplitude ({\sc peak\_to\_peak\_g} in {\it Gaia}) distribution as a function of intensity-averaged  $G$ magnitude ({\sc int\_average\_g}) for RRLs in the {\it Gaia} DR2 and DR3 catalogs. 
We display these distributions for RRab and RRc using column-normalized 2D histograms, highlighting the median $G$ amplitude per magnitude bin to display any existing trends as a function of magnitude. 
The figure shows a roughly constant $G-$amplitude distribution for most of the mean magnitude range and a limit after which the median per bin exhibits a clear decrease. 
For RRab stars, this limit lies around $G\sim20$ for both the {\it Gaia} DR2 and DR3 catalogs and the decrease is smooth.
For RRc stars, a steeper decline is observed at $G\sim20.25$.
Therefore, we consider $G=20$\,mag as a reasonable upper limit for DESI's input catalog completeness. 
This limit corresponds to $d_{\rm H}\sim80$\,kpc when using Equation~\ref{eq:magfeh_relation_rrls} and adopting the median color excess $E_{B-V}$ of the RRLs in our sample (0.04\,mag) from \citet{Schlafly2011} and a representative outer halo [Fe/H] of $-2.0$\,dex.

The distance (magnitude) limit described above resembles the completeness estimation of the PS-1 RRL catalog, for which \citet{Sesar2017} report a completeness of 80\% at 80\,kpc for RRLs with absolute Galactic latitudes $|b| > 15$\,deg. 
Similarly, following a joint probabilistic approach, \citet{Mateu2020} empirically assessed the 2D and 3D completeness of the {\it Gaia} DR2 and the PS-1 RRL catalogs, exploring each survey's full magnitude range.
When analyzing stars at $|b|>20$\,deg, their results indicate that the median completeness of the {\it Gaia} DR2 catalog is of about 92\% and 80\% for RRab and RRc at $G=19.5$ ($\sim65$\,kpc), respectively, and these values drop to 85\% and 67\% at $G=20.0$ ($\sim80$\,kpc). 
For PS-1, the median completeness reaches 80\% and 75\% for RRab and RRc stars at $G=19.5$, and 77\% and 72\% at $G=20.0$.
The work of \citet{Mateu2020} was recently extended to {\it Gaia} DR3 by \citet{Mateu2024}, who found that the completeness drops to $<80$\% and $<70$\% for RRab and RRc stars beyond $\sim$60-80\,kpc.
These results confirm the choice of $G=20.0$ as a sensible limit up to which the DESI input catalogs are reasonably complete. 

The size of the subsample of RRLs used for the MW mass determination reduces to 110 after removing stars with mean $G$ magnitudes fainter than 20. 
In Section~\ref{sec:selection_cuts_biases}, we quantify the effect of including these stars in our mass measurement, as well as the effects of selecting stars from specific programs and DESI target sets.

\section{The model}
\label{sec:model}

We estimate the mass of the MW from the kinematic information of distant tracers observed by DESI using a publicly available  
Python implementation of the code developed by \citet{Shen2022}\footnote{\scalebox{1.0}{Available at \url{https://github.com/al-jshen/gmestan-examples}.}}. 
This methodology allows us to estimate the total mass and the cumulative mass profile of the MW even in the presence of limited or incomplete datasets. 
The code is an extension of the Galactic Mass Estimator (GME) algorithm originally developed by \citet{Eadie2015, Eadie2017}; it takes advantage of Hamiltonian Monte Carlo \citep[HMC;][]{Duane1987} sampling and in particular, its fast and scalable extension the No-U-Turn sampler \citep[NUTS;][]{Hoffman2014}.
This enhancement to GME enables a full probabilistic treatment of all phase-space parameters, that takes into account their uncertainties and includes correlation information between the parameters in a computationally tractable manner.
This, in turn, allows the user to infer thousands of parameters in reasonable timescales. 
This characteristic makes the code particularly useful for handling large datasets with tracers missing reliable proper motion information --- the unknown velocity component of the tracers can be treated as a nuisance parameter in the Bayesian model.
This is particularly useful in our case as a significant fraction of our halo tracers beyond $>50$\,kpc, are missing or have highly uncertain proper motions.

\subsection{Model assumptions, the distribution function, \\and enclosed mass} 

In this model, the Galaxy is considered a spherically symmetric system with a total gravitational potential $\Phi(r)$ as a function of Galactocentric distance $r$ of the form

\begin{equation}
\begin{array}{lc}
\Phi(r) = \Phi_0\ r^{-\gamma}
\end{array}
\label{eq:phi}
\end{equation}
where $\Phi_0$ is a scale factor, and both $\Phi_0$ and $\gamma$ are free parameters.  
Moreover, this methodology assumes a tracer population in dynamical equilibrium with a radial distribution $\rho(r)$ described by a simple power law of slope $\alpha$, that is

\begin{equation}
\begin{array}{lc}
\rho \propto r^{-\alpha}.
\end{array}
\label{eq:rho}
\end{equation}
While this model for the gravitational potential and density profile is overly simplistic, we show through simulations that it can accurately and reliably recover the true mass of the Galaxy (see Section~\ref{sec:mocks}).

It is noteworthy that ample observational evidence indicate the presence of a break in the radial density profile of RRLs and BHBs \citep[see e.g.,][]{Keller2008,Medina2018,Thomas2018,fukushima_stellar_2019,Stringer2021,Chen2023,Medina2024,Amarante2024}. 
This break in their radial distribution is thought to be connected with the origin of the halo \citep[see e.g.,][]{Pillepich2014,deason_galactic_2018}, separating the inner halo, composed of both stars formed in-situ and stars originated from the accretion of satellite systems, from the outer halo, which formed predominantly from the accretion of satellites \citep[e.g.,][]{Naidu2020}. 
The exact position of this break, typically found between 20 and 50\,kpc from the Galactic center, seems to depend on  
the tracer used, the region of the sky surveyed,
and the adopted halo shape (e.g., spherical or triaxial). 
In terms of power-law slopes, the outer halo distribution is commonly found to be steeper than that of the inner regions (see Table 6 in \citealt{Medina2024}). 
We note that a previous study by \citet{Wang2015} provided an expression for the distribution function based on a double power law tracer profile and a NFW potential profile. 
However, these authors still encounter biases in their reported mass estimations, which they attribute to violations of the steady state assumption or a wrong functional form of the distribution function. 
For this work, we consider Equation~\ref{eq:rho} an adequate simplification, as we focus on the most distant stars ($R_{\rm GC} > 50$\,kpc) in the DESI catalog, and these regions can typically be described by simple power laws.

Similar to previous iterations of GME, our model assumes a constant velocity anisotropy parameter $\beta$ for the tracer population. 
This parameter is defined as

\begin{equation}
\begin{array}{lc}
\beta = 1 - \frac{\sigma_\theta^2+\sigma_\phi^2}{2\ \sigma_r^2} 
\end{array}
\label{eq:beta}
\end{equation}
where $\sigma_\theta$, $\sigma_\phi$, and $\sigma_r$ correspond to the azimuthal, polar, and radial  velocity dispersions, respectively.
Thus, by definition, systems with $\beta=0$ are considered isotropic in velocity, whereas $\beta<0$ and $\beta>0$ represent rotationally and radially dominated system, respectively. 
Additionally, the model is restricted to $\alpha>{\rm max(\beta\ (2-\gamma) \ + \gamma/2 ), \ 3)}$, where the first term is adopted from \citet{Evans1997} and the second term is a sensible restriction on the analytical solution (that, in principle, does not represent a physically-motivated restriction). 
As noted by \citet{Shen2022}, however, the derived $\alpha$ does not represent the true physical distribution of halo stars (it is impacted by various biases, as described by these authors) and is primarily considered a nuisance parameter in the model. 
Lastly, $\Phi_0$ is restricted to ensure that the relative energy of the tracers is positive.

In summary, the model has four parameters that define the distribution function of the tracers: $\bm \theta = (\Phi_0,\gamma, \alpha, \beta)$. 
With these parameters, the probability of any particle being at a given location in phase space (i.e., the  distribution function $\mathcal{F}$), is defined as:

\begin{equation}
\begin{array}{lc}
\mathcal{F} (\mathcal{E}, L; \alpha, \beta, \gamma, \Phi_0) = \\\dfrac{ L^{-2\beta} \mathcal{E}^{\frac{\beta(\gamma-2)}{\gamma}+\frac{\alpha}{\gamma}-\frac{3}{2} }\ \Gamma(\frac{\alpha}{\gamma}-\frac{2\beta}{\gamma}+1)   }{  \sqrt{8\pi^32^{-2\beta}} \Phi_0^{ \frac{-2\beta}{\gamma}+\frac{\alpha}{\gamma} }\ \Gamma(\frac{\beta(\gamma-2)}{\gamma}+\frac{\alpha}{\gamma}-\frac{1}{2})\ \Gamma(1-\beta) }, 
\end{array}
\label{eq:DF}
\end{equation}
where $L$ is the total angular momentum of a given Galactic tracer, $\mathcal{E}$ is the relative energy, and $\Gamma$ is the gamma function.
For detailed derivation of $\mathcal{F}$ and a description of its implementation in our model, we refer the reader to \citet{Binney2008}, \citet{Evans1997}, \citet{Eadie2017}, and  \citet{Shen2022}.

Lastly, given the gravitational potential and its dependence on $\Phi_0$ and $\gamma$, the mass enclosed within a radius $r$ can be written as
\begin{equation}
\begin{array}{lc}
\dfrac{M(<r)}{10^{12}\,{\rm M}_\odot } = 2.325 \times 10^{-3}\ \gamma \ (\dfrac{\Phi_0}{10^4\ {\rm km}^2\ {\rm s}^{-2}})(\dfrac{r}{ {\rm kpc}})^{1-\gamma}\ 
\end{array}
\label{eq:mass}
\end{equation}

\noindent where we have used $G=1$ units for computational purposes, leading to the different numerical factors shown above.

We highlight that the enclosed masses derived with our method rely on the validity of the model assumptions in the distance range covered by our data.
Thus, our mass estimates for regions beyond the extent of the data used (i.e., within $R_{\rm GC}=50$\,kpc and beyond $\sim100$\,kpc) are considered extrapolations of our model (as further discussed in Sections~\ref{sec:mocks} and ~\ref{sec:ResultsAndDiscussion}).

\subsection{Hierarchical Bayesian Model}
\label{sec:hierarchicalbayesianmodel}

Although a comprehensive description of the hierarchical Bayesian model employed in this work is provided by \citet{Eadie2017} and \citet{Shen2022}, we reproduce it here for completeness.

In this approach, the measurements of heliocentric distance ($d_{\rm H}$), line-of-sight velocity ($v_{\rm los}$), right ascension ($\alpha_*$\footnote{This notation is used to avoid confusions between the symbol traditionally used for right ascensions, $\alpha$, and the slope of the number density profile in this work.}), declination ($\delta$), and proper motions ($\mu_{\alpha_*}\cos\delta$ and $\mu_\delta$) are treated as samples drawn from distributions that depend on the \textit{true} but unknown values (treated as parameters) and the known measured uncertainties, assuming normal and multivariate normal distributions.

A single tracer has a set of measurements $d_{\rm H}$, $v_{\rm los}$, and $\bm y = (\alpha_*, \delta, \mu_{\alpha_*}\cos\delta, \mu_\delta)$. We assume that the measurements of the first two components are independent of $\bm y$, and that they follow normal distributions with the following means and variances
\begin{equation}
\begin{array}{lc}
D_{\rm H} \sim  \mathcal{N}({d_{\rm H}}^{\dagger}, \Delta {d_{\rm H}}^2 )\\
V_{\rm los} \sim  \mathcal{N}({v_{\rm los}}^{\dagger}, \Delta {v_{\rm los}}^2 )
\end{array}
\end{equation}
where ${d_{\rm H}}^{\dagger}$ and ${v_{\rm los}}^{\dagger}$ repesent the true but unknown values, and $\Delta d_{\rm H}$ and $\Delta v_{\rm los}$ are the known measurement uncertainties used to define the variances (e.g., the standard deviation of $v_{los}$ is $\Delta v_{los}$).

Meanwhile, the vector of measurements $\bm y$ are modeled as draws from a multivariate normal with mean $\bm \vartheta$ and covariance matrix $\bm \Sigma$:
\begin{equation}
\bm Y \sim  \mathcal{N}(\bm\vartheta, \bm \Sigma)
\label{eq:model_multivariatenormal}
\end{equation}
where
\begin{equation}
\begin{array}{lc}
\bm\vartheta = ({\alpha_*}^{\dagger}, \delta^{\dagger}, \mu_{\alpha_*}\cos\delta^{\dagger}, {\mu_\delta}^{\dagger})
\end{array}
\label{eq:model_truevalues}
\end{equation}
represents the true but unknown values of right ascension, declination, and proper motion components, and where $\bm \Sigma$ is the defined by the known measurement uncertainties and their known covariance structure.

Under these assumptions, the probabilities of measuring $d_{\rm H}$, $v_{\rm los}$, and $\bm y$ are:
\begin{equation}
\begin{array}{lc}
p(D_{\rm H}=d_{\rm H}|{d_{\rm H}}^{\dagger}, \Delta {d_{\rm H}})=\frac{1}{\sqrt{2\pi\Delta {d_{\rm H}}^2 }}\exp({ -\frac{(d_{\rm H} - {{d_{\rm H}}^{\dagger}})^2}{2\Delta {d_{\rm H} }^2} })\\
p(V_{\rm los}=v_{\rm los}|{v_{\rm los}}^{\dagger}, \Delta {v_{\rm los}})=\frac{1}{\sqrt{2\pi\Delta {v_{\rm los}}^2 }}\exp({ -\frac{(v_{\rm los} - {{v_{\rm los}}^{\dagger}})^2}{2\Delta {v_{\rm los} }^2} })\\
p(\gm Y=\bm y|\bm \vartheta, \bm \Sigma )= \\{(\frac{1}{2\pi})}^{3} |\bm\Sigma |^{-1/2} \exp{ ( -\frac{1}{2} \ (\bm y-\bm \vartheta)^{T}\ \bm\Sigma^{-1}\ (\bm y-\bm \vartheta) ) }
\end{array}
\label{eq:prob_multivariatenormal}
\end{equation}
and the total likelihood (our measurement model) is then
\begin{equation}
\begin{aligned}
\mathcal{L}(d_{H} & ,v_{los},\bm y|\bm \Delta, \bm \vartheta) = \\
& \prod_{i=1}^N\ p(d_{{\rm H},i}| {d_{{\rm H},i}}^{\dagger}, \Delta d_{{\rm H},i} ) \times p(v_{{\rm los}, i}| {v_{{\rm los}_i}}^{\dagger}, \Delta v_{{\rm los},i} ) \\
& \quad \times p(\bm y_i|\bm \vartheta_i, \bm \Sigma_i)
\end{aligned}
\label{eq:likelihood}
\end{equation}
where $i$ corresponds to $i^{th}$ tracer in the sample. Because the likelihood is in the Heliocentric frame, but the model is in the Galactocentric frame, then a coordinate transformation $h$ is applied to the parameters $({d_{{\rm H},i}}^{\dagger}, {v_{{\rm los}_i}}^{\dagger}, \bm \vartheta$) before passing them to the prior (the distribution function defined by equation~\ref{eq:DF}). 

Thus, the full posterior distribution for a sample containing $N$ tracers is:
\begin{equation}
\begin{aligned}
p(& \bm \theta|\bm y, \bm \Delta) \propto p(\alpha)\ p(\beta)\ p(\Phi_0, \gamma)\\ &\times\prod_{i=1}^N p(d_{{\rm H},i}| {d_{{\rm H},i}}^{\dagger}, \Delta d_{{\rm H},i} ) \ p(v_{{\rm los}, i}| {v_{{\rm los},i}}^{\dagger}, \Delta v_{{\rm los},i} )\ \\ &\times p(\bm y_i|\bm \vartheta_i, \bm \Sigma_i)\  p(h(\bm \vartheta_i)|\bm \theta)\\ 
\end{aligned}
\label{eq:posterior}
\end{equation}
\noindent where the function $h$ is the transformation from Heliocentric to Galactocentric coordinates, 
$p(h(\bm \vartheta_i)|\bm \theta)$ corresponds to the distribution function (Equation~\ref{eq:DF}, the prior), and $p(\alpha), p(\beta)$ and $p(\Phi_0,\gamma)$ are the hyperpriors. The joint hyperprior $p(\Phi_0, \gamma)$ accounts for the previously observed correlation between $\Phi_0$ and $\gamma$ (see Appendix G in \citealt{Shen2022}).

When proper motions are not available for a given tracer, they are treated as nuisance parameters in the model (i.e., they are not measured but inferred). In this case, priors are set for the true but unknown proper motion components, so that the last two terms of equation~\ref{eq:posterior} are written as $ p(\mu_{\alpha_*, i}\cos\delta_i, \mu_{\delta,i})\ p({\alpha_*}_i, \delta_i|{{\alpha_*}_i}^{\dagger}, {\delta_i}^{\dagger}, \bm \Sigma_{{\alpha_*, \delta}_i})\  p(h(\bm \vartheta_i)|\bm \theta_i)$ instead.

We recognize that this approach does not  
explicitly consider the survey's limitations to recover the true phase-space distribution of the tracers (e.g., due to the survey's completeness, footprint, and effects of our selection cuts). 
In Sections~\ref{sec:mocks} and ~\ref{sec:selection_cuts_biases} we discuss the biases in the obtained mass introduced by the most relevant aspects of the survey design and selection cuts made.
Implementing a more complex model that considers the survey's selection function in its design represents 
an avenue of improvement for future work (Slizewski et al. in prep).

For more details of the model design and examples of usage, we refer the reader to \citet{Eadie2015}, \citet{Eadie2016}, \citet{Eadie2017}, \citet{Eadie2019}, and \citet{Slizewski2022}. 
The Python implementation of GME and NUTS and another application to halo stars is further described by \citet{Shen2022}.

The total number of parameters in our model is a combination of those required to model the phase-space information of the tracers (six parameters per tracer) and four parameters used for the distribution function. 
Thus, our model is defined by a total of 664 parameters for the RRL sample (for 110 RRLs) and 2014 parameters for the BHB sample (for 335 BHBs). 
Based on \citet{Shen2022}, we adopt normal distributions as hyperpriors for the parameters corresponding to the distribution function, that is $\alpha\sim\mathcal{N}(4.0,0.1)$, $\beta\sim\mathcal{N}(0.5,0.05)$, and $(\Phi_0,\ \gamma)\sim\mathcal{N}({\bf \mu },{\bf \Sigma})$, 
with ${\bf \mu} = (80, 0.4)$ and

\begin{equation}
    {\bf \Sigma} = \begin{bmatrix}
    47.610 & 0.165 \\
    0.165 & 0.002 \\
\end{bmatrix}.
\end{equation}

We note that, as described by \citet{Shen2022}, modifying the priors on $\alpha$ and $\beta$ in the model does not significantly affect the resulting posterior distributions of $\Phi_0$ and $\gamma$.

We sample the parameter space using four chains with 1000 warm-up draws and 2000 post-warm-up sampling steps each. 
The convergence of the Markov chains is monitored by ensuring that well defined posterior distributions are produced and the metric of convergence R-hat reaches values close to 1.

\begin{figure}
    \centering    
    \includegraphics[width=0.45\textwidth]{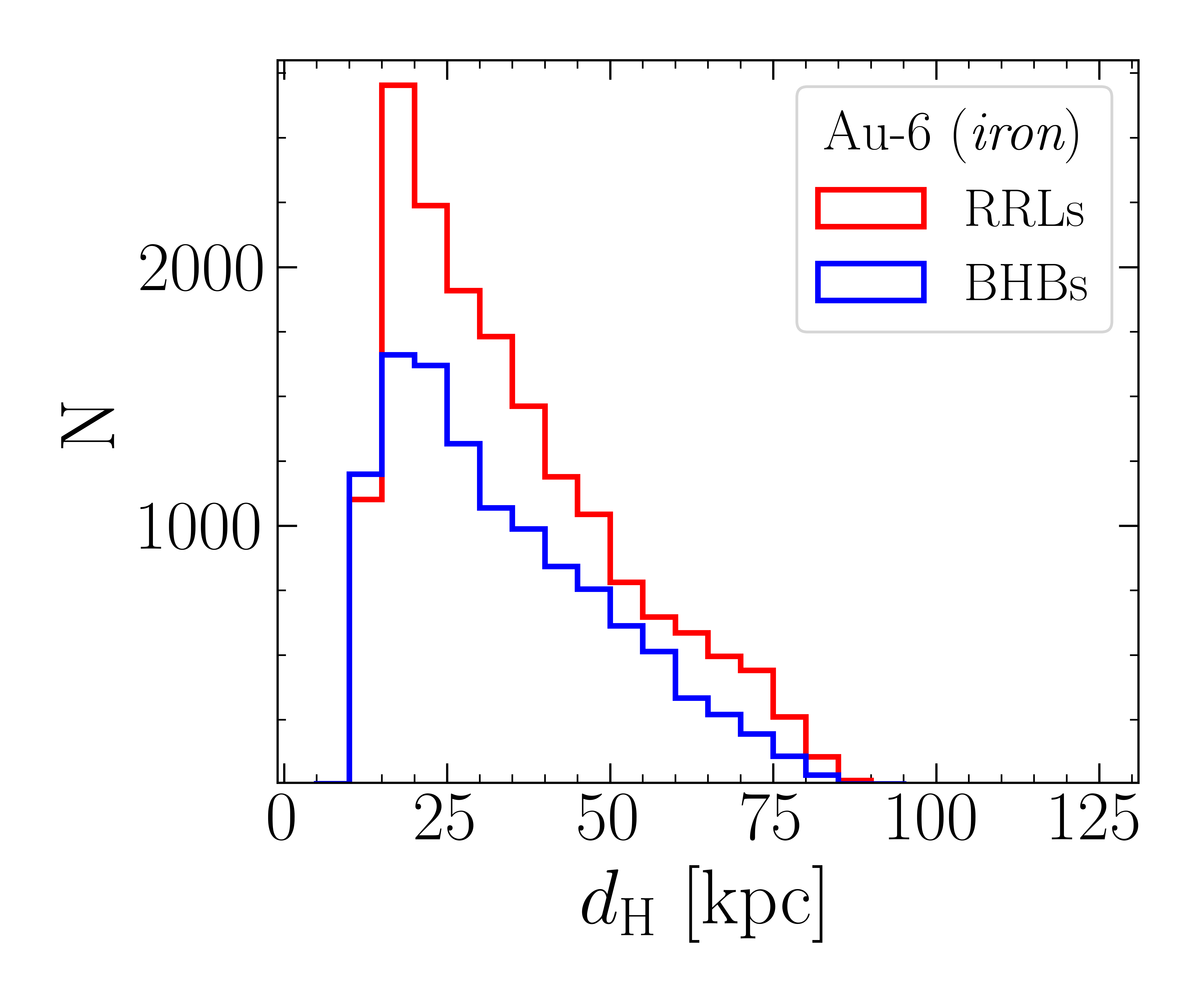}
    \includegraphics[width=0.45\textwidth]{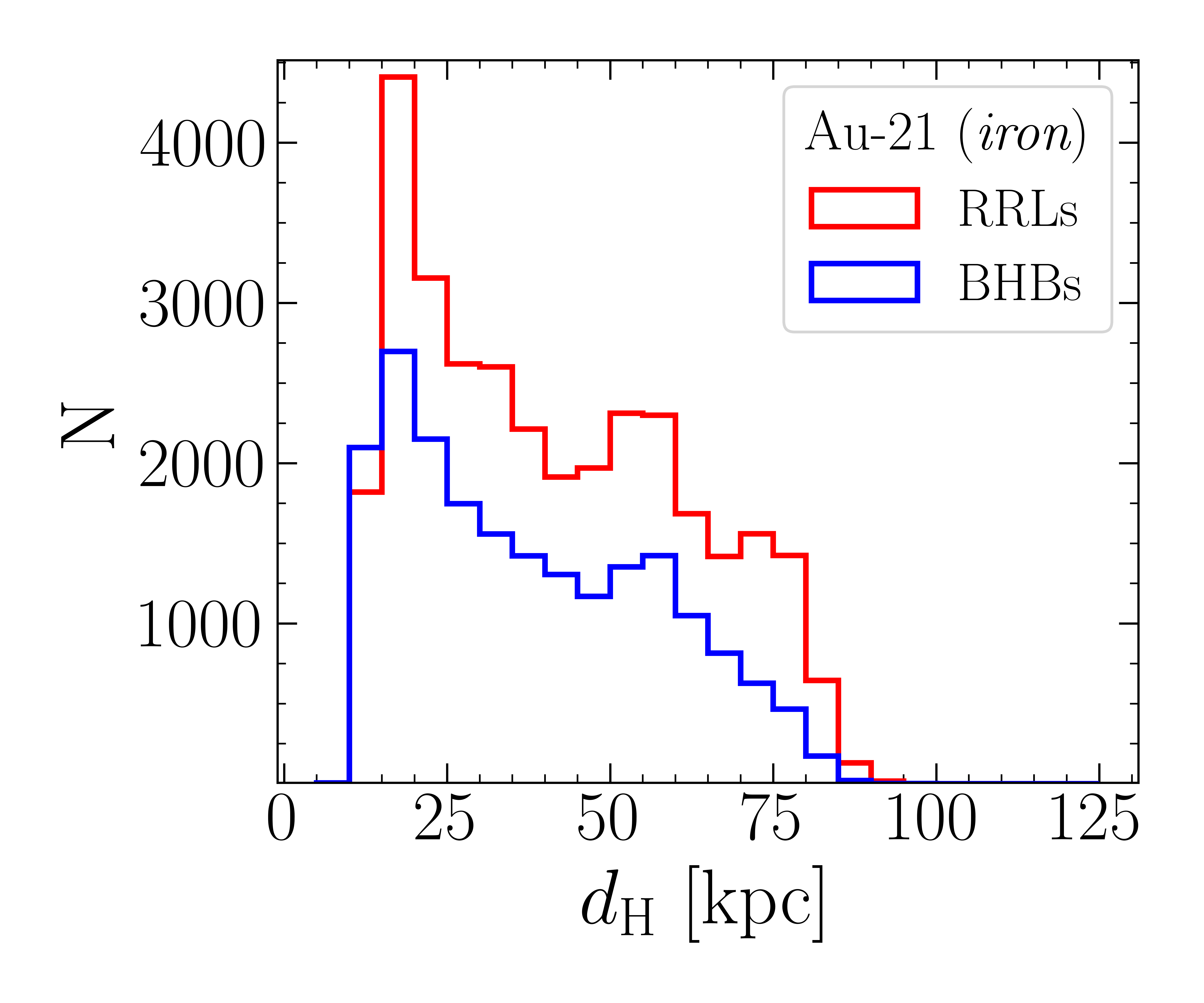}
    \caption{Heliocentric distance distribution of the AuriDESI BHB and RRL catalogs in Au-6 (top) and Au-21 (bottom). 
    The depicted distributions correspond to the stars selected as part of the DESI DR1-like catalogs (the {\it tiled} footprint). 
    Given the design of the AuriDESI simulations (which contain stars with magnitudes $16<g<20$), the stars in these samples cover heliocentric distances ranging from $\sim10$\,kpc out to $\sim$80\,kpc. 
    }
    \label{fig:mocks_dh_distribution}
\end{figure}

\begin{figure*}
    \centering    
    \includegraphics[width=0.495\textwidth]{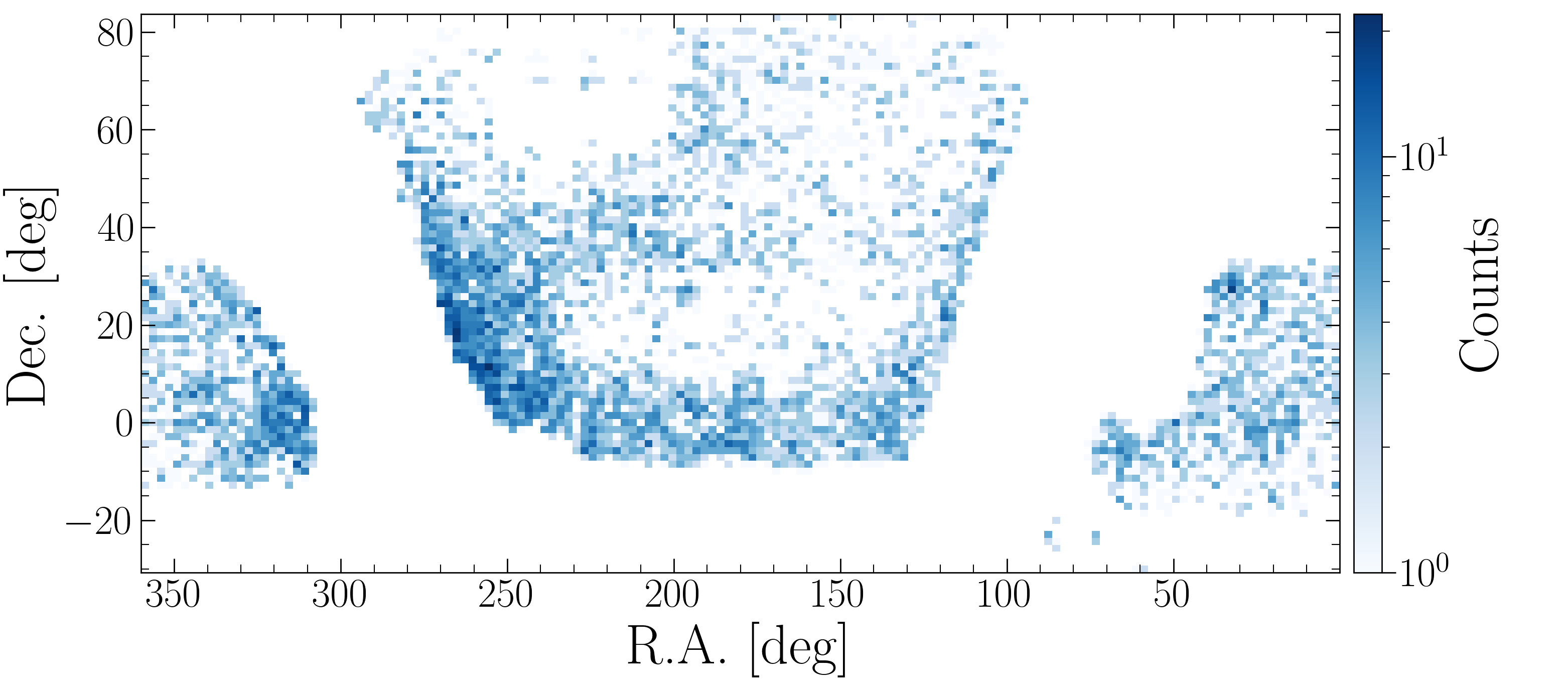}
    \includegraphics[width=0.495\textwidth]{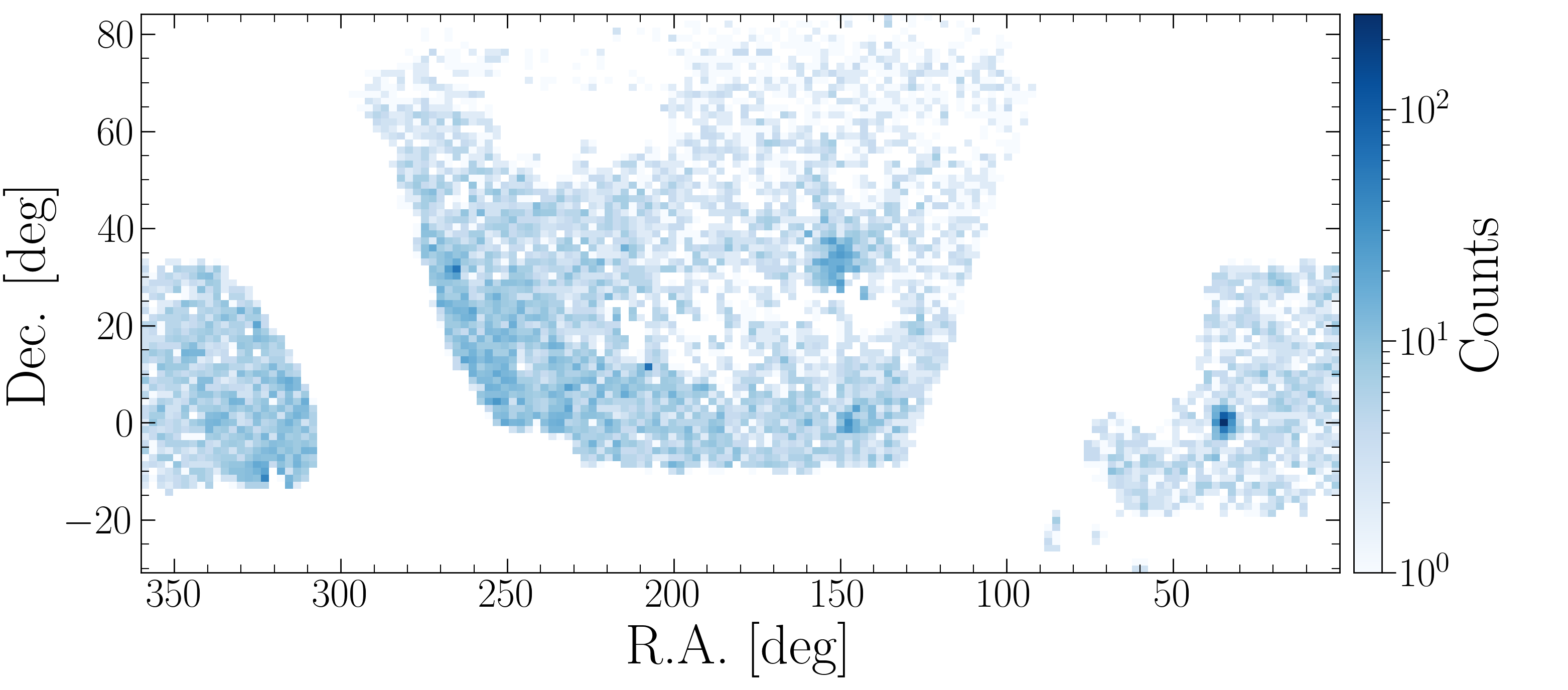}\\
    \includegraphics[width=0.495\textwidth]{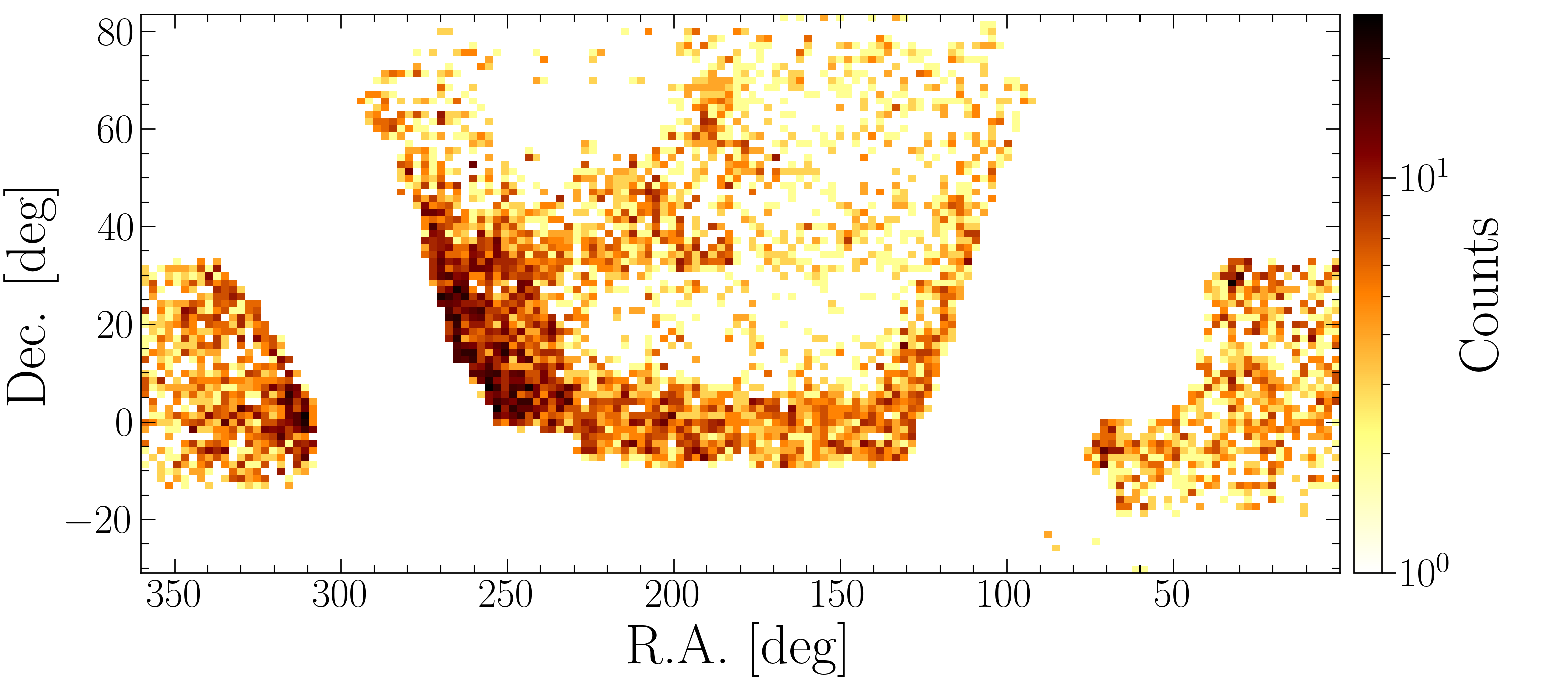}
    \includegraphics[width=0.495\textwidth]{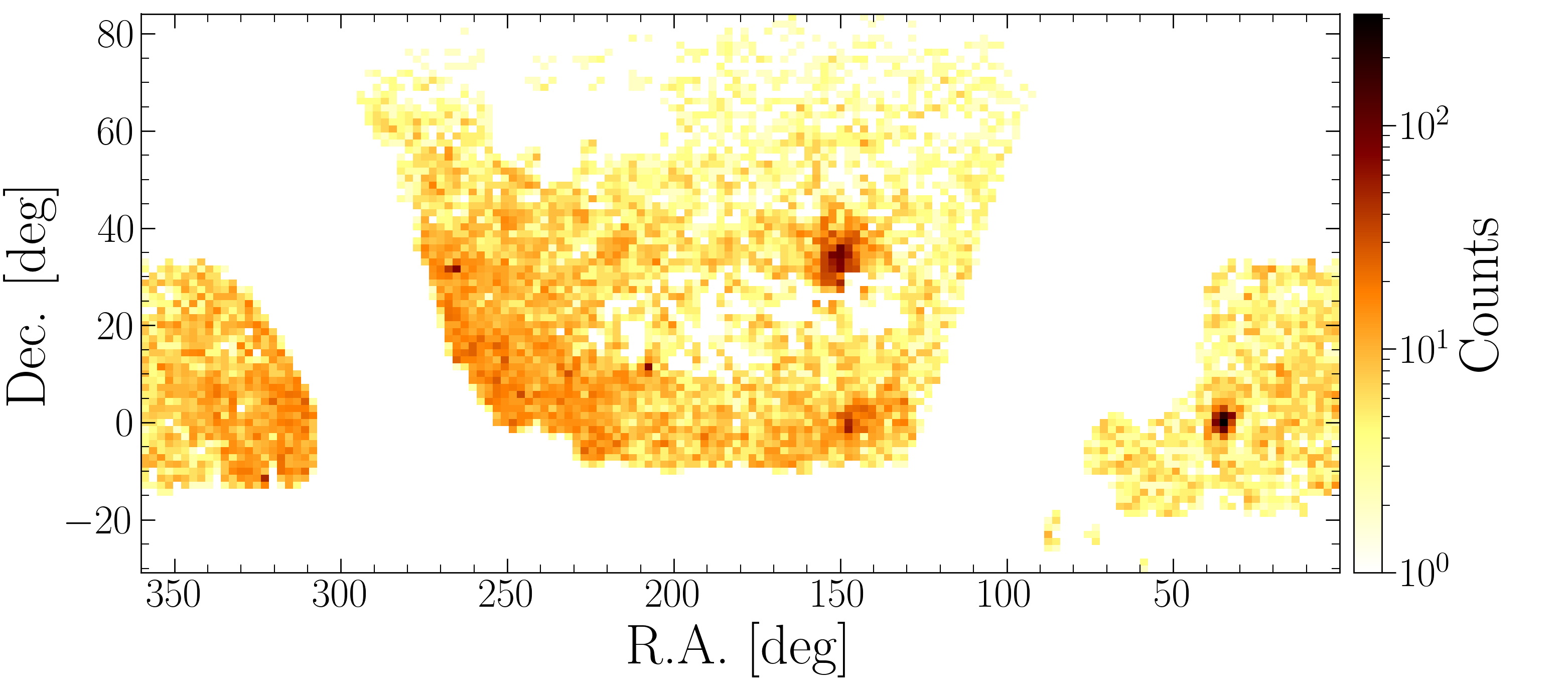}
    \caption{Spatial distribution of the tiled BHB (top panels) and RRL (bottom panels) samples in the AuriDESI mock galaxies Au-6 (left panels) and Au-21 (right panels), shown as 2D histograms. 
    The footprint of these stars shown correspond to the {\it tiled} catalogs, which resemble the spatial coverage of DESI DR1 (see Section~\ref{sec:bhb_rrl_in_mocks}). 
    }
    \label{fig:mocks_distribution}
\end{figure*}

\section{Method validation: Application to AuriDESI}
\label{sec:mocks}

To evaluate the effectiveness of the method to recover the cumulative mass profile of the Galaxy using DESI data, we test our method on simulated data of MW-like halos (with $M_{200}$ between 1--2$\times10^{12}$\,M$_\odot$) from the Auriga project \citep[][]{Grand2017,Grand2024}, adapted to match the key properties of the DESI survey.

\subsection{AuriDESI}
\label{sec:auridesi}

For this work, we employ the AuriDESI mock catalogs \citep[][]{Kizhuprakkat2024}, which mimic the design and selection function of the DESI survey and its observed stars. 
To make AuriDESI, six Auriga halos are mock observed (denoted Au-6, Au-16, Au-21, Au-23, Au-24, and Au-27).
Each halo is observed from four different perspectives within the disk, which differ on the angular position of the solar observer in the plane of the disk of the host galaxy (30, 120, 210, and 300\,deg relative to the bar, where 30\,deg represents the position of the Sun in the MW). 
Here, we focus our attention on two of the Auriga halos, Au-6 and Au-21, both observed at a 30\,deg angle. 
Au-6 has disk structural parameters that closely match those of the MW, while Au-21 has a larger and much more massive central galaxy \citep{Grand2017, Grand2018}.
In addition, Au-6 has a quiet assembly history with few stellar streams or massive disrupting mergers, while Au-21 has a much more active accretion history \citep{Simpson2018,Monachesi2019,Fattahi2020,Pu2025,Riley2025}.
These halos represent extremes for the degree of perturbation that accreted halos of MW-like galaxies can exhibit.

\subsection{BHBs and RRLs in AuriDESI}
\label{sec:bhb_rrl_in_mocks}

Horizontal branch stars in AuriDESI are selected from the `photometric target' catalogs presented in \citet{Kizhuprakkat2024}. 
These catalogs contain all mock stars that meet the MWS selection criteria, i.e., stars within the survey footprint in the magnitude range of DESI's main survey, $16<r<19$ (observed in bright time), and fainter stars in the range $19<r<20$ from DESI's secondary program (observed in dark time).
The extension toward fainter magnitudes allows us to compare the stellar tracers in the mocks and in real data out to larger distances than if we were to use main survey simulated data only, while avoiding magnitudes where incompleteness of these tracers is expected to be non-negligible.
For a detailed description of the targeting strategy of DESI MWS, we refer the reader to the summaries provided by \citet{Cooper2023}, \citet{Kizhuprakkat2024}, and \citet{Koposov2024}.  

Additional cuts are used for the selection of horizontal branch stars in AuriDESI, based on their position in the Kiel diagram ($\log g$ vs. \teff).
We use the PARSEC stellar evolution models \citep[v1.2S;][]{Bressan2012, Tang:2014, Chen:2014, Chen:2015} and always consider the true stellar parameters (i.e.~not convolved with observational uncertainties) since the sources of contamination in the real data are not captured in the AuriDESI methodology (e.g.~variable star light curves that mimic RRL). 
Stars that match the following criteria\footnote{This definition is stricter than in \citet{Kizhuprakkat2024}, which used a broader selection in only $\log g$ and \teff that had contamination from young stars (Riley et al.~in prep).} are considered horizontal branch stars:

\begin{equation}
\begin{aligned}
5500&<T_{\rm eff}\ {\rm [K]}<9750\\
1.4059 + T_{\rm eff}/4250 &< \log g\ {\rm [dex]} < 0.7765 + T_{\rm eff}/3863\\
0.4 &< M_g < 0.85\\
8 &< t_\text{form} {\rm [Gyr]},
\end{aligned}
\label{eq:mock_HB}
\end{equation}
where $M_g$ is the absolute magnitude in the DECam $g$-band and $t_\text{form}$ is the age of the star.

In the PARSEC models, no labels are assigned to distinguish between BHBs and RRLs. 
Thus, mock BHB and RRL samples are drawn adopting a separation at $\teff=7300$\,K, where stars cooler than this limit are then considered RRLs, and hotter stars are labeled as BHBs. 
This limit is selected as it provides a clean separation between BHBs and RRLs in Kiel space in our observed catalogs (see Section~\ref{sec:RRL-BHB-samples}). 
Following this methodology, a total of 20,285 BHBs and 30,327 RRLs are selected from Au-6, and 24,770 BHBs and 39,888 RRLs are selected from Au-21.

\subsubsection{The distance and velocity of BHBs and RRLs in AuriDESI}
\label{sec:mock_distances}

The mock heliocentric distance measurements of BHBs and RRLs are computed using their true heliocentric distances convolved with artificial errors typical of these stars, assuming Gaussian uncertainties around the true values. 
For this, uncertainties of 5\% in the distances of BHBs and RRLs are adopted as a representative errors. 
Figure~\ref{fig:mocks_dh_distribution} shows the heliocentric distance distribution of BHBs and RRLs in Au-6 and Au-21. 
The figure depicts a smooth decline in the number of BHBs and RRLs as a function of distance in the case of Au-6. 
Conversely, the distance distribution of both BHBs and RRLs in Au-21 show overdensities at $d_{\rm H}>50$\,kpc, which reflects the different dynamical state of this halo in comparison with Au-6 and the presence of massive satellites in the mock.

When analyzing the full samples, we find notable discrepancy between the ratio of the BHBs and RRLs in the mock data ($\sim$2:3 in both Au-6 and Au-21), with respect to the ratios observed in the real DESI catalogs ($\sim$1:1; Section~\ref{sec:data}). 
Moreover, the ratio between BHBs and RRLs, both in AuriDESI and in our real observations, is a function of heliocentric distance. 
If only field stars are considered, this ratio is found to be $\sim$3:4 in the mocks at $d_{\rm H}<60$\,kpc and drops to $\sim$2:5 in the range 60-90\,kpc.
For the real data, the BHB-to-RRL ratio remains close to 1:1 over most of the aforementioned distance range, with the exception of $40<d_{\rm H}\ {\rm [kpc]}<70$, where the number of field BHBs is significantly higher than the number of RRLs (5:2).  
We will explore the origin of this discrepancy in future versions of the AuriDESI mock catalogs (Riley et al.~in prep).

The mock line-of-sight velocity measurements are computed convolving the true line-of-sight velocity with assigned uncertainties. 
These uncertainties are designed to mimic the distribution of line-of-sight velocities of the real samples, but do so in different ways.
For BHBs, we assign uncertainties based on $g-r$ vs.~$r$ in a manner similar to \citet{Kizhuprakkat2024}, but account for white dwarfs that artificially inflated the errors for blue stars in that work.
For RRLs, we sample (with replacement) uncertainties from the real DESI RRL catalog, as the uncertainties are primarily a function of number of DESI observations rather than photometry \citep{Medina2025a}.

\begin{figure*}
    \centering    
     \includegraphics[width=0.495\textwidth]{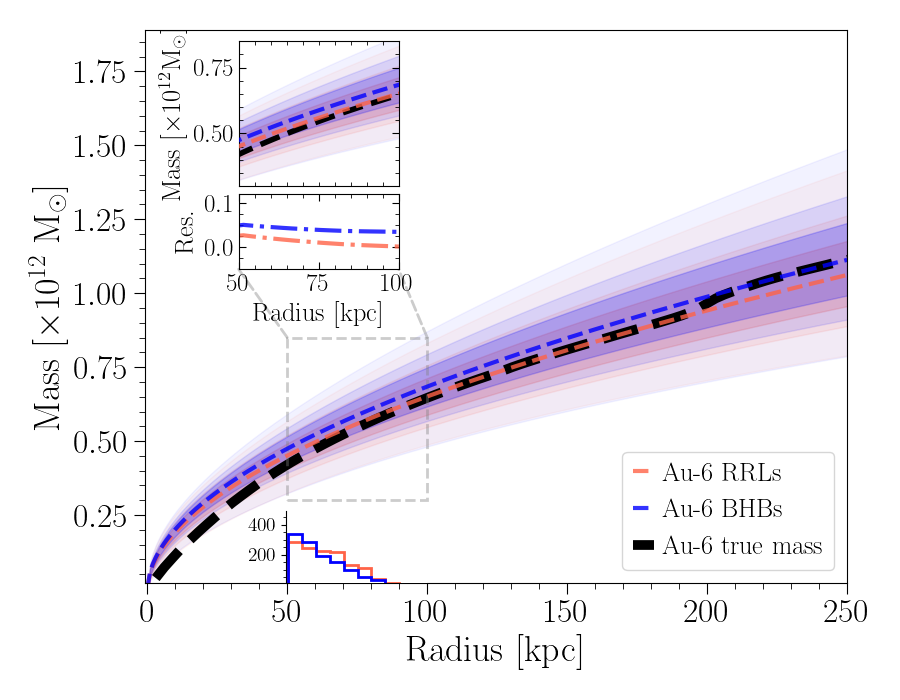}
    \includegraphics[width=0.495\textwidth]{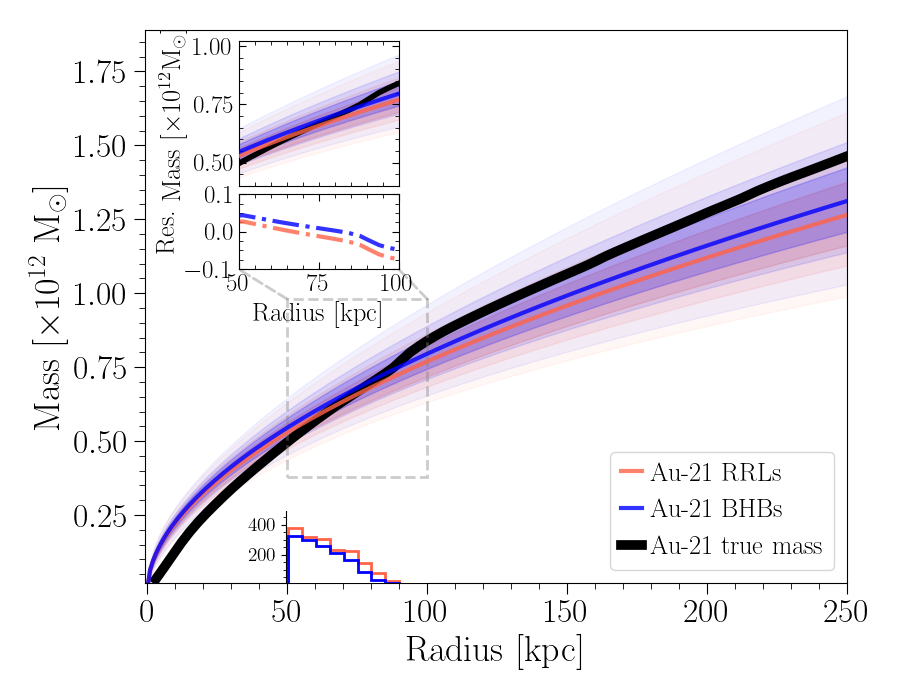}
    \caption{
    Cumulative mass profile of the AuriDESI galaxy Au-6 (left) and Au-21 (right), measured using our GME analysis and the mock BHB and RRL samples (in blue and red, respectively) as described in Section~\ref{sec:aurigamass}. 
   In both panels, shaded regions represent the confidence intervals computed from the 2.5, 12.5, 25.0, 75.0, 87.5, and 97.5 percentiles of the enclosed mass distribution derived from our mock BHBs and RRLs.
    The true enclosed mass of the AuriDESI mocks is shown with a black line in each panel. 
    Enlargements of the region between 50 and 100\,kpc from the center of the galaxies are provided in the top panels of the insets, whereas the differences between the true masses and those derived from the mock samples (Res.) are shown in the bottom panels.
    In this plot, we note the resemblance of the resulting mass profiles shape from BHBs and RRLs at all radii.  
    For Au-6, these masses slightly overestimate the enclosed mass in the range of distances of the tracers ($50<R_{\rm GC}\ {\rm [kpc]}<80$).  
    Beyond $\sim200$\,kpc the model is unable to reproduce sudden increase of the mock galaxy's mass. 
    For Au-21 our results match the mock true mass at $50<R_{\rm GC}\ {\rm [kpc]}<80$ at a $0.10\times10^{12}$\,M$_\odot$ level, with larger inconsistencies outside that distance range (although within our reported confidence intervals).
    }
    \label{fig:mocks_cmp_Rcut50}
\end{figure*}

\subsubsection{Sample selection and spatial coverage of BHBs and RRLs in AuriDESI}
\label{sec:mock_coverage}

To quantify the performance of our model, we use two subsets of AuriDESI BHBs and RRLs per mock simulated galaxy; one restricted to the regions observed at the time of DESI DR1 (i.e., mock stars that fall within the DR1 tiles; hereafter, the {\it tiled} sample) and one covering a more homogeneous footprint (hereafter, the {\it full} mock sample). 
Additionally, to mimic the number of BHBs and RRLs  and their number ratio in our real catalog\footnote{The number of observed BHBs and RRLs are a factor of three and five smaller than those in Au-6, respectively, and a factor of five and six smaller than in Au-21.}, we randomly select 6000 stars from each tiled and full mock catalogs.

Figure~\ref{fig:mocks_distribution} shows the spatial distribution of the BHBs and RRLs in the tiled footprint of Au-6 and Au-21 as 2D histograms. 
For Au-6, the figure shows a largely homogeneous spatial distribution and with no clear overdensities of BHBs nor RRLs. 
For Au-21, the figure displays the presence of massive satellites as overdensities (e.g., at right ascension $\sim$35 and $\sim$150\,deg). 
From the histograms shown in Figure~\ref{fig:mocks_dh_distribution}, it is clear that the number of mock RRLs is larger than that of BHBs for most of the distance range covered (before downsampling). 
Moreover, from the numbers displayed in the histogram (and even after all the selection cuts and downsampling are applied), the amount of both BHBs and RRLs in the mock catalogs is a factor of 3-5 times larger than the one observed in the real catalogs.

Given that the true mass distribution of Au-6 and Au-21 is known from the simulation, we test our ability to recover their CMPs using our GME methodology, applying the same cuts to the mocks as those applied to the real data (see Section ~\ref{sec:cuts}). 
Notably, this entails only considering stars with $R_{\rm GC}>50$\,kpc in the simulation. 
We note that the upper distance limit of the mock coincides with the limit at which our RRL sample is assumed to be complete ($\sim 80$\,kpc). 
Stars in the AuriDESI simulations can be classified as members of intact satellites and/or streams, and stars that are phase-mixed \citep[see e.g.,][]{Riley2025, Shipp2025}. 
For the rest of our analysis, we remove stars that are bound to any satellite whether intact or disrupting.

The removal of stars in satellites only marginally reduces the size of the mock BHB and RRL catalogs in Au-6 (99.7 and 99.6\% of the stars are formed in situ or considered already accreted, respectively), while the raw Au-21 is more affected by this cut (94 and 92\%).
Moreover, as previously mentioned, the number of stars in the simulations significantly exceeds that of the real catalogs. 
To overcome the computational challenge of running the code on a large number of stars, which prevents the Markov chains from converging, we split the catalogs into 16 smaller subsamples (each) containing $\sim$60-215 stars,  selected randomly without repetition of stars. 
We then compute the best model parameters from the median of the combined posterior distributions, and their uncertainties as the 16th and 84th percentiles.

\subsection{Measuring the mass of the Auriga mocks}
\label{sec:aurigamass}

Figure~\ref{fig:mocks_cmp_Rcut50} displays the CMP obtained from the mock BHB and RRL samples, and their comparison with the true cumulative mass of Au-6 and Au-21.
For both simulations, the figure shows the excellent agreement on the measured mass from BHBs and RRLs across the entire 0--250\,kpc distance range. 
In spite of their similarities, however, we observe consistently (but almost inappreciably) higher masses resulting from the BHBs. 
We note that, for Au-6, the method is able to estimate the galactic mass with remarkable accuracy and precision in the distance range of the tracers used  ($50<R_{\rm GC}\ {\rm [kpc]}<90$) and extrapolated out to $\sim200$\,kpc. 
However, the method is unable to recover Au-6's mass increase in the inner regions and beyond 200\,kpc, where its enclosed mass deviates from its overall increase trend. 
For Au-21, our determined mass is overestimated within $\sim 100$\,kpc and underestimated beyond that radius. 
This distance represents the limit between two mass regimes in the halo of Au-21, separated by a break in its true mass profile caused by the accretion of a relatively massive satellite.

To test the effect of including stars at smaller distances from the galactic center, we explore varying the lower limit in galactocentric distance adopted to 30 and 40\,kpc, for both Au-6 and Au-21. 
In addition, we study the effect of changing the adopted spatial distribution by also considering stars in the full homogeneous footprint  (instead of only the footprint of DR1). 
The results of our analysis are provided in Tables~\ref{tab:mock_validation} and \ref{tab:mock_validationH21}.
From the values reported in the tables, we see that the masses inferred from BHBs and RRLs are not only consistent with each other for a given CMP (at all radii), but also when changing the lower limit in distance.

\begin{table*}\small
\caption{
Results of the mass determination of the AuriDESI galaxy Au-6 using GME. 
We report the true mass at different galactocentric radii, and the resulting mass when using our mock BHB and RRL samples.
Our methodology is tested using stars mimicking the spatial distribution of DESI DR1 and stars homogeneously distributed, and varying the lower distance limit of the tracer sample (30, 40, and 50\,kpc).
}
\label{tab:mock_validation}
\begin{center}
\begin{tabular}{|c|c|c|c|c|c|c|c|}
\hline
Dataset & $M$($<50$ kpc) & $M$($<100$ kpc) & $M$($<150$ kpc) & $M$($<200$ kpc) & $M$($<250$ kpc) \\
& [$\times 10^{12}$\,M$_\odot$] & [$\times 10^{12}$\,M$_\odot$] & [$\times 10^{12}$\,M$_\odot$] & [$\times 10^{12}$\,M$_\odot$]  & [$\times 10^{12}$\,M$_\odot$] \\
\hline
True mass & $0.41$ & $0.65$ & $0.81$ & $0.97$ & $1.11$ \\
\hline
\multicolumn{6}{|c|}{{\it Tiled} sample}\\
\hline
$R_{\rm GC}>30$\,kpc (BHBs) & $0.36^{+ 0.08}_{-0.10}$ & $0.54^{+ 0.10}_{-0.14}$ & $0.68^{+ 0.13}_{-0.18}$ & $0.81^{+ 0.16}_{-0.22}$ & $0.92^{+ 0.20}_{-0.27}$ \\
\ \ \ \ \ \ \ \ \ \ \ \ \ \ \ \ \ \ \   (RRLs) & $0.34^{+ 0.06}_{-0.09}$ & $0.52^{+ 0.09}_{-0.14}$ & $0.67^{+ 0.12}_{-0.19}$ & $0.79^{+ 0.16}_{-0.25}$ & $0.90^{+ 0.20}_{-0.32}$ \\
$R_{\rm GC}>40$\,kpc (BHBs) & $0.41^{+ 0.12}_{-0.11}$ & $0.60^{+ 0.16}_{-0.17}$ & $0.75^{+ 0.19}_{-0.22}$ & $0.88^{+ 0.22}_{-0.27}$ & $1.00^{+ 0.25}_{-0.32}$ \\
\ \ \ \ \ \ \ \ \ \ \ \ \ \ \ \ \ \ \   (RRLs) & $0.38^{+ 0.10}_{-0.12}$ & $0.56^{+ 0.13}_{-0.17}$ & $0.70^{+ 0.16}_{-0.22}$ & $0.83^{+ 0.19}_{-0.26}$ & $0.94^{+ 0.22}_{-0.31}$ \\
$R_{\rm GC}>50$\,kpc (BHBs) & $0.47^{+ 0.15}_{-0.12}$ & $0.68^{+ 0.20}_{-0.19}$ & $0.85^{+ 0.25}_{-0.25}$ & $0.99^{+ 0.29}_{-0.31}$ & $1.11^{+ 0.33}_{-0.37}$ \\
\ \ \ \ \ \ \ \ \ \ \ \ \ \ \ \ \ \ \   (RRLs) & $0.45^{+ 0.13}_{-0.12}$ & $0.65^{+ 0.16}_{-0.18}$ & $0.81^{+ 0.20}_{-0.24}$ & $0.94^{+ 0.24}_{-0.30}$ & $1.06^{+ 0.27}_{-0.35}$ \\
\hline
\multicolumn{6}{|c|}{{\it Full} sample}\\
\hline
$R_{\rm GC}>50$\,kpc (BHBs) & $0.48^{+ 0.12}_{-0.12}$ & $0.69^{+ 0.17}_{-0.19}$ & $0.86^{+ 0.21}_{-0.25}$ & $1.01^{+ 0.25}_{-0.31}$ & $1.13^{+ 0.29}_{-0.37}$ \\
\ \ \ \ \ \ \ \ \ \ \ \ \ \ \ \ \ \ \  (RRLs) & $0.46^{+ 0.13}_{-0.11}$ & $0.66^{+ 0.18}_{-0.18}$ & $0.82^{+ 0.22}_{-0.24}$ & $0.96^{+ 0.26}_{-0.30}$ & $1.08^{+ 0.30}_{-0.36}$ \\

\hline
\end{tabular}
\end{center}
\end{table*}

\begin{table*}\small
\caption{
Same as Table~\ref{tab:mock_validation} but for Au-21.}
\label{tab:mock_validationH21}
\begin{center}
\begin{tabular}{|c|c|c|c|c|c|c|c|}
\hline
Dataset & $M$($<50$ kpc) & $M$($<100$ kpc) & $M$($<150$ kpc) & $M$($<200$ kpc) & $M$($<250$ kpc) \\
& [$\times 10^{12}$\,M$_\odot$] & [$\times 10^{12}$\,M$_\odot$] & [$\times 10^{12}$\,M$_\odot$] & [$\times 10^{12}$\,M$_\odot$]  & [$\times 10^{12}$\,M$_\odot$] \\
\hline
True mass & $0.48$ & $0.84$ & $1.07$ & $1.27$ & $1.46$ \\\hline
\multicolumn{6}{|c|}{{\it Tiled} sample}\\
\hline
$R_{\rm GC}>30$\,kpc (BHBs) & $0.50^{+ 0.07}_{-0.09}$ & $0.73^{+ 0.11}_{-0.14}$ & $0.91^{+ 0.16}_{-0.20}$ & $1.07^{+ 0.20}_{-0.25}$ & $1.21^{+ 0.24}_{-0.31}$ \\
\ \ \ \ \ \ \ \ \ \ \ \ \ \ \ \ \ \ \   (RRLs) & $0.49^{+ 0.07}_{-0.08}$ & $0.72^{+ 0.11}_{-0.13}$ & $0.90^{+ 0.15}_{-0.18}$ & $1.05^{+ 0.20}_{-0.23}$ & $1.19^{+ 0.24}_{-0.28}$ \\
$R_{\rm GC}>40$\,kpc (BHBs) & $0.52^{+ 0.09}_{-0.10}$ & $0.76^{+ 0.13}_{-0.16}$ & $0.94^{+ 0.18}_{-0.23}$ & $1.10^{+ 0.22}_{-0.29}$ & $1.25^{+ 0.27}_{-0.34}$ \\
\ \ \ \ \ \ \ \ \ \ \ \ \ \ \ \ \ \ \   (RRLs) & $0.51^{+ 0.08}_{-0.09}$ & $0.74^{+ 0.12}_{-0.15}$ & $0.93^{+ 0.16}_{-0.20}$ & $1.08^{+ 0.20}_{-0.26}$ & $1.22^{+ 0.24}_{-0.31}$ \\
$R_{\rm GC}>50$\,kpc (BHBs) & $0.55^{+ 0.09}_{-0.10}$ & $0.80^{+ 0.14}_{-0.17}$ & $0.99^{+ 0.19}_{-0.23}$ & $1.16^{+ 0.24}_{-0.29}$ & $1.31^{+ 0.28}_{-0.35}$ \\
\ \ \ \ \ \ \ \ \ \ \ \ \ \ \ \ \ \ \   (RRLs) & $0.53^{+ 0.09}_{-0.10}$ & $0.77^{+ 0.14}_{-0.17}$ & $0.96^{+ 0.19}_{-0.23}$ & $1.12^{+ 0.23}_{-0.29}$ & $1.26^{+ 0.28}_{-0.35}$ \\
\hline
\multicolumn{6}{|c|}{{\it Full} sample}\\
\hline
$R_{\rm GC}>50$\,kpc (BHBs) & $0.54^{+ 0.10}_{-0.10}$ & $0.79^{+ 0.15}_{-0.17}$ & $0.98^{+ 0.20}_{-0.23}$ & $1.15^{+ 0.25}_{-0.29}$ & $1.30^{+ 0.29}_{-0.35}$ \\
\ \ \ \ \ \ \ \ \ \ \ \ \ \ \ \ \ \ \  (RRLs) & $0.54^{+ 0.09}_{-0.10}$ & $0.79^{+ 0.13}_{-0.16}$ & $0.98^{+ 0.18}_{-0.23}$ & $1.15^{+ 0.22}_{-0.29}$ & $1.30^{+ 0.27}_{-0.34}$ \\
\hline
\end{tabular}
\end{center}
\end{table*}

A notable finding present in Table~\ref{tab:mock_validation} is the underestimation of Au-6's cumulative mass at a $\sim20$\% level at all radii when the lower limit in distance is adjusted. 
A potential explanation for this disagreement is the distinct rise of Au-6's true CMP at $R_{\rm GC}<50$\,kpc (a rise that differs from the steady mass increase between 50 and 200\,kpc and that of a homogeneous spherical mass distribution). 
In this case, given the significantly higher number of tracers within this limit (see Figure~\ref{fig:mocks_dh_distribution}), the hierarchical Bayesian code attempts to model predominantly the distribution function of these stars and the mass increase at small radii, which results in an underestimation of the enclosed mass beyond 50\,kpc.
In other words, given the assumptions of the model (spherical symmetry, single power law for the density of tracers, and constant velocity anisotropy), it is not possible for the GME method to correctly reproduce the mass distribution of the galaxy across different mass regimes. 
Lastly, in terms of spatial distribution (for the $R_{\rm GC}>50$\,kpc cut), we do not observe noticeable changes in the derived mass depending on the footprint considered ({\it tiled} vs. {\it full} footprint).
This outcome is likely a consequence of the spherical symmetry assumption of the model. 
These results are also visible in Table~\ref{tab:mock_validationH21} for Au-21, and offer a reasonable explanation for the underestimation of its mass $>100$\,kpc and its behavior within 100\,kpc.

\begin{figure*}
    \centering    \includegraphics[width=0.49\textwidth]{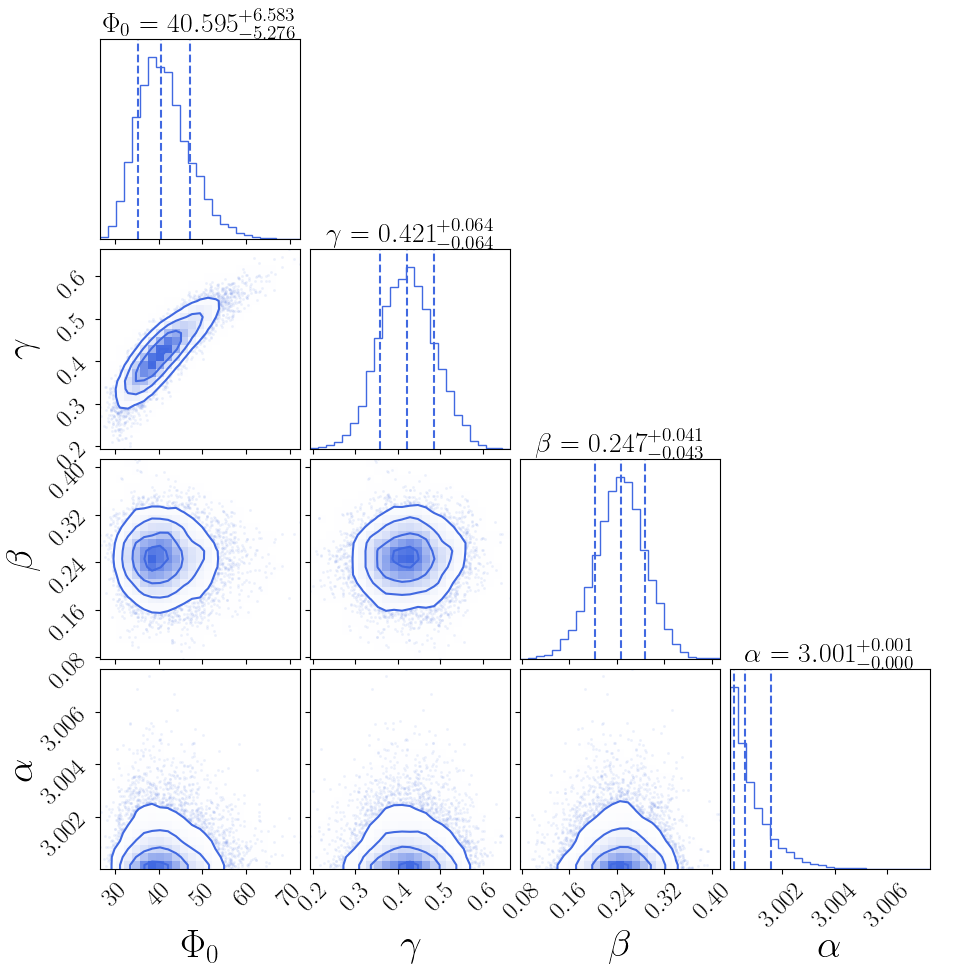}
   \centering    \includegraphics[width=0.49\textwidth]{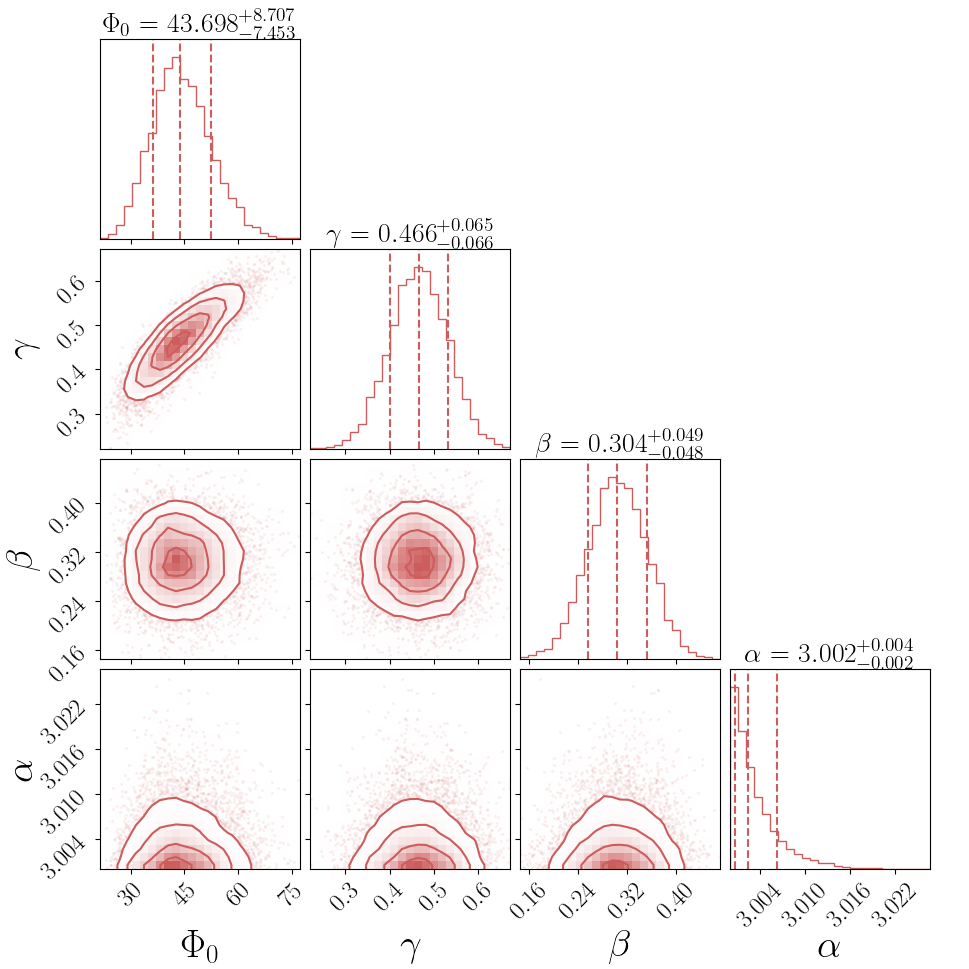}
    \caption{Corner plots showing the marginal posterior probabilities and correlations for the mass estimation model parameters ($\Phi_0$ in units of 10$^4$\,km$^2$\,s$^{-2}$, $\gamma$, $\beta$, $\alpha$), obtained when using the BHB (left) and RRL (right) samples from DESI DR1, as described in Section~\ref{sec:model}. 
    Vertical dashed lines represent the 16th, 50th, and 84th percentiles of the corresponding posterior distributions. 
    }
    \label{fig:cornerplots}
\end{figure*}

\section{Results and discussion} 
\label{sec:ResultsAndDiscussion}

The resulting marginal posterior distributions of the distribution function parameters of our (observed) DESI DR1 BHB and RRL samples, represented with corner plots, are displayed in Figure~\ref{fig:cornerplots}. 
We determine the best-fit values as the 50th percentile of the marginalized distributions and their uncertainties from their 16th and 84th percentiles, and report the resulting values in Table~\ref{tab:best_params}.

From the posterior probabilities we observe a sharply cut skewed distribution of the density profile slope $\alpha$ values for both BHBs and RRLs. 
Interestingly, similar to the results of \citet{Shen2022}, these distributions are clustered close to 3 (the statistical lower bound allowed by the model) and display small variations around this value. 
As noted by these authors, this behavior is likely caused by model mismatches, and represents an avenue of improvement for the underlying model. 
We highlight, however, that this parameter plays a role of a nuisance parameter in the model and does not directly affect the computation of the MW mass.

\begin{figure*}
    \centering    \includegraphics[width=0.49\textwidth]{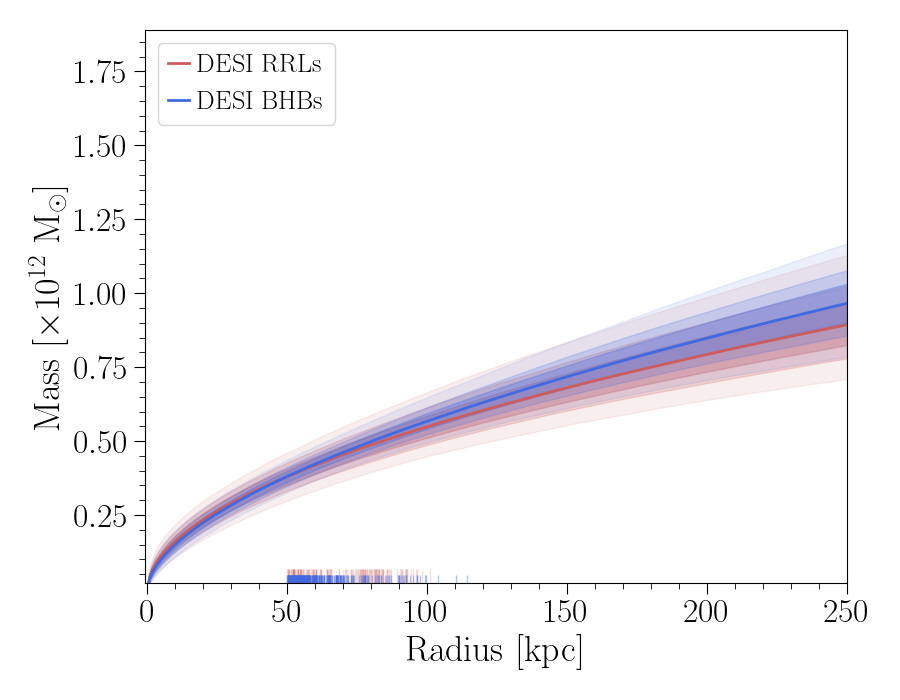}
        \centering    \includegraphics[width=0.49\textwidth]{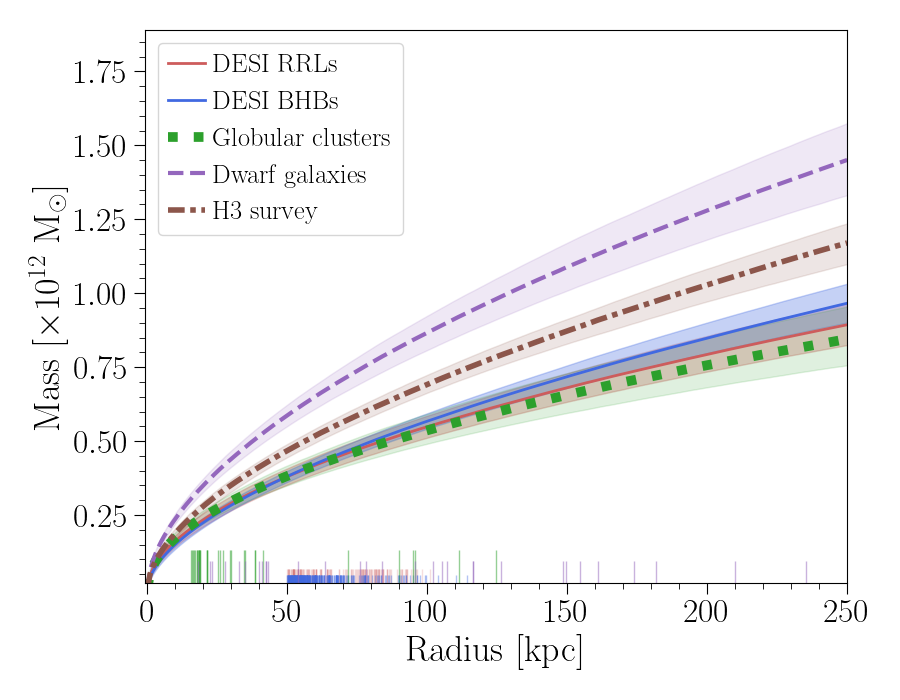}
    \caption{{\it Left}: Cumulative mass profile (Galactocentric distance vs. enclosed mass) of the Galaxy, derived from our BHB and RRL samples. 
    Vertical lines represent the distance of the individual tracers used for our mass modeling. 
    {\it Right}: Same as {\it left} but including the cumulative mass profiles obtained from using the same methodology on different dynamic tracers, namely globular clusters \citep[taken from][]{Eadie2019}, dwarf galaxies \citep[from][]{Slizewski2022}, and giant stars from the H3 survey \citep[from][]{Shen2022}. 
    As in Figure~\ref{fig:mocks_cmp_Rcut50}, shaded regions in the left panel represent the confidence intervals computed from the 2.5, 12.5, 25.0, 75.0, 87.5, and 97.5 percentiles of the enclosed mass distribution for a given tracer. 
    For better readability, we only display the 25.0 and 75.0 percentiles as confidence intervals in the right panel.
    The plots show that, albeit less numerous, globular clusters and dwarf galaxies are able to trace the halo out to greater distances than DESI's BHB and RRL samples. 
    }
    \label{fig:CMP-rrlbhb}
\end{figure*}

Our results also suggest a somewhat isotropic velocity distribution for the outer halo BHBs and RRLs, with $\beta$ values of $0.247_{-0.043}^{+0.041}$ and  $0.304_{-0.048}^{+0.049}$, respectively. 
These results are likely, at least to some extent, affected by the inclusion of stars associated with the GSE merger event, which is characterized by stars on highly radial orbits \citep[see e.g.,][]{Belokurov2018} and that can extend out to large distances in the halo \citep[see e.g.,][]{Medina2025a}. 
Our $\beta$ estimates are broadly comparable to those of \citet{Shen2022} using the same methodology ($\beta=0.36_{-0.04}^{+0.04}$), and those from other works with tracers at large radii including, e.g., the use of K giants from the LAMOST survey and {\it Gaia} ($\beta<0.4$ beyond 50\,kpc; \citealt{Bird2019}), BHBs and K giants from LAMOST, SEGUE, and {\it Gaia} (with $\beta$ values that decline after $R_{\rm GC}>20$\,kpc down to 0.1-0.7 depending on the tracers; \citealt{Bird2021}), and RRLs in DESI accounting for the presence of GSE ($\beta\sim0.35$ beyond $\sim50$\,kpc; \citealt{Medina2025a}).

The posterior probabilities of $\Phi_0$ and $\gamma$ are highly correlated and well defined. 
For BHBs, we find $\Phi_0=40.595_{-5.276}^{+6.583}\times10^4$\,km$^{2}$\,s$^{-2}$, whereas for RRLs the value is slightly larger, $43.698_{-7.453}^{+8.707}\times10^4$\,km$^{2}$\,s$^{-2}$. 
We note that both of our values of $\Phi_0$ are smaller than the values 
derived by \citet{Slizewski2022} and \citet{Shen2022} (56.48 and $50.80_{-6.51}^{+6.98}\times10^4$\,km$^{2}$\,s$^{-2}$, respectively). 
The slope of the gravitational potential $\gamma$ found by these authors ($0.43$) lies between the values we find for BHBs and RRLs in DESI, namely $0.421_{-0.064}^{+0.064}$ and $0.466_{-0.066}^{+0.065}$.

\subsection{Galactic mass estimation with DESI}

The CMPs derived using our BHB and RRL samples, and determined using $\Phi_0$ and $\gamma$, are shown in Figure~\ref{fig:CMP-rrlbhb}. 
The figure also shows confidence intervals computed from the 2.5, 12.5, 25.0, 75.0, 87.5, and 97.5 percentiles of the enclosed mass distribution, and vertical markers representing the distance of each of the tracers used.
We note that, while both BHBs and RRLs are distributed in the 50-100\,kpc region, the CMPs are inferred out to 250\,kpc. 
We stress that our derived mass estimates beyond 100 kpc (based on the validity of our model assumptions at $R_{\rm GC}>100$\,kpc) are an extrapolation of our CMPs within that radius.

\begin{table*}\small
\caption{
Best fit distribution-function parameters obtained with GME when using our BHB and RRL samples. 
}
\label{tab:best_params}
\begin{center}
\begin{tabular}{|c|c|c|c|c|}
\hline
Sample &   $\Phi_0$ &  $\gamma$ &   $\beta$ &  $\alpha$ \\
 &   [$\times10^4$\,km$^2$\,s$^{-2}$] &   &    &  \\
\hline
    BHBs & $40.595_{-5.276}^{+6.583}$ & $0.421_{-0.064}^{+0.064}$ & $0.247_{-0.043}^{+0.041}$ & $3.001_{-0.000}^{+0.001}$ \\
   RRLs & $43.698_{-7.453}^{+8.707}$ & $0.466_{-0.066}^{+0.065}$ & $0.304_{-0.048}^{+0.049}$ & $3.002_{-0.002}^{+0.004}$ \\
\hline
\end{tabular}
\end{center}
\end{table*}

In Table~\ref{tab:bhb_rrl_results}, we report the resulting enclosed mass from the BHB and RRL tracers for Galactocentric distances of 100, 150, 200, and 250\,kpc. 
Tables~\ref{tab:CMPs_appendix_BHBs} and \ref{tab:CMPs_appendix_RRLs} in the Appendix provide more detailed CMPs, also out to 250\,kpc, at 10 kpc increments. 
From the CMPs and the values provided in Table~\ref{tab:bhb_rrl_results}, we find that the enclosed mass obtained from BHBs and RRLs are in good agreement with each other for distances $<50$\,kpc (a region that we do not trace directly with our stars) and between 50 and 100\,kpc. 
For the latter, we obtain an enclosed mass of $0.58_{-0.10}^{+0.12}\times10^{12}$M$_\odot$ and $0.65_{-0.08}^{+0.09}\times10^{12}$M$_\odot$, respectively.
Beyond 100\,kpc, the difference between both profiles becomes more clear overall, although they lie within each other's reported uncertainties.  
In this regard, we highlight the increase in size of the credible intervals beyond 100\,kpc, which reflects the extrapolated nature of the models beyond the extent of our data. 
Indeed, within 250\,kpc we find a total mass of $1.14_{-0.19}^{+0.22}\times10^{12}$M$_\odot$ for BHBs and $0.94_{-0.20}^{+0.25}\times10^{12}$M$_\odot$ for RRLs.

\begin{table*}\small
\caption{
Enclosed mass of the MW derived using the DESI DR1 RRL and BHB catalogs.  
}
\label{tab:bhb_rrl_results}
\begin{center}
\begin{tabular}{|c|cH|c|c|c|c|c|c|c|}
\hline
              Tracer &  $M$($<50$ kpc) & $M$($<75$ kpc) &       $M$($<100$ kpc) &  $M$($<150$ kpc) &       $M$($<200$ kpc) &  $M$($<250$ kpc) &    $M$($<r_{200}$) & $r_{200}$ & N\\
               &        [$\times 10^{12}$\,M$_\odot$] &        [$\times 10^{12}$\,M$_\odot$] &       [$\times 10^{12}$\,M$_\odot$] &       [$\times 10^{12}$\,M$_\odot$] &       [$\times 10^{12}$\,M$_\odot$] &       [$\times 10^{12}$\,M$_\odot$] &   [$\times 10^{12}$\,M$_\odot$] & [kpc] & \\
\hline
          RRL stars & $0.38^{+0.08}_{-0.08}$ & $0.47^{+0.10}_{-0.09}$ & $0.55^{+0.12}_{-0.10}$ & $0.68^{+0.15}_{-0.13}$ & $0.79^{+0.19}_{-0.15}$ & $0.89^{+0.24}_{-0.18}$ & $0.78^{+0.19}_{-0.15}$ & $194.0^{+18.3}_{-15.9}$ & $101$ \\
          BHB stars & $0.38^{+0.06}_{-0.05}$ & $0.48^{+0.07}_{-0.06}$ & $0.57^{+0.08}_{-0.07}$ & $0.72^{+0.12}_{-0.11}$ & $0.85^{+0.16}_{-0.14}$ & $0.97^{+0.20}_{-0.18}$ & $0.85^{+0.16}_{-0.14}$ & $199.4^{+15.1}_{-14.0}$ & $321$ \\
\hline
\end{tabular}
\end{center}
\end{table*}

Similar to previous works, we estimate the mass enclosed within the virial radius $r_{200}$\footnote{$r_{200}$ is defined as the radius at which the local mass density is 200 times larger than the critical density of the universe. }. 
To compute $r_{200}$, we adopt the methodology described in Appendix~\ref{sec:appendix_rvir}, which relies on the assumption of a spherically symmetric mass distribution. 
We then compute the mass enclosed within $r_{200}$, $M_{200}=M(<r_{200})$, for the CMPs of our BHB and RRL samples. 
From these definitions, we obtain $r_{200}=199.4^{+15.1}_{-14.0}$ and $194.0^{+18.3}_{-15.9}$\,kpc for BHBs and RRLs (respectively), from which $M_{200}=0.85^{+0.16}_{-0.14}$ and $0.78^{+0.19}_{-0.15}\times10^{12}$\,M$_\odot$. 
In Section~\ref{sec:comparisonLiterature}, we compare our results with measurements of $M_{200}$ or with the virial mass $M_{\rm vir}$ (i.e., the mass within the virial radius $r_{\rm vir}$) reported by previous works, obtained from different methods and using a variety of tracers. 
For this comparison, we emphasize that our mass estimates out to $r_{\rm vir}$ (and beyond) are the result of the extrapolation of our model, and consequently, should be interpreted with caution.

\subsection{Effect of the selection cuts and other biases}
\label{sec:selection_cuts_biases}

The selection cuts described in Section~\ref{sec:cuts} affect the size of the input catalogs of our method and therefore, their spatial and velocity distributions. 
The most significant cuts, in terms of number of stars removed, are the lower limit on $R_{\rm GC}$ considered, the exclusion of stars associated with the Sgr stream, and the completeness limit for RRLs.

The lower limit in Galactocentric distance in the model also affects significantly the number of BHBs and RRLs tracing the potential. 
In fact, a total of 940 (15\%) and 1224 (19\%) RRLs and BHBs have $R_{\rm GC}$ between 30 and 50\,kpc, in a region that also contains a large number of GSE stars \citep[][]{Medina2025a}.
Modifying the lower limit to 30, 40, and even 45\,kpc, we encounter chains in the NUTS sampling that consistently fail to converge, independent of the sample sizes and priors adopted. 
Potential explanations for this issue include the need to incorporate the gravitational potential of the LMC in the model, which would substantially affect tracers in these regions \citep[see][]{Slizewski2022}, or the need for a model adjustment, to account for the change in slope of number density profiles  of stellar tracers  in the MW halo at Galactocentric distances ranging from $\sim20$-$50$\,kpc (see Table 6 by \citealt{Medina2024} and references therein). 
From our tests in AuriDESI simulations, however, we find that adding stars closer to the galactic center might yield biases of $\sim20$\%. 

Of the full RRL and BHB samples, 10 and 14\% show positions and kinematics consistent with those of Sgr based on the criteria of Section~\ref{sec:cuts}, respectively. 
\citet{Shen2022} reported that the exclusion of stars in Sgr produce a 30\% decrease in the estimated mass with respect to that from the full sample. 
In our case, including Sgr RRLs results in less radial orbits overall ($\beta=0.24_{-0.08}^{+0.07}$) and a 41, 35, and 32\% decrease for the enclosed mass within 100, 200, and 250\,kpc, and 29, 17, and 13\% decrease for BHBs at the same distances. 
We also inspected the effect of excluding every star with $|\tilde{B}_{\odot}|<7.5$\,deg\footnote{This corresponds to the angular distance beyond which the fraction of stars selected from the Sgr cuts in Section~\ref{sec:cuts} drops from $\geq35$\% to $\leq20$\%.}, to avoid biases caused by the selection of Sgr stars based on their distances and velocities.
This experiment leads to a mass estimates $\lesssim10$\% lower than the values reported in Table~\ref{tab:bhb_rrl_results} for both BHBs and RRLs.

A total of 78 RRLs are removed from our sample with our completeness limit cut, at $G=20.0$ ($\sim80$\,kpc).
Including these stars in our sample also makes the orbits of the stars less radial overall ($\beta=0.26_{-0.08}^{+0.07}$) and represents a $\sim$15\% decrease for the enclosed mass within 100, 200, and 250\,kpc. 
Additionally, we quantify the effect of exclusively using stars from specific programs (e.g., stars observed in bright or dark time) and target sets (primary or secondary targets) with different observing efficiencies in DESI. 
These tests lead to 
discrepancies at a 10--30\% level for both the BHB and the RRL sample when compared with the values in Table~\ref{tab:bhb_rrl_results}.

Another bias that might affect our mass measurements relates to the findings of \citet{Han2016}, who studied the intrinsic uncertainty and accuracy in the determination of the masses of MW sized halos from the Aquarius simulation \citep[][]{Springel2008} using a distribution function method. 
These authors found that stars are farther from equilibrium than dark matter particles, and that using them to estimate the masses of halos yield a systematic bias of $\sim$20\%, whereas dark matter particles are an accurate tracer to recover the halo mass (at a 5\% level). 
As stated by \citet{Han2016}, the larger bias when using star tracers is a consequence of them containing the 1\% most bound particles of their host subhalo at the time of star formation, and these stars are the most resistant to tidal stripping and phase mixing\footnote{We note, however, that such bias was not reported by \citet{Hattori2021} when using a distribution function method to derive the mass of a MW-like galaxy from the Fire-2 Latte simulation suite.}. 
Interestingly, \citet{Han2016} showed that there is a net (definite) bias in the distribution function method developed by a previous work using the same simulations and the particle tagging method (\citealt{Wang2015}; see e.g., \citealt{Cooper2010}), even for a model that adopts a double power law tracer density for the distribution function. 
In their work, they report that, separate from this net bias, there is an extra source of stochastic (random) bias associated with any steady-state dynamical model, due to deviation from steady state. 
The level of stochastic bias is small in the Aquarius halos, but is found to be much larger when using halos from the Millennium simulation or the Apostle simulations \citep[see e.g.,][]{Wang2018}, as large as a factor of two using halo stars as tracers.

\begin{figure}
    \centering     
\includegraphics[width=0.49\textwidth]{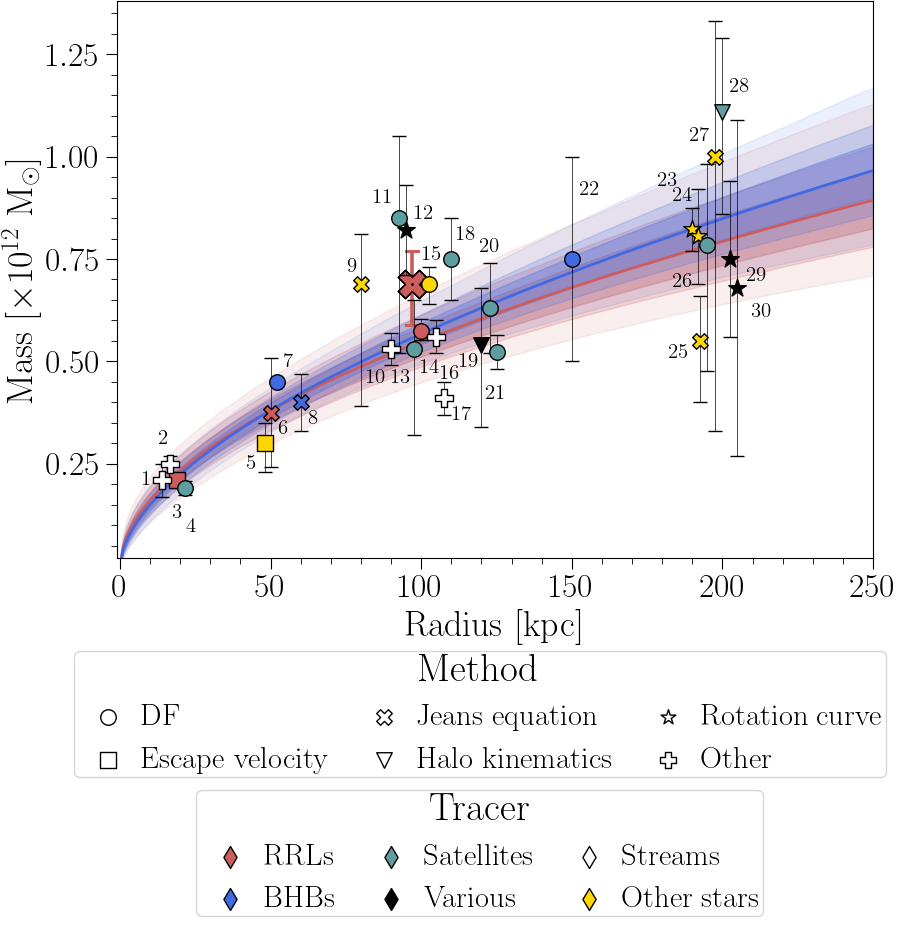}
    \caption{  
    Adapted from \citet{Bobylev2023} and \citet{Bayer2025}. 
    Enclosed mass of the MW derived from different methods and tracers.
    In this case, we highlight specific works from the literature, following:
1--\citep{Kuepper2015},
2--\citep{Malhan2019},
3--\citep{Prudil2022},
4--\citep{Posti2019},
5--\citep{Williams2017},
6--\citep{Ablimit2017},
7--\citep{Williams2015},
8--\citep{Xue2008},
9--\citep{Gnedin2010},
10--\citep{Bayer2025},
11--\citep{Vasiliev2019},
12--\citep{McMillan2017},
13--\citep{Eadie2019},
14--\citep{Hattori2021},
15--\citep{Shen2022},
16--\citep{Vasiliev2021},
17--\citep{Gibbons2014},
18--\citep{CorreaMagnus2022},
19--\citep{Battaglia2005},
20--\citep{Eadie2017},
21--\citep{Eadie2016},
22--\citep{Deason2012b},
23--\citep{Ablimit2020},
24--\citep{Zhou2023},
25--\citep{Bird2022},
26--\citep{Wang2022},
27--\citep{Bird2022},
28--\citep{Sun2023},
29--\citep{Bajkova2016},
30--\citep{Bhattacharjee2014}.
The mass enclosed within 100\,kpc using the Jeans equation method implemented in NIMBLE (see Appendix~\ref{sec:nimble_cmp}) using DESI DR1 RRLs is shown with a significantly larger marker. 
Small shifts in distance were added to the markers at 
20, 50, 100, 125, and 200\,kpc to improve the readability of the comparison. 
These shifts range between 1.5 and 5\,kpc, depending on the level of crowding at a given distance.
    }
    \label{fig:literature-comparison_cmplit}
\end{figure}

\begin{figure*}
    \centering   
    \includegraphics[width=1.02\textwidth]{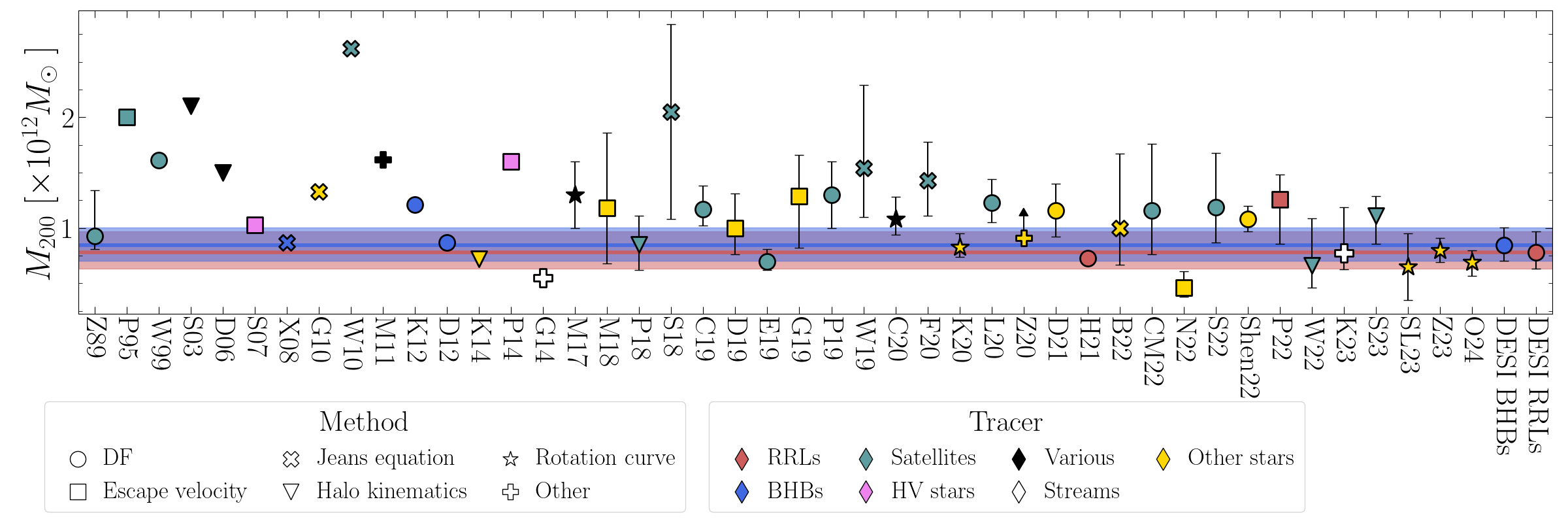}
    \caption{ Comparison of our results (DESI BHBs and DESI RRLs) with MW mass measurements from the literature. We compare our results with those obtained with different methods, namely  distribution function (DF), escape velocity, Jeans equation, halo kinematics, rotation curve, among others. 
    Different colors represent the tracers used in these works, and include  BHBs, RRLs, globular clusters and/or dwarf galaxies (shown jointly as ``Satellites''), high-velocity (HV) stars, a combination of tracers (``Various''), streams, a stars other than RRLs and BHBs. 
    The codes used to refer to previous works is as follows: 
\citet{Zaritsky1989} (Z89; assuming radial orbits), 
\citet{Peebles1995} (P95), 
\citet{Wilkinson1999} (W99), 
\citet{Sakamoto2003} (S03), 
\citet{Dehnen2006} (D03), 
\citet{Smith2007} (S07), 
\citet{Xue2008} (X08), 
\citet{Gnedin2010} (G10),
\citet{Watkins2010} (W10), 
\citet{Mcmillan2011} (M11), 
\citet{Kafle2012} (K12), 
\citet{Deason2012b} (D12), 
\citet{Kafle2014} (K14), 
\citet{Piffl2014} (P14), 
\citet{Gibbons2014} (G14), 
\citet{McMillan2017} (M17), 
\citet{Monari2018} (M18), 
\citet{Patel2018} (P18; result including Sgr),
\citet{Sohn2018} (S18), 
\citet{Callingham2019} (C19), 
\citet{Deason2019} (D19), 
\citet{Eadie2019} (E19), 
\citet{Grand2019} (G19), 
\citet{Posti2019} (P19), 
\citet{Watkins2019} (W19), 
\citet{Cautun2020} (C20), 
\citet{Fritz2020} (F20),
\citet{Karukes2020} (K20),
\citet{Li2020} (L20), 
\citet{Zaritsky2020} (Z20; lower limit based on the timing argument), 
\citet{Deason2021} (D21), 
\citet{Hattori2021} (H21), 
\citet{Bird2022} (B22), 
\citet{CorreaMagnus2022} (CM22), 
\citet{Necib2022b} (N22b),
\citet{Slizewski2022} (S22),
\citet{Shen2022} (Shen22), 
\citet{Prudil2022} (P22),
\citet{Wang2022} (W22), 
\citet{Koposov2023} (K23), 
\citet{Sun2023} (S23), 
\citet{SylosLabini2023} (SL23), 
\citet{Zhou2023} (Z23), 
\citet{Ou2024} (O24).
    }
    \label{fig:literature-comparison}
\end{figure*}

\subsection{The enclosed mass inferred from different tracers}
\label{sec:comparisonmethods}

Figure~\ref{fig:CMP-rrlbhb} shows a comparison of our results with CMPs from works following a similar methodology but with different tracers, namely \citet{Eadie2019} (globular clusters), \citet{Slizewski2022} (dwarf galaxies), and \citet{Shen2022} (K giants). 
Our results confirm that the hierarchical Bayesian framework to estimate the MW mass is sensitive to the tracer choice, as suggested by \citet{Slizewski2022}. 
The plot shows that dwarf galaxies cover a substantially larger range of distances than BHBs and RRLs, and the CMP obtained using them as tracers is significantly more massive than our estimations (at all radii). 
From the findings of \citet{Slizewski2022}, this could be attributed to their choice of input galaxies for the method  (dwarf galaxies in unbound orbits or affected by the potential of a massive LMC can lead to overestimations of the MW mass; see e.g., \citealt{Erkal2020}).
Our BHB results are most similar to the estimation made with K giants as tracers, and our RRL results are most similar to those from globular clusters. 
Both of our CMPs lie in between the enclosed mass derived by \citet{Eadie2019} and \citet{Shen2022} beyond 50\,kpc, consistent with their results within uncertainties.

\subsection{Comparison with the literature} 
\label{sec:comparisonLiterature}

To put our results in a more general context, we first compare our CMPs with mass measurements from the literature determined using different 
methods 
(distribution function, 
escape velocity, 
Jeans equation, 
halo kinematics, 
rotation curve, 
and others) and 
tracers
(BHBs, RRLs, MW satellite systems, among others). 
In Figure~\ref{fig:literature-comparison_cmplit} we display our results together with estimates of the enclosed mass of the MW at different radii from the literature.
The figure shows that mass estimates from the literature are consistent (overall) with each other within 100 kpc (in particular at $\sim 20$ and $\sim 50$\,kpc, where they are in agreement with our CMPs). 
Beyond that radius, there is a larger scatter in the mass estimates. 
We find that, within the distance range covered by the DESI BHBs and RRLs (50--100\,kpc), our results display good agreement with the works by \citet{Vasiliev2021} and \citet{Bayer2025} using the Sgr stream as the tracer and with the distribution function-based estimate of \citet{Eadie2019} using globular clusters.
For distances where our CMPs are extrapolated, and in particular at $>150$\,kpc, 
our results are consistent with works using the rotation curve method (\citealt{Bajkova2016}, \citealt{Ablimit2020}, and \citealt{Zhou2023}), 
and with the mass derived by \citealt{Wang2022} using a action-based distribution function methodology and globular clusters as tracers.

As described above, different methodologies and tracers can lead to discrepant mass estimates at a given radius. 
As a test of the effect of using different methodologies with a similar set of tracers of the potential, we followed an alternative approach to measure the CMP of the MW. 
For this test, we employ the 
publicly available NIMBLE (Non-parametrIc jeans Modeling with B-spLinEs) code \citep{Rehemtulla+22}, a spherical Jeans modeling tool to estimate the cumulative mass distribution of a system from (observationally biased) 6D phase space information.
We find that masses derived with NIMBLE are systematically higher than those from GME (by 25\%, for RRLs with $R_{\rm GC}$ between 50 and 100\,kpc). 
A detailed description of the implementation of this methodology to our data and its results is presented in Appendix~\ref{sec:nimble}.

Figure~\ref{fig:literature-comparison} illustrates the result of the comparison between our estimates of the virial mass $M_{200}$ and those from the literature. 
In the figure, we show a compilation of works dating over three decades back that reflect mass measurements with uncomfortably large scatter and uncertainties. 
Similar to Figure~\ref{fig:literature-comparison_cmplit}, we consider results obtained with a variety of methodologies 
and tracers. 
Our results lie at the lower end of the mass measurements depicted in Figure~\ref{fig:literature-comparison}. 
It is worth noticing that, from the figure, our $M_{200}$ measurements are (overall) lower than masses derived using satellites as tracers (by $\sim40\%$, irrespective of the method), lower than masses derived using the escape velocity method (by about 50\%).
We also note our remarkable agreement with rotation curve-based results, and the consistent $M_{200}$ compared with that recently obtained by  \citet{Koposov2023} through potential modelling of using stellar streams.

When compared with previous distribution function-based results, our measured $M_{200}$ tend to lie at the bottom end of the distribution. 
It should be noted, however, that the choice of a tracer (stellar vs. satellite) seems to play a bigger role in this regard 
(as previously mentioned). 
If we focus the comparison of $M_{200}$ on works using BHBs as tracers exclusively (\citealt{Xue2008} and \citealt{Deason2012b}), we find an excellent agreement (with the exception of \citealt{Kafle2012}, who used a distribution function method). 
This is also the case for the mass within 100\,kpc reported by \citet{Bayer2025}  using BHB stars in the Sgr stream. 
This comparison yields remarkably similar results 
(we find $M(<100\ {\rm kpc})=0.57^{+0.08}_{-0.07}\times10^{12}$\,M$_\odot$ vs. their reported $0.53\pm0.04\times10^{12}$\,M$_\odot$).
For RRLs, we only compare our $M_{200}$ results with the estimate from \citet{Prudil2022}, based on the escape velocity method. 
Since, as mentioned earlier in this section, our estimates tend to be lower than those from the escape velocity method, we do not draw firm conclusions from this comparison.

\section{Conclusions}
\label{sec:conclusions}

In this work, we employed full 6D information of blue horizontal-branch stars (BHBs) and RR Lyrae stars (RRLs) observed in the first year of operations of the Dark Energy Spectroscopic Instrument (DESI) to infer the mass of the Milky Way (MW). 
To this end, we used the Bayesian framework developed by \citet{Eadie2017} and later extended by \citet{Shen2022}, taking full advantage of the ubiquity of BHBs and RRLs in the halo and their nature as precise distance indicators.
With catalogs composed of a total of 5,290 BHBs  and 6,240 RRLs, covering distances out to $\sim100$\,kpc, we combine DESI's line-of-sight velocities with proper motions from the {\it Gaia} space mission and derive a cumulative mass profile of the Galaxy 
extrapolated out to 250\,kpc.

With mock DESI observations based on Auriga simulations, we test the effectiveness (precision and accuracy) of the method in recovering the Galactic mass subject to the survey's selection function (AuriDESI). 
In particular, we use two realizations of AuriDESI that represent different merger histories: one with a relatively quiescent accretion history (Au-6) and one with an ongoing disruption of massive satellites (Au-21). 
Using the same selection cuts as the ones used for DESI's real data, but allowing for variations in the lower distance limit of the tracers, we find that the masses inferred from AuriDESI BHBs and RRLs are largely indistinguishable from each other every time, for every set of input data. 
The output mass is underestimated for most of the setups analyzed that include in the analysis stars in the range $30-50$\,kpc from the simulated galaxy's center (up to $\sim$25\%). 
Using a 50\,kpc lower limit, we are able to recover the enclosed mass of the more quiet halo at a level as good as $\sim1$\% at 100\,kpc, and $\sim5$\% at 250\,kpc.  
For the mock galaxy with a more active recent accretion history, the level of agreement is of $\sim5$\% at 100\,kpc and $\sim10$\% at 250\,kpc. 
We also test the effects of 
incomplete spatial coverage of the tracers
in the inferred mass, and find no significant changes with respect to the cases of full (homogeneous) spatial distribution.

Our sample of selected BHBs and RRLs used to constrain the MW mass consists of 330 and 110 field stars (respectively), 
with Galactocentric distances $>50$\,kpc. 
Within 100\,kpc, we report an enclosed mass of 
$M(<100\ {\rm kpc}) = 0.57^{+0.08}_{-0.07}\times10^{12}$\,M$_\odot$  
and 
$M(<100\ {\rm kpc}) = 0.55^{+0.12}_{-0.10}\times10^{12}$\,M$_\odot$  
when using BHBs and RRLs. 
The best-fit model parameters estimated from our BHB sample yield an extrapolated $M_{200}$ of $0.85^{+0.16}_{-0.14}\times10^{12}$\,M$_\odot$, where $r_{200} = 199.4^{+15.1}_{-14.0}$\,kpc. 
For RRLs, a mass of $M_{200} = 0.78^{+0.19}_{-0.15}\times10^{12}$\,M$_\odot$ 
is enclosed within  $r_{200}=194.0^{+18.3}_{-15.9}$\,kpc.
When compared with previous results, these values tend to be, in general, lower than estimates obtained using dwarf galaxies and/or globular clusters as mass tracers, and in good agreement with masses estimated via rotation curves or stellar stream modelling.
Our inferred cumulative mass profiles are also reasonably consistent with literature estimates   
reported at different distances (in particular for BHBs and RRLs), although with comparatively smaller confidence intervals than works at similar distances (especially for $R_{\rm GC}\geq100$\,kpc). 
These results highlight the importance of spectroscopic surveys in estimating the MW mass, in particular via velocity determinations of a large number of outer halo distance indicators at increasingly larger distances from the Galactic center.

A notable limitation of the current methodology relates to the efficient use of computational resources, to enable the estimation of the Galactic mass with large catalogs without the need to split the tracers into smaller subsamples. 
In this regard, in spite of the significant reduction in computation time of the NUTS algorithm \citep[][]{Shen2022} with respect to the original GME \citep[][]{Eadie2015,Eadie2019}, 
further improvements in the methodology are still required. 
Another way of extending on the mass measurements presented in this work concerns the assumptions of the model, adapting it to allow for higher flexibility in the modeling process. 
For instance, the assumption of spherical symmetry and the use of a single power law to describe the halo's density profile represent an oversimplification that does not account for the existing evidence of a halo that is more complex in shape \citep[triaxial and with a break in the number density profile; see e.g.,][]{Deason2011b, Zinn2014,Medina2018,Thomas2018,han_stellar_2022,Amarante2024,Medina2024}, as mentioned by  \citet{Shen2022}.
Furthermore, improvement in the modeling might include the adoption of a non-constant velocity anisotropy parameter, and a careful treatment of the dynamical state of halo stars (linked to the assumption of dynamical equilibrium) and high velocity stars (e.g., by selecting them as outliers in the model based on the tracers' velocity distribution).
Other biases introduced by the design of our model include the treatment of survey completeness (which can cause differences in the inferred masses of 10--30\%)
and the methods used to remove stars from massive streams (e.g., Sgr, with mass differences of $\lesssim10$\%). 
A more flexible MW model with an improved treatment of the survey's selection function will be implemented in 
future iterations of the code used for this work, in pursuit of the goals of DESI MWS (Slizewski et al. in prep).

In this work, we demonstrated the effectiveness of the positions and velocities of stars measured by DESI to trace the mass of the MW out to large radii.  
Our results confirm the important role of observing large numbers of remote distance indicators in the outer halo, and the power of state-of-the-art methodologies to measure the MW mass and are directly applicable to other large spectroscopic surveys in the current era. 
These results will only be refined with the even larger samples of distance tracers and data-processing improvements forthcoming in future DESI data releases and the advent of complementary large sky spectroscopic surveys mapping the outer Galactic halo \citep[e.g., 4MOST;][]{deJong2019}.

\section*{Data availability}
The data shown in all the figures of this manuscript will be made available in Zenodo (\href{https://doi.org/10.5281/zenodo.16915647}{https://doi.org/10.5281/zenodo.16915647}).

\acknowledgments

We acknowledge highly valuable feedback from Jeff Shen (Princeton University) and Anika Slizewski (University of Toronto). 
G.E.M. and T.S.L. acknowledge financial support from Natural Sciences and Engineering Research Council of Canada (NSERC) through grant RGPIN-2022-04794.
G.E.M. acknowledges support from an Arts \& Science Postdoctoral Fellowship at the University of Toronto. 
G.M.E. acknowledges financial support from NSERC through grant RGPIN-2020-04554. 
A.H.R. was supported by a fellowship funded by the Wenner Gren Foundation, a Research Fellowship from the Royal Commission for the Exhibition of 1851, and by STFC through grant ST/T000244/1.
M.V. and N.R. acknowledge support from  NASA-ATP grants 80NSSC20K0509 and 80NSSC24K0938.   
S.K. acknowledges support from Science \& Technology Facilities Council (STFC) (grant ST/Y001001/1).

This material is based upon work supported by the U.S. Department of Energy (DOE), Office of Science, Office of High-Energy Physics, under Contract No. DE–AC02–05CH11231, and by the National Energy Research Scientific Computing Center, a DOE Office of Science User Facility under the same contract. Additional support for DESI was provided by the U.S. National Science Foundation (NSF), Division of Astronomical Sciences under Contract No. AST-0950945 to the NSF’s National Optical-Infrared Astronomy Research Laboratory; the Science and Technology Facilities Council of the United Kingdom; the Gordon and Betty Moore Foundation; the Heising-Simons Foundation; the French Alternative Energies and Atomic Energy Commission (CEA); the National Council of Humanities, Science and Technology of Mexico (CONAHCYT); the Ministry of Science, Innovation and Universities of Spain (MICIU/AEI/10.13039/501100011033), and by the DESI Member Institutions: \url{https://www.desi.lbl.gov/collaborating-institutions}. Any opinions, findings, and conclusions or recommendations expressed in this material are those of the author(s) and do not necessarily reflect the views of the U. S. National Science Foundation, the U. S. Department of Energy, or any of the listed funding agencies.

The authors are honored to be permitted to conduct scientific research on Iolkam Du’ag (Kitt Peak), a mountain with particular significance to the Tohono O’odham Nation.

For the purpose of open access, the author has applied a Creative Commons Attribution (CC BY) licence to any Author Accepted Manuscript version arising from this submission.

This research has made use of the SIMBAD database, operated at CDS, Strasbourg, France \citep{Simbad}.
This research has made use of NASA’s Astrophysics Data System Bibliographic Services.

{\it Software:} 
{\code{numpy} \citep{numpy},
\code{pandas} \citep{pandas2022}, 
\code{scipy} \citep{2020SciPy-NMeth},
\code{matplotlib} \citep{matplotlib}, 
\code{astropy} \citep{astropy,astropy:2018, astropy:2022}.
}


\bibliography{references} 

\begin{thebibliography}{}
\expandafter\ifx\csname natexlab\endcsname\relax\def\natexlab#1{#1}\fi
\providecommand{\url}[1]{\href{#1}{#1}}
\providecommand{\dodoi}[1]{doi:~\href{http://doi.org/#1}{\nolinkurl{#1}}}
\providecommand{\doeprint}[1]{\href{http://ascl.net/#1}{\nolinkurl{http://ascl.net/#1}}}
\providecommand{\doarXiv}[1]{\href{https://arxiv.org/abs/#1}{\nolinkurl{https://arxiv.org/abs/#1}}}

\bibitem[{{Ablimit} \& {Zhao}(2017)}]{Ablimit2017}
{Ablimit}, I., \& {Zhao}, G. 2017, \apj, 846, 10, \dodoi{10.3847/1538-4357/aa83b2}

\bibitem[{{Ablimit} {et~al.}(2020){Ablimit}, {Zhao}, {Flynn}, \& {Bird}}]{Ablimit2020}
{Ablimit}, I., {Zhao}, G., {Flynn}, C., \& {Bird}, S.~A. 2020, \apjl, 895, L12, \dodoi{10.3847/2041-8213/ab8d45}

\bibitem[{{Allende Prieto} {et~al.}(2006){Allende Prieto}, {Beers}, {Wilhelm}, {Newberg}, {Rockosi}, {Yanny}, \& {Lee}}]{AllendePrieto2006}
{Allende Prieto}, C., {Beers}, T.~C., {Wilhelm}, R., {et~al.} 2006, \apj, 636, 804, \dodoi{10.1086/498131}

\bibitem[{{Amarante} {et~al.}(2024){Amarante}, {Koposov}, \& {Laporte}}]{Amarante2024}
{Amarante}, J. A.~S., {Koposov}, S.~E., \& {Laporte}, C. F.~P. 2024, arXiv e-prints, arXiv:2404.09825, \dodoi{10.48550/arXiv.2404.09825}

\bibitem[{{Anand} {et~al.}(2024){Anand}, {Guy}, {Bailey}, {Moustakas}, {Aguilar}, {Ahlen}, {Bolton}, {Brodzeller}, {Brooks}, {Claybaugh}, {Cole}, {de la Macorra}, {Dey}, {Fanning}, {Forero-Romero}, {Gazta{\~n}aga}, {Gontcho A Gontcho}, {Gutierrez}, {Honscheid}, {Howlett}, {Juneau}, {Kirkby}, {Kisner}, {Kremin}, {Lambert}, {Landriau}, {Le Guillou}, {Manera}, {Meisner}, {Miquel}, {Mueller}, {Niz}, {Palanque-Delabrouille}, {Percival}, {Poppett}, {Prada}, {Raichoor}, {Rezaie}, {Rossi}, {Sanchez}, {Schlafly}, {Schlegel}, {Schubnell}, {Sprayberry}, {Tarl{\'e}}, {Warner}, {Weaver}, {Zhou}, \& {Zou}}]{Anand2024}
{Anand}, A., {Guy}, J., {Bailey}, S., {et~al.} 2024, \aj, 168, 124, \dodoi{10.3847/1538-3881/ad60c2}

\bibitem[{{Apfel} {et~al.}(2025){Apfel}, {McKinnon}, {Rockosi}, {Guhathakurta}, \& {Johnston}}]{Apfel2025}
{Apfel}, M., {McKinnon}, K., {Rockosi}, C.~M., {Guhathakurta}, P., \& {Johnston}, K.~V. 2025, \apj, 988, 225, \dodoi{10.3847/1538-4357/ade43d}

\bibitem[{{Astropy Collaboration} {et~al.}(2013){Astropy Collaboration}, {Robitaille}, {Tollerud}, {Greenfield}, {Droettboom}, {Bray}, {Aldcroft}, {Davis}, {Ginsburg}, {Price-Whelan}, {Kerzendorf}, {Conley}, {Crighton}, {Barbary}, {Muna}, {Ferguson}, {Grollier}, {Parikh}, {Nair}, {Unther}, {Deil}, {Woillez}, {Conseil}, {Kramer}, {Turner}, {Singer}, {Fox}, {Weaver}, {Zabalza}, {Edwards}, {Azalee Bostroem}, {Burke}, {Casey}, {Crawford}, {Dencheva}, {Ely}, {Jenness}, {Labrie}, {Lim}, {Pierfederici}, {Pontzen}, {Ptak}, {Refsdal}, {Servillat}, \& {Streicher}}]{astropy}
{Astropy Collaboration}, {Robitaille}, T.~P., {Tollerud}, E.~J., {et~al.} 2013, \aap, 558, A33, \dodoi{10.1051/0004-6361/201322068}

\bibitem[{{Astropy Collaboration} {et~al.}(2018){Astropy Collaboration}, {Price-Whelan}, {Sip{\H{o}}cz}, {G{\"u}nther}, {Lim}, {Crawford}, {Conseil}, {Shupe}, {Craig}, {Dencheva}, {Ginsburg}, {VanderPlas}, {Bradley}, {P{\'e}rez-Su{\'a}rez}, {de Val-Borro}, {Aldcroft}, {Cruz}, {Robitaille}, {Tollerud}, {Ardelean}, {Babej}, {Bach}, {Bachetti}, {Bakanov}, {Bamford}, {Barentsen}, {Barmby}, {Baumbach}, {Berry}, {Biscani}, {Boquien}, {Bostroem}, {Bouma}, {Brammer}, {Bray}, {Breytenbach}, {Buddelmeijer}, {Burke}, {Calderone}, {Cano Rodr{\'\i}guez}, {Cara}, {Cardoso}, {Cheedella}, {Copin}, {Corrales}, {Crichton}, {D'Avella}, {Deil}, {Depagne}, {Dietrich}, {Donath}, {Droettboom}, {Earl}, {Erben}, {Fabbro}, {Ferreira}, {Finethy}, {Fox}, {Garrison}, {Gibbons}, {Goldstein}, {Gommers}, {Greco}, {Greenfield}, {Groener}, {Grollier}, {Hagen}, {Hirst}, {Homeier}, {Horton}, {Hosseinzadeh}, {Hu}, {Hunkeler}, {Ivezi{\'c}}, {Jain}, {Jenness}, {Kanarek}, {Kendrew}, {Kern}, {Kerzendorf}, {Khvalko}, {King}, {Kirkby}, {Kulkarni},
  {Kumar}, {Lee}, {Lenz}, {Littlefair}, {Ma}, {Macleod}, {Mastropietro}, {McCully}, {Montagnac}, {Morris}, {Mueller}, {Mumford}, {Muna}, {Murphy}, {Nelson}, {Nguyen}, {Ninan}, {N{\"o}the}, {Ogaz}, {Oh}, {Parejko}, {Parley}, {Pascual}, {Patil}, {Patil}, {Plunkett}, {Prochaska}, {Rastogi}, {Reddy Janga}, {Sabater}, {Sakurikar}, {Seifert}, {Sherbert}, {Sherwood-Taylor}, {Shih}, {Sick}, {Silbiger}, {Singanamalla}, {Singer}, {Sladen}, {Sooley}, {Sornarajah}, {Streicher}, {Teuben}, {Thomas}, {Tremblay}, {Turner}, {Terr{\'o}n}, {van Kerkwijk}, {de la Vega}, {Watkins}, {Weaver}, {Whitmore}, {Woillez}, {Zabalza}, \& {Astropy Contributors}}]{astropy:2018}
{Astropy Collaboration}, {Price-Whelan}, A.~M., {Sip{\H{o}}cz}, B.~M., {et~al.} 2018, \aj, 156, 123, \dodoi{10.3847/1538-3881/aabc4f}

\bibitem[{{Astropy Collaboration} {et~al.}(2022){Astropy Collaboration}, {Price-Whelan}, {Lim}, {Earl}, {Starkman}, {Bradley}, {Shupe}, {Patil}, {Corrales}, {Brasseur}, {N{\"o}the}, {Donath}, {Tollerud}, {Morris}, {Ginsburg}, {Vaher}, {Weaver}, {Tocknell}, {Jamieson}, {van Kerkwijk}, {Robitaille}, {Merry}, {Bachetti}, {G{\"u}nther}, {Aldcroft}, {Alvarado-Montes}, {Archibald}, {B{\'o}di}, {Bapat}, {Barentsen}, {Baz{\'a}n}, {Biswas}, {Boquien}, {Burke}, {Cara}, {Cara}, {Conroy}, {Conseil}, {Craig}, {Cross}, {Cruz}, {D'Eugenio}, {Dencheva}, {Devillepoix}, {Dietrich}, {Eigenbrot}, {Erben}, {Ferreira}, {Foreman-Mackey}, {Fox}, {Freij}, {Garg}, {Geda}, {Glattly}, {Gondhalekar}, {Gordon}, {Grant}, {Greenfield}, {Groener}, {Guest}, {Gurovich}, {Handberg}, {Hart}, {Hatfield-Dodds}, {Homeier}, {Hosseinzadeh}, {Jenness}, {Jones}, {Joseph}, {Kalmbach}, {Karamehmetoglu}, {Ka{\l}uszy{\'n}ski}, {Kelley}, {Kern}, {Kerzendorf}, {Koch}, {Kulumani}, {Lee}, {Ly}, {Ma}, {MacBride}, {Maljaars}, {Muna}, {Murphy}, {Norman},
  {O'Steen}, {Oman}, {Pacifici}, {Pascual}, {Pascual-Granado}, {Patil}, {Perren}, {Pickering}, {Rastogi}, {Roulston}, {Ryan}, {Rykoff}, {Sabater}, {Sakurikar}, {Salgado}, {Sanghi}, {Saunders}, {Savchenko}, {Schwardt}, {Seifert-Eckert}, {Shih}, {Jain}, {Shukla}, {Sick}, {Simpson}, {Singanamalla}, {Singer}, {Singhal}, {Sinha}, {Sip{\H{o}}cz}, {Spitler}, {Stansby}, {Streicher}, {{\v{S}}umak}, {Swinbank}, {Taranu}, {Tewary}, {Tremblay}, {de Val-Borro}, {Van Kooten}, {Vasovi{\'c}}, {Verma}, {de Miranda Cardoso}, {Williams}, {Wilson}, {Winkel}, {Wood-Vasey}, {Xue}, {Yoachim}, {Zhang}, {Zonca}, \& {Astropy Project Contributors}}]{astropy:2022}
{Astropy Collaboration}, {Price-Whelan}, A.~M., {Lim}, P.~L., {et~al.} 2022, \apj, 935, 167, \dodoi{10.3847/1538-4357/ac7c74}

\bibitem[{{Bajkova} \& {Bobylev}(2016)}]{Bajkova2016}
{Bajkova}, A.~T., \& {Bobylev}, V.~V. 2016, Astronomy Letters, 42, 567, \dodoi{10.1134/S1063773716090012}

\bibitem[{{Barbosa} {et~al.}(2022){Barbosa}, {Santucci}, {Rossi}, {Limberg}, {P{\'e}rez-Villegas}, \& {Perottoni}}]{Barbosa2022}
{Barbosa}, F.~O., {Santucci}, R.~M., {Rossi}, S., {et~al.} 2022, \apj, 940, 30, \dodoi{10.3847/1538-4357/ac983f}

\bibitem[{{Battaglia} {et~al.}(2005){Battaglia}, {Helmi}, {Morrison}, {Harding}, {Olszewski}, {Mateo}, {Freeman}, {Norris}, \& {Shectman}}]{Battaglia2005}
{Battaglia}, G., {Helmi}, A., {Morrison}, H., {et~al.} 2005, \mnras, 364, 433, \dodoi{10.1111/j.1365-2966.2005.09367.x}

\bibitem[{{Bayer} {et~al.}(2025){Bayer}, {Starkenburg}, {Thomas}, {Martin}, {Helmi}, {Bystr{\"o}m}, {de Boer}, {Fern{\'a}ndez Alvar}, {Gwyn}, {Ibata}, {Jablonka}, {Kordopatis}, {Matsuno}, {McConnachie}, {Medina}, {S{\'a}nchez-Janssen}, \& {Sestito}}]{Bayer2025}
{Bayer}, M., {Starkenburg}, E., {Thomas}, G.~F., {et~al.} 2025, arXiv e-prints, arXiv:2502.17319, \dodoi{10.48550/arXiv.2502.17319}

\bibitem[{{Belokurov} {et~al.}(2018){Belokurov}, {Erkal}, {Evans}, {Koposov}, \& {Deason}}]{Belokurov2018}
{Belokurov}, V., {Erkal}, D., {Evans}, N.~W., {Koposov}, S.~E., \& {Deason}, A.~J. 2018, \mnras, 478, 611, \dodoi{10.1093/mnras/sty982}

\bibitem[{{Belokurov} {et~al.}(2014){Belokurov}, {Koposov}, {Evans}, {Pe{\~n}arrubia}, {Irwin}, {Smith}, {Lewis}, {Gieles}, {Wilkinson}, {Gilmore}, {Olszewski}, \& {Niederste-Ostholt}}]{Belokurov2014}
{Belokurov}, V., {Koposov}, S.~E., {Evans}, N.~W., {et~al.} 2014, \mnras, 437, 116, \dodoi{10.1093/mnras/stt1862}

\bibitem[{{Bhardwaj} {et~al.}(2023){Bhardwaj}, {Marconi}, {Rejkuba}, {de Grijs}, {Singh}, {Braga}, {Kanbur}, {Ngeow}, {Ripepi}, {Bono}, {De Somma}, \& {Dall'Ora}}]{Bhardwaj2023}
{Bhardwaj}, A., {Marconi}, M., {Rejkuba}, M., {et~al.} 2023, \apjl, 944, L51, \dodoi{10.3847/2041-8213/acba7f}

\bibitem[{{Bhattacharjee} {et~al.}(2014){Bhattacharjee}, {Chaudhury}, \& {Kundu}}]{Bhattacharjee2014}
{Bhattacharjee}, P., {Chaudhury}, S., \& {Kundu}, S. 2014, \apj, 785, 63, \dodoi{10.1088/0004-637X/785/1/63}

\bibitem[{{Binney} \& {Tremaine}(2008)}]{Binney2008}
{Binney}, J., \& {Tremaine}, S. 2008, {Galactic Dynamics: Second Edition}

\bibitem[{{Bird} {et~al.}(2019){Bird}, {Xue}, {Liu}, {Shen}, {Flynn}, \& {Yang}}]{Bird2019}
{Bird}, S.~A., {Xue}, X.-X., {Liu}, C., {et~al.} 2019, \aj, 157, 104, \dodoi{10.3847/1538-3881/aafd2e}

\bibitem[{{Bird} {et~al.}(2021){Bird}, {Xue}, {Liu}, {Shen}, {Flynn}, {Yang}, {Zhao}, \& {Tian}}]{Bird2021}
---. 2021, \apj, 919, 66, \dodoi{10.3847/1538-4357/abfa9e}

\bibitem[{{Bird} {et~al.}(2022){Bird}, {Xue}, {Liu}, {Flynn}, {Shen}, {Wang}, {Yang}, {Zhai}, {Zhu}, {Zhao}, \& {Tian}}]{Bird2022}
---. 2022, \mnras, 516, 731, \dodoi{10.1093/mnras/stac2036}

\bibitem[{{Bland-Hawthorn} \& {Gerhard}(2016)}]{Bland-Hawthorn2016}
{Bland-Hawthorn}, J., \& {Gerhard}, O. 2016, \araa, 54, 529, \dodoi{10.1146/annurev-astro-081915-023441}

\bibitem[{{Bobylev} \& {Baykova}(2023)}]{Bobylev2023}
{Bobylev}, V.~V., \& {Baykova}, A.~T. 2023, Astronomy Reports, 67, 812, \dodoi{10.1134/S1063772923080024}

\bibitem[{{Braga} {et~al.}(2021){Braga}, {Crestani}, {Fabrizio}, {Bono}, {Sneden}, {Preston}, {Storm}, {Kamann}, {Latour}, {Lala}, {Lemasle}, {Prudil}, {Altavilla}, {Chaboyer}, {Dall'Ora}, {Ferraro}, {Gilligan}, {Fiorentino}, {Iannicola}, {Inno}, {Kwak}, {Marengo}, {Marinoni}, {Marrese}, {Mart{\'\i}nez-V{\'a}zquez}, {Monelli}, {Mullen}, {Matsunaga}, {Neeley}, {Stetson}, {Valenti}, \& {Zoccali}}]{Braga2021}
{Braga}, V.~F., {Crestani}, J., {Fabrizio}, M., {et~al.} 2021, \apj, 919, 85, \dodoi{10.3847/1538-4357/ac1074}

\bibitem[{{Bressan} {et~al.}(2012){Bressan}, {Marigo}, {Girardi}, {Salasnich}, {Dal Cero}, {Rubele}, \& {Nanni}}]{Bressan2012}
{Bressan}, A., {Marigo}, P., {Girardi}, L., {et~al.} 2012, \mnras, 427, 127, \dodoi{10.1111/j.1365-2966.2012.21948.x}

\bibitem[{{Bystr{\"o}m} {et~al.}(2024){Bystr{\"o}m}, {Koposov}, {Lilleengen}, {Li}, {Bell}, {Beraldo e Silva}, {Carrillo}, {Chandra}, {Gnedin}, {Han}, {Medina}, {Najita}, {Riley}, {Thomas}, {Valluri}, {Aguilar}, {Ahlen}, {Allende Prieto}, {Brooks}, {Claybaugh}, {Cole}, {Dawson}, {de la Macorra}, {Font-Ribera}, {Forero-Romero}, {Gazta{\~n}aga}, {Gontcho}, {Kremin}, {Lambert}, {Landriau}, {Le Guillou}, {Levi}, {Meisner}, {Miquel}, {Moustakas}, {Prada}, {P{\'e}rez-R{\`a}fols}, {Rossi}, {Sanchez}, {Schlegel}, {Schubnell}, {Sprayberry}, {Tarl{\'e}}, {Weaver}, \& {Zou}}]{Bystroem2024}
{Bystr{\"o}m}, A., {Koposov}, S.~E., {Lilleengen}, S., {et~al.} 2024, arXiv e-prints, arXiv:2410.09149, \dodoi{10.48550/arXiv.2410.09149}

\bibitem[{{Callingham} {et~al.}(2019){Callingham}, {Cautun}, {Deason}, {Frenk}, {Wang}, {G{\'o}mez}, {Grand}, {Marinacci}, \& {Pakmor}}]{Callingham2019}
{Callingham}, T.~M., {Cautun}, M., {Deason}, A.~J., {et~al.} 2019, \mnras, 484, 5453, \dodoi{10.1093/mnras/stz365}

\bibitem[{{Cardelli} {et~al.}(1989){Cardelli}, {Clayton}, \& {Mathis}}]{Cardelli1989}
{Cardelli}, J.~A., {Clayton}, G.~C., \& {Mathis}, J.~S. 1989, \apj, 345, 245, \dodoi{10.1086/167900}

\bibitem[{{Catelan} \& {Smith}(2015)}]{Catelan2015}
{Catelan}, M., \& {Smith}, H.~A. 2015, {Pulsating Stars}

\bibitem[{{Cautun} {et~al.}(2020){Cautun}, {Ben{\'\i}tez-Llambay}, {Deason}, {Frenk}, {Fattahi}, {G{\'o}mez}, {Grand}, {Oman}, {Navarro}, \& {Simpson}}]{Cautun2020}
{Cautun}, M., {Ben{\'\i}tez-Llambay}, A., {Deason}, A.~J., {et~al.} 2020, \mnras, 494, 4291, \dodoi{10.1093/mnras/staa1017}

\bibitem[{{Chen} {et~al.}(2023){Chen}, {Li}, {Wang}, {Gong}, {Chen}, \& {Long}}]{Chen2023}
{Chen}, A., {Li}, Z., {Wang}, Y., {et~al.} 2023, \mnras, 525, 3075, \dodoi{10.1093/mnras/stad2296}

\bibitem[{{Chen} {et~al.}(2015){Chen}, {Bressan}, {Girardi}, {Marigo}, {Kong}, \& {Lanza}}]{Chen:2015}
{Chen}, Y., {Bressan}, A., {Girardi}, L., {et~al.} 2015, \mnras, 452, 1068, \dodoi{10.1093/mnras/stv1281}

\bibitem[{{Chen} {et~al.}(2014){Chen}, {Girardi}, {Bressan}, {Marigo}, {Barbieri}, \& {Kong}}]{Chen:2014}
{Chen}, Y., {Girardi}, L., {Bressan}, A., {et~al.} 2014, \mnras, 444, 2525, \dodoi{10.1093/mnras/stu1605}

\bibitem[{{Clementini} {et~al.}(2019){Clementini}, {Ripepi}, {Molinaro}, {Garofalo}, {Muraveva}, {Rimoldini}, {Guy}, {Jevardat de Fombelle}, {Nienartowicz}, {Marchal}, {Audard}, {Holl}, {Leccia}, {Marconi}, {Musella}, {Mowlavi}, {Lecoeur-Taibi}, {Eyer}, {De Ridder}, {Regibo}, {Sarro}, {Szabados}, {Evans}, \& {Riello}}]{Clementini2019}
{Clementini}, G., {Ripepi}, V., {Molinaro}, R., {et~al.} 2019, \aap, 622, A60, \dodoi{10.1051/0004-6361/201833374}

\bibitem[{{Clementini} {et~al.}(2023){Clementini}, {Ripepi}, {Garofalo}, {Molinaro}, {Muraveva}, {Leccia}, {Rimoldini}, {Holl}, {Jevardat de Fombelle}, {Sartoretti}, {Marchal}, {Audard}, {Nienartowicz}, {Andrae}, {Marconi}, {Szabados}, {Evans}, {Lecoeur-Taibi}, {Mowlavi}, {Musella}, \& {Eyer}}]{Clementini2023}
{Clementini}, G., {Ripepi}, V., {Garofalo}, A., {et~al.} 2023, \aap, 674, A18, \dodoi{10.1051/0004-6361/202243964}

\bibitem[{{Cooper} {et~al.}(2010){Cooper}, {Cole}, {Frenk}, {White}, {Helly}, {Benson}, {De Lucia}, {Helmi}, {Jenkins}, {Navarro}, {Springel}, \& {Wang}}]{Cooper2010}
{Cooper}, A.~P., {Cole}, S., {Frenk}, C.~S., {et~al.} 2010, \mnras, 406, 744, \dodoi{10.1111/j.1365-2966.2010.16740.x}

\bibitem[{{Cooper} {et~al.}(2023){Cooper}, {Koposov}, {Allende Prieto}, {Manser}, {Kizhuprakkat}, {Myers}, {Dey}, {G{\"a}nsicke}, {Li}, {Rockosi}, {Valluri}, {Najita}, {Deason}, {Raichoor}, {Wang}, {Ting}, {Kim}, {Carrillo}, {Wang}, {Beraldo e Silva}, {Han}, {Ding}, {S{\'a}nchez-Conde}, {Aguilar}, {Ahlen}, {Bailey}, {Belokurov}, {Brooks}, {Cunha}, {Dawson}, {de la Macorra}, {Doel}, {Eisenstein}, {Fagrelius}, {Fanning}, {Font-Ribera}, {Forero-Romero}, {Gazta{\~n}aga}, {Gontcho a Gontcho}, {Guy}, {Honscheid}, {Kehoe}, {Kisner}, {Kremin}, {Landriau}, {Levi}, {Martini}, {Meisner}, {Miquel}, {Moustakas}, {Nie}, {Palanque-Delabrouille}, {Percival}, {Poppett}, {Prada}, {Rehemtulla}, {Schlafly}, {Schlegel}, {Schubnell}, {Sharples}, {Tarl{\'e}}, {Wechsler}, {Weinberg}, {Zhou}, \& {Zou}}]{Cooper2023}
{Cooper}, A.~P., {Koposov}, S.~E., {Allende Prieto}, C., {et~al.} 2023, \apj, 947, 37, \dodoi{10.3847/1538-4357/acb3c0}

\bibitem[{{Correa Magnus} \& {Vasiliev}(2022)}]{CorreaMagnus2022}
{Correa Magnus}, L., \& {Vasiliev}, E. 2022, \mnras, 511, 2610, \dodoi{10.1093/mnras/stab3726}

\bibitem[{{de Jong} {et~al.}(2019){de Jong}, {Agertz}, {Berbel}, {Aird}, {Alexander}, {Amarsi}, {Anders}, {Andrae}, {Ansarinejad}, {Ansorge}, {Antilogus}, {Anwand-Heerwart}, {Arentsen}, {Arnadottir}, {Asplund}, {Auger}, {Azais}, {Baade}, {Baker}, {Baker}, {Balbinot}, {Baldry}, {Banerji}, {Barden}, {Barklem}, {Barth{\'e}l{\'e}my-Mazot}, {Battistini}, {Bauer}, {Bell}, {Bellido-Tirado}, {Bellstedt}, {Belokurov}, {Bensby}, {Bergemann}, {Bestenlehner}, {Bielby}, {Bilicki}, {Blake}, {Bland-Hawthorn}, {Boeche}, {Boland}, {Boller}, {Bongard}, {Bongiorno}, {Bonifacio}, {Boudon}, {Brooks}, {Brown}, {Brown}, {Br{\"u}ggen}, {Brynnel}, {Brzeski}, {Buchert}, {Buschkamp}, {Caffau}, {Caillier}, {Carrick}, {Casagrande}, {Case}, {Casey}, {Cesarini}, {Cescutti}, {Chapuis}, {Chiappini}, {Childress}, {Christlieb}, {Church}, {Cioni}, {Cluver}, {Colless}, {Collett}, {Comparat}, {Cooper}, {Couch}, {Courbin}, {Croom}, {Croton}, {Daguis{\'e}}, {Dalton}, {Davies}, {Davis}, {de Laverny}, {Deason}, {Dionies}, {Disseau}, {Doel},
  {D{\"o}scher}, {Driver}, {Dwelly}, {Eckert}, {Edge}, {Edvardsson}, {Youssoufi}, {Elhaddad}, {Enke}, {Erfanianfar}, {Farrell}, {Fechner}, {Feiz}, {Feltzing}, {Ferreras}, {Feuerstein}, {Feuillet}, {Finoguenov}, {Ford}, {Fotopoulou}, {Fouesneau}, {Frenk}, {Frey}, {Gaessler}, {Geier}, {Gentile Fusillo}, {Gerhard}, {Giannantonio}, {Giannone}, {Gibson}, {Gillingham}, {Gonz{\'a}lez-Fern{\'a}ndez}, {Gonzalez-Solares}, {Gottloeber}, {Gould}, {Grebel}, {Gueguen}, {Guiglion}, {Haehnelt}, {Hahn}, {Hansen}, {Hartman}, {Hauptner}, {Hawkins}, {Haynes}, {Haynes}, {Heiter}, {Helmi}, {Aguayo}, {Hewett}, {Hinton}, {Hobbs}, {Hoenig}, {Hofman}, {Hook}, {Hopgood}, {Hopkins}, {Hourihane}, {Howes}, {Howlett}, {Huet}, {Irwin}, {Iwert}, {Jablonka}, {Jahn}, {Jahnke}, {Jarno}, {Jin}, {Jofre}, {Johl}, {Jones}, {J{\"o}nsson}, {Jordan}, {Karovicova}, {Khalatyan}, {Kelz}, {Kennicutt}, {King}, {Kitaura}, {Klar}, {Klauser}, {Kneib}, {Koch}, {Koposov}, {Kordopatis}, {Korn}, {Kosmalski}, {Kotak}, {Kovalev}, {Kreckel}, {Kripak}, {Krumpe},
  {Kuijken}, {Kunder}, {Kushniruk}, {Lam}, {Lamer}, {Laurent}, {Lawrence}, {Lehmitz}, {Lemasle}, {Lewis}, {Li}, {Lidman}, {Lind}, {Liske}, {Lizon}, {Loveday}, {Ludwig}, {McDermid}, {Maguire}, {Mainieri}, {Mali}, {Mandel}, {Mandel}, {Mannering}, {Martell}, {Martinez Delgado}, {Matijevic}, {McGregor}, {McMahon}, {McMillan}, {Mena}, {Merloni}, {Meyer}, {Michel}, {Micheva}, {Migniau}, {Minchev}, {Monari}, {Muller}, {Murphy}, {Muthukrishna}, {Nandra}, {Navarro}, {Ness}, {Nichani}, {Nichol}, {Nicklas}, {Niederhofer}, {Norberg}, {Obreschkow}, {Oliver}, {Owers}, {Pai}, {Pankratow}, {Parkinson}, {Paschke}, {Paterson}, {Pecontal}, {Parry}, {Phillips}, {Pillepich}, {Pinard}, {Pirard}, {Piskunov}, {Plank}, {Pl{\"u}schke}, {Pons}, {Popesso}, {Power}, {Pragt}, {Pramskiy}, {Pryer}, {Quattri}, {Queiroz}, {Quirrenbach}, {Rahurkar}, {Raichoor}, {Ramstedt}, {Rau}, {Recio-Blanco}, {Reiss}, {Renaud}, {Revaz}, {Rhode}, {Richard}, {Richter}, {Rix}, {Robotham}, {Roelfsema}, {Romaniello}, {Rosario}, {Rothmaier}, {Roukema}, {Ruchti},
  {Rupprecht}, {Rybizki}, {Ryde}, {Saar}, {Sadler}, {Sahl{\'e}n}, {Salvato}, {Sassolas}, {Saunders}, {Saviauk}, {Sbordone}, {Schmidt}, {Schnurr}, {Scholz}, {Schwope}, {Seifert}, {Shanks}, {Sheinis}, {Sivov}, {Sk{\'u}lad{\'o}ttir}, {Smartt}, {Smedley}, {Smith}, {Smith}, {Sorce}, {Spitler}, {Starkenburg}, {Steinmetz}, {Stilz}, {Storm}, {Sullivan}, {Sutherland}, {Swann}, {Tamone}, {Taylor}, {Teillon}, {Tempel}, {ter Horst}, {Thi}, {Tolstoy}, {Trager}, {Traven}, {Tremblay}, {Tresse}, {Valentini}, {van de Weygaert}, {van den Ancker}, {Veljanoski}, {Venkatesan}, {Wagner}, {Wagner}, {Walcher}, {Waller}, {Walton}, {Wang}, {Winkler}, {Wisotzki}, {Worley}, {Worseck}, {Xiang}, {Xu}, {Yong}, {Zhao}, {Zheng}, {Zscheyge}, \& {Zucker}}]{deJong2019}
{de Jong}, R.~S., {Agertz}, O., {Berbel}, A.~A., {et~al.} 2019, The Messenger, 175, 3, \dodoi{10.18727/0722-6691/5117}

\bibitem[{{Deason} {et~al.}(2011{\natexlab{a}}){Deason}, {Belokurov}, \& {Evans}}]{Deason2011a}
{Deason}, A.~J., {Belokurov}, V., \& {Evans}, N.~W. 2011{\natexlab{a}}, \mnras, 411, 1480, \dodoi{10.1111/j.1365-2966.2010.17785.x}

\bibitem[{{Deason} {et~al.}(2011{\natexlab{b}}){Deason}, {Belokurov}, \& {Evans}}]{Deason2011b}
---. 2011{\natexlab{b}}, \mnras, 416, 2903, \dodoi{10.1111/j.1365-2966.2011.19237.x}

\bibitem[{{Deason} {et~al.}(2012{\natexlab{a}}){Deason}, {Belokurov}, {Evans}, \& {An}}]{Deason2012b}
{Deason}, A.~J., {Belokurov}, V., {Evans}, N.~W., \& {An}, J. 2012{\natexlab{a}}, \mnras, 424, L44, \dodoi{10.1111/j.1745-3933.2012.01283.x}

\bibitem[{{Deason} {et~al.}(2012{\natexlab{b}}){Deason}, {Belokurov}, {Evans}, \& {McCarthy}}]{Deason2012a}
{Deason}, A.~J., {Belokurov}, V., {Evans}, N.~W., \& {McCarthy}, I.~G. 2012{\natexlab{b}}, \apj, 748, 2, \dodoi{10.1088/0004-637X/748/1/2}

\bibitem[{Deason {et~al.}(2018)Deason, Belokurov, \& Koposov}]{deason_galactic_2018}
Deason, A.~J., Belokurov, V., \& Koposov, S.~E. 2018, The Astrophysical Journal, 852, 118, \dodoi{10.3847/1538-4357/aa9d19}

\bibitem[{{Deason} {et~al.}(2019){Deason}, {Fattahi}, {Belokurov}, {Evans}, {Grand}, {Marinacci}, \& {Pakmor}}]{Deason2019}
{Deason}, A.~J., {Fattahi}, A., {Belokurov}, V., {et~al.} 2019, \mnras, 485, 3514, \dodoi{10.1093/mnras/stz623}

\bibitem[{{Deason} {et~al.}(2021){Deason}, {Erkal}, {Belokurov}, {Fattahi}, {G{\'o}mez}, {Grand}, {Pakmor}, {Xue}, {Liu}, {Yang}, {Zhang}, \& {Zhao}}]{Deason2021}
{Deason}, A.~J., {Erkal}, D., {Belokurov}, V., {et~al.} 2021, \mnras, 501, 5964, \dodoi{10.1093/mnras/staa3984}

\bibitem[{{Dehnen} {et~al.}(2006){Dehnen}, {McLaughlin}, \& {Sachania}}]{Dehnen2006}
{Dehnen}, W., {McLaughlin}, D.~E., \& {Sachania}, J. 2006, \mnras, 369, 1688, \dodoi{10.1111/j.1365-2966.2006.10404.x}

\bibitem[{{DESI Collaboration} {et~al.}(2016{\natexlab{a}}){DESI Collaboration}, {Aghamousa}, {Aguilar}, {Ahlen}, {Alam}, {Allen}, {Allende Prieto}, {Annis}, {Bailey}, {Balland}, {Ballester}, {Baltay}, {Beaufore}, {Bebek}, {Beers}, {Bell}, {Bernal}, {Besuner}, {Beutler}, {Blake}, {Bleuler}, {Blomqvist}, {Blum}, {Bolton}, {Briceno}, {Brooks}, {Brownstein}, {Buckley-Geer}, {Burden}, {Burtin}, {Busca}, {Cahn}, {Cai}, {Cardiel-Sas}, {Carlberg}, {Carton}, {Casas}, {Castander}, {Cervantes-Cota}, {Claybaugh}, {Close}, {Coker}, {Cole}, {Comparat}, {Cooper}, {Cousinou}, {Crocce}, {Cuby}, {Cunningham}, {Davis}, {Dawson}, {de la Macorra}, {De Vicente}, {Delubac}, {Derwent}, {Dey}, {Dhungana}, {Ding}, {Doel}, {Duan}, {Ealet}, {Edelstein}, {Eftekharzadeh}, {Eisenstein}, {Elliott}, {Escoffier}, {Evatt}, {Fagrelius}, {Fan}, {Fanning}, {Farahi}, {Farihi}, {Favole}, {Feng}, {Fernandez}, {Findlay}, {Finkbeiner}, {Fitzpatrick}, {Flaugher}, {Flender}, {Font-Ribera}, {Forero-Romero}, {Fosalba}, {Frenk}, {Fumagalli}, {Gaensicke},
  {Gallo}, {Garcia-Bellido}, {Gaztanaga}, {Pietro Gentile Fusillo}, {Gerard}, {Gershkovich}, {Giannantonio}, {Gillet}, {Gonzalez-de-Rivera}, {Gonzalez-Perez}, {Gott}, {Graur}, {Gutierrez}, {Guy}, {Habib}, {Heetderks}, {Heetderks}, {Heitmann}, {Hellwing}, {Herrera}, {Ho}, {Holland}, {Honscheid}, {Huff}, {Hutchinson}, {Huterer}, {Hwang}, {Illa Laguna}, {Ishikawa}, {Jacobs}, {Jeffrey}, {Jelinsky}, {Jennings}, {Jiang}, {Jimenez}, {Johnson}, {Joyce}, {Jullo}, {Juneau}, {Kama}, {Karcher}, {Karkar}, {Kehoe}, {Kennamer}, {Kent}, {Kilbinger}, {Kim}, {Kirkby}, {Kisner}, {Kitanidis}, {Kneib}, {Koposov}, {Kovacs}, {Koyama}, {Kremin}, {Kron}, {Kronig}, {Kueter-Young}, {Lacey}, {Lafever}, {Lahav}, {Lambert}, {Lampton}, {Landriau}, {Lang}, {Lauer}, {Le Goff}, {Le Guillou}, {Le Van Suu}, {Lee}, {Lee}, {Leitner}, {Lesser}, {Levi}, {L'Huillier}, {Li}, {Liang}, {Lin}, {Linder}, {Loebman}, {Luki{\'c}}, {Ma}, {MacCrann}, {Magneville}, {Makarem}, {Manera}, {Manser}, {Marshall}, {Martini}, {Massey}, {Matheson}, {McCauley},
  {McDonald}, {McGreer}, {Meisner}, {Metcalfe}, {Miller}, {Miquel}, {Moustakas}, {Myers}, {Naik}, {Newman}, {Nichol}, {Nicola}, {Nicolati da Costa}, {Nie}, {Niz}, {Norberg}, {Nord}, {Norman}, {Nugent}, {O'Brien}, {Oh}, {Olsen}, {Padilla}, {Padmanabhan}, {Padmanabhan}, {Palanque-Delabrouille}, {Palmese}, {Pappalardo}, {P{\^a}ris}, {Park}, {Patej}, {Peacock}, {Peiris}, {Peng}, {Percival}, {Perruchot}, {Pieri}, {Pogge}, {Pollack}, {Poppett}, {Prada}, {Prakash}, {Probst}, {Rabinowitz}, {Raichoor}, {Ree}, {Refregier}, {Regal}, {Reid}, {Reil}, {Rezaie}, {Rockosi}, {Roe}, {Ronayette}, {Roodman}, {Ross}, {Ross}, {Rossi}, {Rozo}, {Ruhlmann-Kleider}, {Rykoff}, {Sabiu}, {Samushia}, {Sanchez}, {Sanchez}, {Schlegel}, {Schneider}, {Schubnell}, {Secroun}, {Seljak}, {Seo}, {Serrano}, {Shafieloo}, {Shan}, {Sharples}, {Sholl}, {Shourt}, {Silber}, {Silva}, {Sirk}, {Slosar}, {Smith}, {Smoot}, {Som}, {Song}, {Sprayberry}, {Staten}, {Stefanik}, {Tarle}, {Sien Tie}, {Tinker}, {Tojeiro}, {Valdes}, {Valenzuela}, {Valluri},
  {Vargas-Magana}, {Verde}, {Walker}, {Wang}, {Wang}, {Weaver}, {Weaverdyck}, {Wechsler}, {Weinberg}, {White}, {Yang}, {Yeche}, {Zhang}, {Zhao}, {Zheng}, {Zhou}, {Zhou}, {Zhu}, {Zou}, \& {Zu}}]{DESI2016a}
{DESI Collaboration}, {Aghamousa}, A., {Aguilar}, J., {et~al.} 2016{\natexlab{a}}, arXiv e-prints, arXiv:1611.00036, \dodoi{10.48550/arXiv.1611.00036}

\bibitem[{{DESI Collaboration} {et~al.}(2016{\natexlab{b}}){DESI Collaboration}, {Aghamousa}, {Aguilar}, {Ahlen}, {Alam}, {Allen}, {Allende Prieto}, {Annis}, {Bailey}, {Balland}, {Ballester}, {Baltay}, {Beaufore}, {Bebek}, {Beers}, {Bell}, {Bernal}, {Besuner}, {Beutler}, {Blake}, {Bleuler}, {Blomqvist}, {Blum}, {Bolton}, {Briceno}, {Brooks}, {Brownstein}, {Buckley-Geer}, {Burden}, {Burtin}, {Busca}, {Cahn}, {Cai}, {Cardiel-Sas}, {Carlberg}, {Carton}, {Casas}, {Castander}, {Cervantes-Cota}, {Claybaugh}, {Close}, {Coker}, {Cole}, {Comparat}, {Cooper}, {Cousinou}, {Crocce}, {Cuby}, {Cunningham}, {Davis}, {Dawson}, {de la Macorra}, {De Vicente}, {Delubac}, {Derwent}, {Dey}, {Dhungana}, {Ding}, {Doel}, {Duan}, {Ealet}, {Edelstein}, {Eftekharzadeh}, {Eisenstein}, {Elliott}, {Escoffier}, {Evatt}, {Fagrelius}, {Fan}, {Fanning}, {Farahi}, {Farihi}, {Favole}, {Feng}, {Fernandez}, {Findlay}, {Finkbeiner}, {Fitzpatrick}, {Flaugher}, {Flender}, {Font-Ribera}, {Forero-Romero}, {Fosalba}, {Frenk}, {Fumagalli}, {Gaensicke},
  {Gallo}, {Garcia-Bellido}, {Gaztanaga}, {Pietro Gentile Fusillo}, {Gerard}, {Gershkovich}, {Giannantonio}, {Gillet}, {Gonzalez-de-Rivera}, {Gonzalez-Perez}, {Gott}, {Graur}, {Gutierrez}, {Guy}, {Habib}, {Heetderks}, {Heetderks}, {Heitmann}, {Hellwing}, {Herrera}, {Ho}, {Holland}, {Honscheid}, {Huff}, {Hutchinson}, {Huterer}, {Hwang}, {Illa Laguna}, {Ishikawa}, {Jacobs}, {Jeffrey}, {Jelinsky}, {Jennings}, {Jiang}, {Jimenez}, {Johnson}, {Joyce}, {Jullo}, {Juneau}, {Kama}, {Karcher}, {Karkar}, {Kehoe}, {Kennamer}, {Kent}, {Kilbinger}, {Kim}, {Kirkby}, {Kisner}, {Kitanidis}, {Kneib}, {Koposov}, {Kovacs}, {Koyama}, {Kremin}, {Kron}, {Kronig}, {Kueter-Young}, {Lacey}, {Lafever}, {Lahav}, {Lambert}, {Lampton}, {Landriau}, {Lang}, {Lauer}, {Le Goff}, {Le Guillou}, {Le Van Suu}, {Lee}, {Lee}, {Leitner}, {Lesser}, {Levi}, {L'Huillier}, {Li}, {Liang}, {Lin}, {Linder}, {Loebman}, {Luki{\'c}}, {Ma}, {MacCrann}, {Magneville}, {Makarem}, {Manera}, {Manser}, {Marshall}, {Martini}, {Massey}, {Matheson}, {McCauley},
  {McDonald}, {McGreer}, {Meisner}, {Metcalfe}, {Miller}, {Miquel}, {Moustakas}, {Myers}, {Naik}, {Newman}, {Nichol}, {Nicola}, {Nicolati da Costa}, {Nie}, {Niz}, {Norberg}, {Nord}, {Norman}, {Nugent}, {O'Brien}, {Oh}, {Olsen}, {Padilla}, {Padmanabhan}, {Padmanabhan}, {Palanque-Delabrouille}, {Palmese}, {Pappalardo}, {P{\^a}ris}, {Park}, {Patej}, {Peacock}, {Peiris}, {Peng}, {Percival}, {Perruchot}, {Pieri}, {Pogge}, {Pollack}, {Poppett}, {Prada}, {Prakash}, {Probst}, {Rabinowitz}, {Raichoor}, {Ree}, {Refregier}, {Regal}, {Reid}, {Reil}, {Rezaie}, {Rockosi}, {Roe}, {Ronayette}, {Roodman}, {Ross}, {Ross}, {Rossi}, {Rozo}, {Ruhlmann-Kleider}, {Rykoff}, {Sabiu}, {Samushia}, {Sanchez}, {Sanchez}, {Schlegel}, {Schneider}, {Schubnell}, {Secroun}, {Seljak}, {Seo}, {Serrano}, {Shafieloo}, {Shan}, {Sharples}, {Sholl}, {Shourt}, {Silber}, {Silva}, {Sirk}, {Slosar}, {Smith}, {Smoot}, {Som}, {Song}, {Sprayberry}, {Staten}, {Stefanik}, {Tarle}, {Sien Tie}, {Tinker}, {Tojeiro}, {Valdes}, {Valenzuela}, {Valluri},
  {Vargas-Magana}, {Verde}, {Walker}, {Wang}, {Wang}, {Weaver}, {Weaverdyck}, {Wechsler}, {Weinberg}, {White}, {Yang}, {Yeche}, {Zhang}, {Zhao}, {Zheng}, {Zhou}, {Zhou}, {Zhu}, {Zou}, \& {Zu}}]{DESI2016b}
---. 2016{\natexlab{b}}, arXiv e-prints, arXiv:1611.00037, \dodoi{10.48550/arXiv.1611.00037}

\bibitem[{{DESI Collaboration} {et~al.}(2022){DESI Collaboration}, {Abareshi}, {Aguilar}, {Ahlen}, {Alam}, {Alexander}, {Alfarsy}, {Allen}, {Allende Prieto}, {Alves}, {Ameel}, {Armengaud}, {Asorey}, {Aviles}, {Bailey}, {Balaguera-Antol{\'\i}nez}, {Ballester}, {Baltay}, {Bault}, {Beltran}, {Benavides}, {BenZvi}, {Berti}, {Besuner}, {Beutler}, {Bianchi}, {Blake}, {Blanc}, {Blum}, {Bolton}, {Bose}, {Bramall}, {Brieden}, {Brodzeller}, {Brooks}, {Brownewell}, {Buckley-Geer}, {Cahn}, {Cai}, {Canning}, {Capasso}, {Carnero Rosell}, {Carton}, {Casas}, {Castander}, {Cervantes-Cota}, {Chabanier}, {Chaussidon}, {Chuang}, {Circosta}, {Cole}, {Cooper}, {da Costa}, {Cousinou}, {Cuceu}, {Davis}, {Dawson}, {de la Cruz-Noriega}, {de la Macorra}, {de Mattia}, {Della Costa}, {Demmer}, {Derwent}, {Dey}, {Dey}, {Dhungana}, {Ding}, {Dobson}, {Doel}, {Donald-McCann}, {Donaldson}, {Douglass}, {Duan}, {Dunlop}, {Edelstein}, {Eftekharzadeh}, {Eisenstein}, {Enriquez-Vargas}, {Escoffier}, {Evatt}, {Fagrelius}, {Fan}, {Fanning},
  {Fawcett}, {Ferraro}, {Ereza}, {Flaugher}, {Font-Ribera}, {Forero-Romero}, {Frenk}, {Fromenteau}, {G{\"a}nsicke}, {Garcia-Quintero}, {Garrison}, {Gazta{\~n}aga}, {Gerardi}, {Gil-Mar{\'\i}n}, {Gontcho a Gontcho}, {Gonzalez-Morales}, {Gonzalez-de-Rivera}, {Gonzalez-Perez}, {Gordon}, {Graur}, {Green}, {Grove}, {Gruen}, {Gutierrez}, {Guy}, {Hahn}, {Harris}, {Herrera}, {Herrera-Alcantar}, {Honscheid}, {Howlett}, {Huterer}, {Ir{\v{s}}i{\v{c}}}, {Ishak}, {Jelinsky}, {Jiang}, {Jimenez}, {Jing}, {Joyce}, {Jullo}, {Juneau}, {Kara{\c{c}}ayl{\i}}, {Karamanis}, {Karcher}, {Karim}, {Kehoe}, {Kent}, {Kirkby}, {Kisner}, {Kitaura}, {Koposov}, {Kov{\'a}cs}, {Kremin}, {Krolewski}, {L'Huillier}, {Lahav}, {Lambert}, {Lamman}, {Lan}, {Landriau}, {Lane}, {Lang}, {Lange}, {Lasker}, {Le Guillou}, {Leauthaud}, {Le Van Suu}, {Levi}, {Li}, {Magneville}, {Manera}, {Manser}, {Marshall}, {Martini}, {McCollam}, {McDonald}, {Meisner}, {Mena-Fern{\'a}ndez}, {Meneses-Rizo}, {Mezcua}, {Miller}, {Miquel}, {Montero-Camacho}, {Moon},
  {Moustakas}, {Mueller}, {Mu{\~n}oz-Guti{\'e}rrez}, {Myers}, {Nadathur}, {Najita}, {Napolitano}, {Neilsen}, {Newman}, {Nie}, {Ning}, {Niz}, {Norberg}, {Noriega}, {O'Brien}, {Obuljen}, {Palanque-Delabrouille}, {Palmese}, {Zhiwei}, {Pappalardo}, {PENG}, {Percival}, {Perruchot}, {Pogge}, {Poppett}, {Porredon}, {Prada}, {Prochaska}, {Pucha}, {P{\'e}rez-Fern{\'a}ndez}, {P{\'e}rez-R{\`a}fols}, {Rabinowitz}, {Raichoor}, {Ramirez-Solano}, {Ram{\'\i}rez-P{\'e}rez}, {Ravoux}, {Reil}, {Rezaie}, {Rocher}, {Rockosi}, {Roe}, {Roodman}, {Ross}, {Rossi}, {Ruggeri}, {Ruhlmann-Kleider}, {Sabiu}, {Safonova}, {Said}, {Saintonge}, {Salas Catonga}, {Samushia}, {Sanchez}, {Saulder}, {Schaan}, {Schlafly}, {Schlegel}, {Schmoll}, {Scholte}, {Schubnell}, {Secroun}, {Seo}, {Serrano}, {Sharples}, {Sholl}, {Silber}, {Silva}, {Sirk}, {Siudek}, {Smith}, {Sprayberry}, {Staten}, {Stupak}, {Tan}, {Tarl{\'e}}, {Tie}, {Tojeiro}, {Ure{\~n}a-L{\'o}pez}, {Valdes}, {Valenzuela}, {Valluri}, {Vargas-Maga{\~n}a}, {Verde}, {Walther}, {Wang}, {Wang},
  {Weaver}, {Weaverdyck}, {Wechsler}, {Wilson}, {Yang}, {Yu}, {Yuan}, {Y{\`e}che}, {Zhang}, {Zhang}, {Zhao}, {Zhou}, {Zhou}, {Zou}, {Zou}, {Zou}, {Zu}, \& {DESI Collaboration}}]{DESI2022}
{DESI Collaboration}, {Abareshi}, B., {Aguilar}, J., {et~al.} 2022, \aj, 164, 207, \dodoi{10.3847/1538-3881/ac882b}

\bibitem[{{DESI Collaboration} {et~al.}(2024){DESI Collaboration}, {Adame}, {Aguilar}, {Ahlen}, {Alam}, {Alexander}, {Allende Prieto}, {Alvarez}, {Alves}, {Anand}, {Andrade}, {Armengaud}, {Avila}, {Aviles}, {Awan}, {Bahr-Kalus}, {Bailey}, {Baltay}, {Bault}, {Behera}, {BenZvi}, {Beutler}, {Bianchi}, {Blake}, {Blum}, {Bonici}, {Brieden}, {Brodzeller}, {Brooks}, {Buckley-Geer}, {Burtin}, {Calderon}, {Canning}, {Carnero Rosell}, {Cereskaite}, {Cervantes-Cota}, {Chabanier}, {Chaussidon}, {Chaves-Montero}, {Chebat}, {Chen}, {Chen}, {Claybaugh}, {Cole}, {Cuceu}, {Davis}, {Dawson}, {de la Macorra}, {de Mattia}, {Deiosso}, {Dey}, {Dey}, {Ding}, {Doel}, {Edelstein}, {Eftekharzadeh}, {Eisenstein}, {Elbers}, {Elliott}, {Fagrelius}, {Fanning}, {Ferraro}, {Ereza}, {Findlay}, {Flaugher}, {Font-Ribera}, {Forero-S{\'a}nchez}, {Forero-Romero}, {Frenk}, {Garcia-Quintero}, {Garrison}, {Gazta{\~n}aga}, {Gil-Mar{\'\i}n}, {Gontcho}, {Gonzalez-Morales}, {Gonzalez-Perez}, {Gordon}, {Green}, {Gruen}, {Gsponer}, {Gutierrez}, {Guy},
  {Hadzhiyska}, {Hahn}, {Hanif}, {Herrera-Alcantar}, {Honscheid}, {Howlett}, {Huterer}, {Ir{\v{s}}i{\v{c}}}, {Ishak}, {Joyce}, {Juneau}, {Kara{\c{c}}ayl{\i}}, {Kehoe}, {Kent}, {Kirkby}, {Kong}, {Koposov}, {Kremin}, {Krolewski}, {Lahav}, {Lai}, {Lan}, {Landriau}, {Lang}, {Lasker}, {Le Goff}, {Le Guillou}, {Leauthaud}, {Levi}, {Li}, {Lodha}, {Magneville}, {Manera}, {Margala}, {Martini}, {Matthewson}, {Maus}, {McDonald}, {Medina-Varela}, {Meisner}, {Mena-Fern{\'a}ndez}, {Miquel}, {Moon}, {Moore}, {Moustakas}, {Mudur}, {Mueller}, {Mu{\~n}oz-Guti{\'e}rrez}, {Myers}, {Nadathur}, {Napolitano}, {Neveux}, {Newman}, {Nguyen}, {Nie}, {Niz}, {Noriega}, {Padmanabhan}, {Paillas}, {Palanque-Delabrouille}, {Pan}, {Penmetsa}, {Percival}, {Pieri}, {Pinon}, {Poppett}, {Porredon}, {Prada}, {P{\'e}rez-Fern{\'a}ndez}, {P{\'e}rez-R{\`a}fols}, {Rabinowitz}, {Raichoor}, {Ram{\'\i}rez-P{\'e}rez}, {Ramirez-Solano}, {Rashkovetskyi}, {Ravoux}, {Rezaie}, {Rich}, {Rocher}, {Rockosi}, {Roe}, {Rosado-Marin}, {Ross}, {Rossi}, {Ruggeri},
  {Ruhlmann-Kleider}, {Samushia}, {Sanchez}, {Saulder}, {Schlafly}, {Schlegel}, {Schubnell}, {Seo}, {Shafieloo}, {Sharples}, {Silber}, {Slosar}, {Smith}, {Sprayberry}, {Tan}, {Tarl{\'e}}, {Taylor}, {Trusov}, {Vaisakh}, {Valcin}, {Valdes}, {Valogiannis}, {Vargas-Maga{\~n}a}, {Verde}, {Walther}, {Wang}, {Wang}, {Weaver}, {Weaverdyck}, {Wechsler}, {Weinberg}, {White}, \& {Wilson}}]{DESI_DR1_cosmology_2024}
{DESI Collaboration}, {Adame}, A.~G., {Aguilar}, J., {et~al.} 2024, arXiv e-prints, arXiv:2411.12022, \dodoi{10.48550/arXiv.2411.12022}

\bibitem[{{DESI Collaboration} {et~al.}(2025{\natexlab{a}}){DESI Collaboration}, {Abdul-Karim}, {Adame}, {Aguado}, {Aguilar}, {Ahlen}, {Alam}, {Aldering}, {Alexander}, {Alfarsy}, {Allen}, {Allende Prieto}, {Alves}, {Anand}, {Andrade}, {Armengaud}, {Avila}, {Aviles}, {Awan}, {Bailey}, {Baleato Lizancos}, {Ballester}, {Bault}, {Bautista}, {BenZvi}, {Beraldo e Silva}, {Bermejo-Climent}, {Beutler}, {Bianchi}, {Blake}, {Blum}, {Bolton}, {Bonici}, {Brieden}, {Brodzeller}, {Brooks}, {Buckley-Geer}, {Burtin}, {Canning}, {Carnero Rosell}, {Carr}, {Carrilho}, {Casas}, {Castander}, {Cereskaite}, {Cervantes-Cota}, {Chaussidon}, {Chaves-Montero}, {Chen}, {Chen}, {Claybaugh}, {Cole}, {Cooper}, {Cousinou}, {Cuceu}, {Davis}, {Dawson}, {de Belsunce}, {de la Cruz}, {de la Macorra}, {de Mattia}, {Deiosso}, {Della Costa}, {Demina}, {Demirbozan}, {DeRose}, {Dey}, {Dey}, {Ding}, {Ding}, {Doel}, {Douglass}, {Dowicz}, {Ebina}, {Edelstein}, {Eisenstein}, {Elbers}, {Emas}, {Escoffier}, {Fagrelius}, {Fan}, {Fanning}, {Fawcett},
  {Fern\'andez-Garc\'ia}, {Ferraro}, {Findlay}, {Font-Ribera}, {Forero-Romero}, {Forero-S\'anchez}, {Frenk}, {G{\"a}nsicke}, {Galbany}, {Garc\'ia-Bellido}, {Garcia-Quintero}, {Garrison}, {Gazta{\~n}aga}, {Gil-Mar\'in}, {Gnedin}, {Gontcho}, {Gonzalez-Morales}, {Gonzalez-Perez}, {Gordon}, {Graur}, {Green}, {Gruen}, {Gsponer}, {Guandalin}, {Gutierrez}, {Guy}, {Hahn}, {Han}, {Han}, {He}, {Herrera-Alcantar}, {Honscheid}, {Hou}, {Howlett}, {Huterer}, {Ir\v\{s\}i\v\{c\}}, {Ishak}, {Jacques}, {Jimenez}, {Jing}, {Joachimi}, {Joudaki}, {Joyce}, {Jullo}, {Juneau}, {Kara{\c{c}}ayl{\i}}, {Karim}, {Kehoe}, {Kent}, {Khederlarian}, {Kirkby}, {Kisner}, {Kitaura}, {Kizhuprakkat}, {Kong}, {Koposov}, {Kremin}, {Krolewski}, {Lahav}, {Lai}, {Lamman}, {Lan}, {Landriau}, {Lang}, {Lange}, {Lasker}, {Le Goff}, {Le Guillou}, {Leauthaud}, {Levi}, {Li}, {Li}, {Lodha}, {Lokken}, {Luo}, {Magneville}, {Manera}, {Manser}, {Margala}, {Martini}, {Maus}, {McCullough}, {McDonald}, {Medina}, {Medina-Varela}, {Meisner},
  {Mena-Fern\textbackslash'andez}, {Menegas}, {Mezcua}, {Miquel}, {Montero-Camacho}, {Moon}, {Moustakas}, {Munoz-Guti\'errez}, {Munoz-Santos}, {Myers}, {Myles}, {Nadathur}, {Najita}, {Napolitano}, {Newman}, {Nikakhtar}, {Nikutta}, {Niz}, {Noriega}, {Padmanabhan}, {Paillas}, {Palanque-Delabrouille}, {Palmese}, {Pan}, {Pan}, {Parkinson}, {Peacock}, {Percival}, {P\'erez-Fern\'andez}, {P\'erez-R\'afols}, \& {Peterson}}]{DESI_DR1_2025}
{DESI Collaboration}, {Abdul-Karim}, M., {Adame}, A.~G., {et~al.} 2025{\natexlab{a}}, arXiv e-prints, arXiv:2503.14745, \dodoi{10.48550/arXiv.2503.14745}

\bibitem[{{DESI Collaboration} {et~al.}(2025{\natexlab{b}}){DESI Collaboration}, {Abdul-Karim}, {Aguilar}, {Ahlen}, {Alam}, {Allen}, {Allende Prieto}, {Alves}, {Anand}, {Andrade}, {Armengaud}, {Aviles}, {Bailey}, {Baltay}, {Bansal}, {Bault}, {Behera}, {BenZvi}, {Bianchi}, {Blake}, {Brieden}, {Brodzeller}, {Brooks}, {Buckley-Geer}, {Burtin}, {Calderon}, {Canning}, {Carnero Rosell}, {Carrilho}, {Casas}, {Castander}, {Cereskaite}, {Charles}, {Chaussidon}, {Chaves-Montero}, {Chebat}, {Chen}, {Claybaugh}, {Cole}, {Cooper}, {Cuceu}, {Dawson}, {de la Macorra}, {de Mattia}, {Deiosso}, {Della Costa}, {Demina}, {Dey}, {Dey}, {Ding}, {Doel}, {Edelstein}, {Eisenstein}, {Elbers}, {Fagrelius}, {Fanning}, {Fern\'andez-Garc\'ia}, {Ferraro}, {Font-Ribera}, {Forero-Romero}, {Frenk}, {Garcia-Quintero}, {Garrison}, {Gazta{\~n}aga}, {Gil-Mar\'in}, {Gontcho}, {Gonzalez}, {Gonzalez-Morales}, {Gordon}, {Green}, {Gutierrez}, {Guy}, {Hadzhiyska}, {Hahn}, {He}, {Herbold}, {Herrera-Alcantar}, {Ho}, {Honscheid}, {Howlett}, {Huterer},
  {Ishak}, {Juneau}, {Kamble}, {Kara{\c{c}}ayl{\i}}, {Kehoe}, {Kent}, {Kim}, {Kirkby}, {Kisner}, {Koposov}, {Kremin}, {Krolewski}, {Lahav}, {Lamman}, {Landriau}, {Lang}, {Lasker}, {Le Goff}, {Le Guillou}, {Leauthaud}, {Levi}, {Li}, {Li}, {Lodha}, {Lokken}, {Lozano-Rodr\'iguez}, {Magneville}, {Manera}, {Martini}, {Matthewson}, {Meisner}, {Mena-Fern\'andez}, {Menegas}, {Mergulhao}, {Miquel}, {Moustakas}, {Munoz-Guti\'errez}, {Munoz-Santos}, {Myers}, {Nadathur}, {Naidoo}, {Napolitano}, {Newman}, {Niz}, {Noriega}, {Paillas}, {Palanque-Delabrouille}, {Pan}, {Peacock}, {Pellejero Ibanez}, {Percival}, {P\'erez-Fern\'andez}, {P\'erez-R\'afols}, {Pieri}, {Poppett}, {Prada}, {Rabinowitz}, {Raichoor}, {Ram\'irez-P\'erez}, {Rashkovetskyi}, {Ravoux}, {Rich}, {Rocher}, {Rockosi}, {Rohlf}, {Rom\'an-Herrera}, {Ross}, {Rossi}, {Ruggeri}, {Ruhlmann-Kleider}, {Samushia}, {Sanchez}, {Sanders}, {Schlegel}, {Schubnell}, {Seo}, {Shafieloo}, {Sharples}, {Silber}, {Sinigaglia}, {Sprayberry}, {Tan}, {Tarl\'e}, {Taylor}, {Turner},
  {Urena-L\'opez}, {Vaisakh}, {Valdes}, {Valogiannis}, {Vargas-Magana}, {Verde}, {Walther}, {Weaver}, {Weinberg}, {White}, {Wolfson}, {Y\'eche}, {Yu}, {Zaborowski}, {Zarrouk}, {Zhai}, {Zhang}, {Zhao}, {Zhao}, {Zhou}, \& {Zou}}]{desi2025ii_bao}
{DESI Collaboration}, {Abdul-Karim}, M., {Aguilar}, J., {et~al.} 2025{\natexlab{b}}, arXiv e-prints, arXiv:2503.14738, \dodoi{10.48550/arXiv.2503.14738}

\bibitem[{{Duane} {et~al.}(1987){Duane}, {Kennedy}, {Pendleton}, \& {Roweth}}]{Duane1987}
{Duane}, S., {Kennedy}, A.~D., {Pendleton}, B.~J., \& {Roweth}, D. 1987, Physics Letters B, 195, 216, \dodoi{10.1016/0370-2693(87)91197-X}

\bibitem[{{Eadie} \& {Juri{\'c}}(2019)}]{Eadie2019}
{Eadie}, G., \& {Juri{\'c}}, M. 2019, \apj, 875, 159, \dodoi{10.3847/1538-4357/ab0f97}

\bibitem[{{Eadie} \& {Harris}(2016)}]{Eadie2016}
{Eadie}, G.~M., \& {Harris}, W.~E. 2016, \apj, 829, 108, \dodoi{10.3847/0004-637X/829/2/108}

\bibitem[{{Eadie} {et~al.}(2015){Eadie}, {Harris}, \& {Widrow}}]{Eadie2015}
{Eadie}, G.~M., {Harris}, W.~E., \& {Widrow}, L.~M. 2015, \apj, 806, 54, \dodoi{10.1088/0004-637X/806/1/54}

\bibitem[{{Eadie} {et~al.}(2017){Eadie}, {Springford}, \& {Harris}}]{Eadie2017}
{Eadie}, G.~M., {Springford}, A., \& {Harris}, W.~E. 2017, \apj, 835, 167, \dodoi{10.3847/1538-4357/835/2/167}

\bibitem[{{Erkal} {et~al.}(2020){Erkal}, {Belokurov}, \& {Parkin}}]{Erkal2020}
{Erkal}, D., {Belokurov}, V.~A., \& {Parkin}, D.~L. 2020, \mnras, 498, 5574, \dodoi{10.1093/mnras/staa2840}

\bibitem[{{Erkal} {et~al.}(2021){Erkal}, {Deason}, {Belokurov}, {Xue}, {Koposov}, {Bird}, {Liu}, {Simion}, {Yang}, {Zhang}, \& {Zhao}}]{Erkal2021}
{Erkal}, D., {Deason}, A.~J., {Belokurov}, V., {et~al.} 2021, \mnras, 506, 2677, \dodoi{10.1093/mnras/stab1828}

\bibitem[{{Evans} {et~al.}(1997){Evans}, {Hafner}, \& {de Zeeuw}}]{Evans1997}
{Evans}, N.~W., {Hafner}, R.~M., \& {de Zeeuw}, P.~T. 1997, \mnras, 286, 315, \dodoi{10.1093/mnras/286.2.315}

\bibitem[{{Fattahi} {et~al.}(2020){Fattahi}, {Deason}, {Frenk}, {Simpson}, {G{\'o}mez}, {Grand}, {Monachesi}, {Marinacci}, \& {Pakmor}}]{Fattahi2020}
{Fattahi}, A., {Deason}, A.~J., {Frenk}, C.~S., {et~al.} 2020, \mnras, 497, 4459, \dodoi{10.1093/mnras/staa2221}

\bibitem[{{Fermani} \& {Sch{\"o}nrich}(2013)}]{Fermani2013}
{Fermani}, F., \& {Sch{\"o}nrich}, R. 2013, \mnras, 430, 1294, \dodoi{10.1093/mnras/sts703}

\bibitem[{{Fritz} {et~al.}(2020){Fritz}, {Di Cintio}, {Battaglia}, {Brook}, \& {Taibi}}]{Fritz2020}
{Fritz}, T.~K., {Di Cintio}, A., {Battaglia}, G., {Brook}, C., \& {Taibi}, S. 2020, \mnras, 494, 5178, \dodoi{10.1093/mnras/staa1040}

\bibitem[{Fukushima {et~al.}(2019)Fukushima, Chiba, Tanaka, Hayashi, Homma, Okamoto, Komiyama, Tanaka, Arimoto, \& Matsuno}]{fukushima_stellar_2019}
Fukushima, T., Chiba, M., Tanaka, M., {et~al.} 2019, Publications of the Astronomical Society of Japan, 71, 72, \dodoi{10.1093/pasj/psz052}

\bibitem[{{Garofalo} {et~al.}(2022){Garofalo}, {Delgado}, {Sarro}, {Clementini}, {Muraveva}, {Marconi}, \& {Ripepi}}]{Garofalo2022}
{Garofalo}, A., {Delgado}, H.~E., {Sarro}, L.~M., {et~al.} 2022, \mnras, 513, 788, \dodoi{10.1093/mnras/stac735}

\bibitem[{{Gibbons} {et~al.}(2014){Gibbons}, {Belokurov}, \& {Evans}}]{Gibbons2014}
{Gibbons}, S.~L.~J., {Belokurov}, V., \& {Evans}, N.~W. 2014, \mnras, 445, 3788, \dodoi{10.1093/mnras/stu1986}

\bibitem[{{Gnedin} {et~al.}(2010){Gnedin}, {Brown}, {Geller}, \& {Kenyon}}]{Gnedin2010}
{Gnedin}, O.~Y., {Brown}, W.~R., {Geller}, M.~J., \& {Kenyon}, S.~J. 2010, \apjl, 720, L108, \dodoi{10.1088/2041-8205/720/1/L108}

\bibitem[{{Grand} {et~al.}(2019){Grand}, {Deason}, {White}, {Simpson}, {G{\'o}mez}, {Marinacci}, \& {Pakmor}}]{Grand2019}
{Grand}, R. J.~J., {Deason}, A.~J., {White}, S. D.~M., {et~al.} 2019, \mnras, 487, L72, \dodoi{10.1093/mnrasl/slz092}

\bibitem[{{Grand} {et~al.}(2024){Grand}, {Fragkoudi}, {G{\'o}mez}, {Jenkins}, {Marinacci}, {Pakmor}, \& {Springel}}]{Grand2024}
{Grand}, R. J.~J., {Fragkoudi}, F., {G{\'o}mez}, F.~A., {et~al.} 2024, \mnras, 532, 1814, \dodoi{10.1093/mnras/stae1598}

\bibitem[{{Grand} {et~al.}(2017){Grand}, {G{\'o}mez}, {Marinacci}, {Pakmor}, {Springel}, {Campbell}, {Frenk}, {Jenkins}, \& {White}}]{Grand2017}
{Grand}, R. J.~J., {G{\'o}mez}, F.~A., {Marinacci}, F., {et~al.} 2017, \mnras, 467, 179, \dodoi{10.1093/mnras/stx071}

\bibitem[{{Grand} {et~al.}(2018){Grand}, {Helly}, {Fattahi}, {Cautun}, {Cole}, {Cooper}, {Deason}, {Frenk}, {G{\'o}mez}, {Hunt}, {Marinacci}, {Pakmor}, {Simpson}, {Springel}, \& {Xu}}]{Grand2018}
{Grand}, R. J.~J., {Helly}, J., {Fattahi}, A., {et~al.} 2018, \mnras, 481, 1726, \dodoi{10.1093/mnras/sty2403}

\bibitem[{{GRAVITY Collaboration} {et~al.}(2021){GRAVITY Collaboration}, {Abuter}, {Amorim}, {Baub{\"o}ck}, {Berger}, {Bonnet}, {Brandner}, {Cl{\'e}net}, {Davies}, {de Zeeuw}, {Dexter}, {Dallilar}, {Drescher}, {Eckart}, {Eisenhauer}, {F{\"o}rster Schreiber}, {Garcia}, {Gao}, {Gendron}, {Genzel}, {Gillessen}, {Habibi}, {Haubois}, {Hei{\ss}el}, {Henning}, {Hippler}, {Horrobin}, {Jim{\'e}nez-Rosales}, {Jochum}, {Jocou}, {Kaufer}, {Kervella}, {Lacour}, {Lapeyr{\`e}re}, {Le Bouquin}, {L{\'e}na}, {Lutz}, {Nowak}, {Ott}, {Paumard}, {Perraut}, {Perrin}, {Pfuhl}, {Rabien}, {Rodr{\'\i}guez-Coira}, {Shangguan}, {Shimizu}, {Scheithauer}, {Stadler}, {Straub}, {Straubmeier}, {Sturm}, {Tacconi}, {Vincent}, {von Fellenberg}, {Waisberg}, {Widmann}, {Wieprecht}, {Wiezorrek}, {Woillez}, {Yazici}, {Young}, \& {Zins}}]{Gravity2021}
{GRAVITY Collaboration}, {Abuter}, R., {Amorim}, A., {et~al.} 2021, \aap, 647, A59, \dodoi{10.1051/0004-6361/202040208}

\bibitem[{{Guy} {et~al.}(2023){Guy}, {Bailey}, {Kremin}, {Alam}, {Alexander}, {Allende Prieto}, {BenZvi}, {Bolton}, {Brooks}, {Chaussidon}, {Cooper}, {Dawson}, {de la Macorra}, {Dey}, {Dey}, {Dhungana}, {Eisenstein}, {Font-Ribera}, {Forero-Romero}, {Gazta{\~n}aga}, {Gontcho A Gontcho}, {Green}, {Honscheid}, {Ishak}, {Kehoe}, {Kirkby}, {Kisner}, {Koposov}, {Lan}, {Landriau}, {Le Guillou}, {Levi}, {Magneville}, {Manser}, {Martini}, {Meisner}, {Miquel}, {Moustakas}, {Myers}, {Newman}, {Nie}, {Palanque-Delabrouille}, {Percival}, {Poppett}, {Prada}, {Raichoor}, {Ravoux}, {Ross}, {Schlafly}, {Schlegel}, {Schubnell}, {Sharples}, {Tarl{\'e}}, {Weaver}, {Y{\'e}che}, {Zhou}, {Zhou}, \& {Zou}}]{Guy2023}
{Guy}, J., {Bailey}, S., {Kremin}, A., {et~al.} 2023, \aj, 165, 144, \dodoi{10.3847/1538-3881/acb212}

\bibitem[{{Han} {et~al.}(2016){Han}, {Wang}, {Cole}, \& {Frenk}}]{Han2016}
{Han}, J., {Wang}, W., {Cole}, S., \& {Frenk}, C.~S. 2016, \mnras, 456, 1017, \dodoi{10.1093/mnras/stv2522}

\bibitem[{Han {et~al.}(2022)Han, Conroy, Johnson, Speagle, Bonaca, Chandra, Naidu, Ting, Woody, \& Zaritsky}]{han_stellar_2022}
Han, J.~J., Conroy, C., Johnson, B.~D., {et~al.} 2022, The Astronomical Journal, 164, 249, \dodoi{10.3847/1538-3881/ac97e9}

\bibitem[{{Hattori} {et~al.}(2021){Hattori}, {Valluri}, \& {Vasiliev}}]{Hattori2021}
{Hattori}, K., {Valluri}, M., \& {Vasiliev}, E. 2021, \mnras, 508, 5468, \dodoi{10.1093/mnras/stab2898}

\bibitem[{{Hernitschek} {et~al.}(2017){Hernitschek}, {Sesar}, {Rix}, {Belokurov}, {Martinez-Delgado}, {Martin}, {Kaiser}, {Hodapp}, {Chambers}, {Wainscoat}, {Magnier}, {Kudritzki}, {Metcalfe}, \& {Draper}}]{Hernitschek2017}
{Hernitschek}, N., {Sesar}, B., {Rix}, H.-W., {et~al.} 2017, \apj, 850, 96, \dodoi{10.3847/1538-4357/aa960c}

\bibitem[{Hoffman \& Gelman(2014)}]{Hoffman2014}
Hoffman, M.~D., \& Gelman, A. 2014, Journal of Machine Learning Research, 15, 1593

\bibitem[{{Hunt} \& {Vasiliev}(2025)}]{Hunt2025}
{Hunt}, J. A.~S., \& {Vasiliev}, E. 2025, \nar, 100, 101721, \dodoi{10.1016/j.newar.2024.101721}

\bibitem[{{Hunter}(2007)}]{matplotlib}
{Hunter}, J.~D. 2007, Computing in Science and Engineering, 9, 90, \dodoi{10.1109/MCSE.2007.55}

\bibitem[{{Husser} {et~al.}(2013){Husser}, {Wende-von Berg}, {Dreizler}, {Homeier}, {Reiners}, {Barman}, \& {Hauschildt}}]{Husser2013}
{Husser}, T.~O., {Wende-von Berg}, S., {Dreizler}, S., {et~al.} 2013, \aap, 553, A6, \dodoi{10.1051/0004-6361/201219058}

\bibitem[{{Iorio} \& {Belokurov}(2021)}]{Iorio2021}
{Iorio}, G., \& {Belokurov}, V. 2021, \mnras, 502, 5686, \dodoi{10.1093/mnras/stab005}

\bibitem[{{Jeans}(1915)}]{Jeans1915}
{Jeans}, J.~H. 1915, \mnras, 76, 70, \dodoi{10.1093/mnras/76.2.70}

\bibitem[{{Kafle} {et~al.}(2012){Kafle}, {Sharma}, {Lewis}, \& {Bland-Hawthorn}}]{Kafle2012}
{Kafle}, P.~R., {Sharma}, S., {Lewis}, G.~F., \& {Bland-Hawthorn}, J. 2012, \apj, 761, 98, \dodoi{10.1088/0004-637X/761/2/98}

\bibitem[{{Kafle} {et~al.}(2014){Kafle}, {Sharma}, {Lewis}, \& {Bland-Hawthorn}}]{Kafle2014}
---. 2014, \apj, 794, 59, \dodoi{10.1088/0004-637X/794/1/59}

\bibitem[{{Karukes} {et~al.}(2020){Karukes}, {Benito}, {Iocco}, {Trotta}, \& {Geringer-Sameth}}]{Karukes2020}
{Karukes}, E.~V., {Benito}, M., {Iocco}, F., {Trotta}, R., \& {Geringer-Sameth}, A. 2020, \jcap, 2020, 033, \dodoi{10.1088/1475-7516/2020/05/033}

\bibitem[{{Keller} {et~al.}(2008){Keller}, {Murphy}, {Prior}, {DaCosta}, \& {Schmidt}}]{Keller2008}
{Keller}, S.~C., {Murphy}, S., {Prior}, S., {DaCosta}, G., \& {Schmidt}, B. 2008, \apj, 678, 851, \dodoi{10.1086/526516}

\bibitem[{{Kizhuprakkat} {et~al.}(2024){Kizhuprakkat}, {Cooper}, {Riley}, {Koposov}, {Aguilar}, {Ahlen}, {Allende Prieto}, {Brooks}, {Claybaugh}, {Dawson}, {de la Macorra}, {Doel}, {Forero-Romero}, {Frenk}, {Gazta{\~n}aga}, {Gnedin}, {Grand}, {Gontcho A Gontcho}, {Honscheid}, {Kehoe}, {Landriau}, {Manera}, {Meisner}, {Miquel}, {Nie}, {Prada}, {Rezaie}, {Rossi}, {Sanchez}, {Schubnell}, {Seo}, {Tarl{\'e}}, {Valluri}, \& {Zhou}}]{Kizhuprakkat2024}
{Kizhuprakkat}, N., {Cooper}, A.~P., {Riley}, A.~H., {et~al.} 2024, \mnras, 531, 4108, \dodoi{10.1093/mnras/stae1415}

\bibitem[{{Koposov} {et~al.}(2019){Koposov}, {Belokurov}, {Li}, {Mateu}, {Erkal}, {Grillmair}, {Hendel}, {Price-Whelan}, {Laporte}, {Hawkins}, {Sohn}, {del Pino}, {Evans}, {Slater}, {Kallivayalil}, {Navarro}, \& {Orphan Aspen Treasury Collaboration}}]{Koposov:2019}
{Koposov}, S.~E., {Belokurov}, V., {Li}, T.~S., {et~al.} 2019, \mnras, 485, 4726, \dodoi{10.1093/mnras/stz457}

\bibitem[{{Koposov} {et~al.}(2023){Koposov}, {Erkal}, {Li}, {Da Costa}, {Cullinane}, {Ji}, {Kuehn}, {Lewis}, {Pace}, {Shipp}, {Zucker}, {Bland-Hawthorn}, {Lilleengen}, {Martell}, \& {S5 Collaboration}}]{Koposov2023}
{Koposov}, S.~E., {Erkal}, D., {Li}, T.~S., {et~al.} 2023, \mnras, 521, 4936, \dodoi{10.1093/mnras/stad551}

\bibitem[{{Koposov} {et~al.}(2024){Koposov}, {Allende Prieto}, {Cooper}, {Li}, {Beraldo e Silva}, {Kim}, {Carrillo}, {Dey}, {Manser}, {Nikakhtar}, {Riley}, {Rockosi}, {Valluri}, {Aguilar}, {Ahlen}, {Bailey}, {Blum}, {Brooks}, {Claybaugh}, {Cole}, {de la Macorra}, {Dey}, {Forero-Romero}, {Gazta{\~n}aga}, {Guy}, {Kremin}, {Le Guillou}, {Levi}, {Manera}, {Meisner}, {Miquel}, {Moustakas}, {Nie}, {Palanque-Delabrouille}, {Percival}, {Rezaie}, {Rossi}, {Sanchez}, {Schlafly}, {Schubnell}, {Tarl{\'e}}, {Weaver}, \& {Zhou}}]{Koposov2024}
{Koposov}, S.~E., {Allende Prieto}, C., {Cooper}, A.~P., {et~al.} 2024, \mnras, 533, 1012, \dodoi{10.1093/mnras/stae1842}

\bibitem[{{Koposov} {et~al.}(2025){Koposov}, {Li}, {Allende Prieto}, {Medina}, {Sandford}, {Aguado}, {Beraldo e Silva}, {Bystr{\"o}m}, {Cooper}, {Dey}, {Frenk}, {Kizhuprakkat}, {Li}, {Najita}, {Riley}, {Silva}, {Thomas}, {Valluri}, {Aguilar}, {Ahlen}, {Bianchi}, {Brooks}, {Claybaugh}, {Cole}, {Cuceu}, {de la Macorra}, {Della Costa}, {Dey}, {Doel}, {Edelstein}, {Font-Ribera}, {Forero-Romero}, {Gazta{\~n}aga}, {Gontcho}, {Gutierrez}, {Guy}, {Honscheid}, {Jimenez}, {Kehoe}, {Kirkby}, {Kisner}, {Kremin}, {Lahav}, {Landriau}, {Le Guillou}, {Leauthaud}, {Levi}, {Manera}, {Meisner}, {Miquel}, {Moustakas}, {Nadathur}, {Palanque-Delabrouille}, {Percival}, {Prada}, {P{\'e}rez-R{\`a}fols}, {Rossi}, {Sanchez}, {Schlafly}, {Schlegel}, {Seo}, {Sharples}, {Silber}, {Sprayberry}, {Tarl'e}, {Weaver}, {Zhou}, \& {Zou}}]{Koposov2025}
{Koposov}, S.~E., {Li}, T.~S., {Allende Prieto}, C., {et~al.} 2025, arXiv e-prints, arXiv:2505.14787, \dodoi{10.48550/arXiv.2505.14787}

\bibitem[{{K{\"u}pper} {et~al.}(2015){K{\"u}pper}, {Balbinot}, {Bonaca}, {Johnston}, {Hogg}, {Kroupa}, \& {Santiago}}]{Kuepper2015}
{K{\"u}pper}, A. H.~W., {Balbinot}, E., {Bonaca}, A., {et~al.} 2015, \apj, 803, 80, \dodoi{10.1088/0004-637X/803/2/80}

\bibitem[{{Kurucz}(1979)}]{Kurucz1979}
{Kurucz}, R.~L. 1979, \apjs, 40, 1, \dodoi{10.1086/190589}

\bibitem[{{Kurucz}(1993)}]{Kurucz1993}
---. 1993, {SYNTHE spectrum synthesis programs and line data}

\bibitem[{{Li} {et~al.}(2019){Li}, {Koposov}, {Zucker}, {Lewis}, {Kuehn}, {Simpson}, {Ji}, {Shipp}, {Mao}, {Geha}, {Pace}, {Mackey}, {Allam}, {Tucker}, {Da Costa}, {Erkal}, {Simon}, {Mould}, {Martell}, {Wan}, {De Silva}, {Bechtol}, {Balbinot}, {Belokurov}, {Bland-Hawthorn}, {Casey}, {Cullinane}, {Drlica-Wagner}, {Sharma}, {Vivas}, {Wechsler}, {Yanny}, \& {S5 Collaboration}}]{Li2019}
{Li}, T.~S., {Koposov}, S.~E., {Zucker}, D.~B., {et~al.} 2019, \mnras, 490, 3508, \dodoi{10.1093/mnras/stz2731}

\bibitem[{{Li} \& {White}(2008)}]{Li2008}
{Li}, Y.-S., \& {White}, S. D.~M. 2008, \mnras, 384, 1459, \dodoi{10.1111/j.1365-2966.2007.12748.x}

\bibitem[{{Li} {et~al.}(2020){Li}, {Qian}, {Han}, {Li}, {Wang}, \& {Jing}}]{Li2020}
{Li}, Z.-Z., {Qian}, Y.-Z., {Han}, J., {et~al.} 2020, \apj, 894, 10, \dodoi{10.3847/1538-4357/ab84f0}

\bibitem[{{Little} \& {Tremaine}(1987)}]{Little1987}
{Little}, B., \& {Tremaine}, S. 1987, \apj, 320, 493, \dodoi{10.1086/165567}

\bibitem[{{Majewski} {et~al.}(2003){Majewski}, {Skrutskie}, {Weinberg}, \& {Ostheimer}}]{Majewski2003}
{Majewski}, S.~R., {Skrutskie}, M.~F., {Weinberg}, M.~D., \& {Ostheimer}, J.~C. 2003, \apj, 599, 1082, \dodoi{10.1086/379504}

\bibitem[{{Malhan} \& {Ibata}(2019)}]{Malhan2019}
{Malhan}, K., \& {Ibata}, R.~A. 2019, \mnras, 486, 2995, \dodoi{10.1093/mnras/stz1035}

\bibitem[{{Mateu}(2024)}]{Mateu2024}
{Mateu}, C. 2024, Research Notes of the American Astronomical Society, 8, 85, \dodoi{10.3847/2515-5172/ad3540}

\bibitem[{{Mateu} {et~al.}(2020){Mateu}, {Holl}, {De Ridder}, \& {Rimoldini}}]{Mateu2020}
{Mateu}, C., {Holl}, B., {De Ridder}, J., \& {Rimoldini}, L. 2020, \mnras, 496, 3291, \dodoi{10.1093/mnras/staa1676}

\bibitem[{{McMillan}(2011)}]{Mcmillan2011}
{McMillan}, P.~J. 2011, \mnras, 414, 2446, \dodoi{10.1111/j.1365-2966.2011.18564.x}

\bibitem[{{McMillan}(2017)}]{McMillan2017}
---. 2017, \mnras, 465, 76, \dodoi{10.1093/mnras/stw2759}

\bibitem[{{Medina} {et~al.}(2024){Medina}, {Mu{\~n}oz}, {Carlin}, {Vivas}, {Grebel}, {Mart{\'\i}nez-V{\'a}zquez}, \& {Hansen}}]{Medina2024}
{Medina}, G.~E., {Mu{\~n}oz}, R.~R., {Carlin}, J.~L., {et~al.} 2024, arXiv e-prints, arXiv:2402.14055, \dodoi{10.48550/arXiv.2402.14055}

\bibitem[{{Medina} {et~al.}(2018){Medina}, {Mu{\~n}oz}, {Vivas}, {Carlin}, {F{\"o}rster}, {Mart{\'\i}nez}, {Galbany}, {Gonz{\'a}lez-Gait{\'a}n}, {Hamuy}, {de Jaeger}, {Maureira}, \& {San Mart{\'\i}n}}]{Medina2018}
{Medina}, G.~E., {Mu{\~n}oz}, R.~R., {Vivas}, A.~K., {et~al.} 2018, \apj, 855, 43, \dodoi{10.3847/1538-4357/aaad02}

\bibitem[{{Medina} {et~al.}(2025{\natexlab{a}}){Medina}, {Li}, {Koposov}, {Riley}, {Beraldo e Silva}, {Valluri}, {Wang}, {Bystr{\"o}m}, {Gnedin}, {Carlberg}, {Kizhuprakkat}, {Weaver}, {Aguilar}, {Ahlen}, {Bianchi}, {Brooks}, {Claybaugh}, {Cooper}, {de la Macorra}, {Dey}, {Doel}, {Font-Ribera}, {Forero-Romero}, {Gazta{\~n}aga}, {Gontcho}, {Gutierrez}, {Guy}, {Honscheid}, {Ishak}, {Kisner}, {Landriau}, {Le Guillou}, {Meisner}, {Miquel}, {Myers}, {Nadathur}, {Poppett}, {Prada}, {P{\'e}rez-R{\`a}fols}, {Rossi}, {Sanchez}, {Seo}, {Sprayberry}, {Tarl{\'e}}, {Wechsler}, {Zhou}, \& {Zou}}]{Medina2025a}
{Medina}, G.~E., {Li}, T.~S., {Koposov}, S.~E., {et~al.} 2025{\natexlab{a}}, arXiv e-prints, arXiv:2504.02924, \dodoi{10.48550/arXiv.2504.02924}

\bibitem[{{Medina} {et~al.}(2025{\natexlab{b}}){Medina}, {Li}, {Allende Prieto}, {Beraldo e Silva}, {Bystrom}, {Carlberg}, {Koposov}, {Lambert}, {Najita}, {Rockosi}, {Kizhuprakkat}, {Riley}, {Aguilar}, {Ahlen}, {Bianchi}, {Brooks}, {Claybaugh}, {Cooper}, {de la Macorra}, {Dey}, {Doel}, {Forero-Romero}, {Gazta{\~n}aga}, {Gontcho}, {Gutierrez}, {Ishak}, {Kehoe}, {Kisner}, {Landriau}, {Le Guillou}, {Meisner}, {Miquel}, {Prada}, {Perez-Rafols}, {Rossi}, {Sanchez}, {Schlegel}, {Silber}, {Sprayberry}, {Tarle}, {Weaver}, \& {Zhou}}]{Medina2025b}
{Medina}, G.~E., {Li}, T.~S., {Allende Prieto}, C., {et~al.} 2025{\natexlab{b}}, arXiv e-prints, arXiv:2505.10614, \dodoi{10.48550/arXiv.2505.10614}

\bibitem[{{Miller} {et~al.}(2024){Miller}, {Doel}, {Gutierrez}, {Besuner}, {Brooks}, {Gallo}, {Heetderks}, {Jelinsky}, {Kent}, {Lampton}, {Levi}, {Liang}, {Meisner}, {Sholl}, {Silber}, {Sprayberry}, {Aguilar}, {de la Macorra}, {Eisenstein}, {Fanning}, {Font-Ribera}, {Gazta{\~n}aga}, {Gontcho A Gontcho}, {Honscheid}, {Jimenez}, {Joyce}, {Kehoe}, {Kisner}, {Kremin}, {Landriau}, {Le Guillou}, {Magneville}, {Martini}, {Miquel}, {Moustakas}, {Nie}, {Percival}, {Poppett}, {Prada}, {Rossi}, {Schlegel}, {Schubnell}, {Seo}, {Sharples}, {Tarl{\'e}}, {Vargas-Maga{\~n}a}, {Zhou}, \& {the DESI Collaboration}}]{DESI_Corrector_2024}
{Miller}, T.~N., {Doel}, P., {Gutierrez}, G., {et~al.} 2024, \aj, 168, 95, \dodoi{10.3847/1538-3881/ad45fe}

\bibitem[{{Monachesi} {et~al.}(2019){Monachesi}, {G{\'o}mez}, {Grand}, {Simpson}, {Kauffmann}, {Bustamante}, {Marinacci}, {Pakmor}, {Springel}, {Frenk}, {White}, \& {Tissera}}]{Monachesi2019}
{Monachesi}, A., {G{\'o}mez}, F.~A., {Grand}, R. J.~J., {et~al.} 2019, \mnras, 485, 2589, \dodoi{10.1093/mnras/stz538}

\bibitem[{{Monari} {et~al.}(2018){Monari}, {Famaey}, {Carrillo}, {Piffl}, {Steinmetz}, {Wyse}, {Anders}, {Chiappini}, \& {Jan{\ss}en}}]{Monari2018}
{Monari}, G., {Famaey}, B., {Carrillo}, I., {et~al.} 2018, \aap, 616, L9, \dodoi{10.1051/0004-6361/201833748}

\bibitem[{{Naidu} {et~al.}(2020){Naidu}, {Conroy}, {Bonaca}, {Johnson}, {Ting}, {Caldwell}, {Zaritsky}, \& {Cargile}}]{Naidu2020}
{Naidu}, R.~P., {Conroy}, C., {Bonaca}, A., {et~al.} 2020, \apj, 901, 48, \dodoi{10.3847/1538-4357/abaef4}

\bibitem[{{Narloch} {et~al.}(2024){Narloch}, {Hajdu}, {Pietrzy{\'n}ski}, {Gieren}, {Zgirski}, {Wielg{\'o}rski}, {Karczmarek}, {G{\'o}rski}, \& {Graczyk}}]{Narloch2024}
{Narloch}, W., {Hajdu}, G., {Pietrzy{\'n}ski}, G., {et~al.} 2024, \aap, 689, A138, \dodoi{10.1051/0004-6361/202450364}

\bibitem[{{Necib} \& {Lin}(2022)}]{Necib2022b}
{Necib}, L., \& {Lin}, T. 2022, \apj, 926, 189, \dodoi{10.3847/1538-4357/ac4244}

\bibitem[{{Ou} {et~al.}(2024){Ou}, {Eilers}, {Necib}, \& {Frebel}}]{Ou2024}
{Ou}, X., {Eilers}, A.-C., {Necib}, L., \& {Frebel}, A. 2024, \mnras, 528, 693, \dodoi{10.1093/mnras/stae034}

\bibitem[{{Pace} {et~al.}(2022){Pace}, {Erkal}, \& {Li}}]{Pace2022}
{Pace}, A.~B., {Erkal}, D., \& {Li}, T.~S. 2022, \apj, 940, 136, \dodoi{10.3847/1538-4357/ac997b}

\bibitem[{{Patel} {et~al.}(2018){Patel}, {Besla}, {Mandel}, \& {Sohn}}]{Patel2018}
{Patel}, E., {Besla}, G., {Mandel}, K., \& {Sohn}, S.~T. 2018, \apj, 857, 78, \dodoi{10.3847/1538-4357/aab78f}

\bibitem[{{Peebles}(1995)}]{Peebles1995}
{Peebles}, P.~J.~E. 1995, \apj, 449, 52, \dodoi{10.1086/176031}

\bibitem[{{Piffl} {et~al.}(2014){Piffl}, {Scannapieco}, {Binney}, {Steinmetz}, {Scholz}, {Williams}, {de Jong}, {Kordopatis}, {Matijevi{\v{c}}}, {Bienaym{\'e}}, {Bland-Hawthorn}, {Boeche}, {Freeman}, {Gibson}, {Gilmore}, {Grebel}, {Helmi}, {Munari}, {Navarro}, {Parker}, {Reid}, {Seabroke}, {Watson}, {Wyse}, \& {Zwitter}}]{Piffl2014}
{Piffl}, T., {Scannapieco}, C., {Binney}, J., {et~al.} 2014, \aap, 562, A91, \dodoi{10.1051/0004-6361/201322531}

\bibitem[{{Pillepich} {et~al.}(2014){Pillepich}, {Vogelsberger}, {Deason}, {Rodriguez-Gomez}, {Genel}, {Nelson}, {Torrey}, {Sales}, {Marinacci}, {Springel}, {Sijacki}, \& {Hernquist}}]{Pillepich2014}
{Pillepich}, A., {Vogelsberger}, M., {Deason}, A., {et~al.} 2014, \mnras, 444, 237, \dodoi{10.1093/mnras/stu1408}

\bibitem[{{Planck Collaboration} {et~al.}(2016){Planck Collaboration}, {Ade}, {Aghanim}, {Arnaud}, {Ashdown}, {Aumont}, {Baccigalupi}, {Banday}, {Barreiro}, {Bartlett}, {Bartolo}, {Battaner}, {Battye}, {Benabed}, {Beno{\^\i}t}, {Benoit-L{\'e}vy}, {Bernard}, {Bersanelli}, {Bielewicz}, {Bock}, {Bonaldi}, {Bonavera}, {Bond}, {Borrill}, {Bouchet}, {Boulanger}, {Bucher}, {Burigana}, {Butler}, {Calabrese}, {Cardoso}, {Catalano}, {Challinor}, {Chamballu}, {Chary}, {Chiang}, {Chluba}, {Christensen}, {Church}, {Clements}, {Colombi}, {Colombo}, {Combet}, {Coulais}, {Crill}, {Curto}, {Cuttaia}, {Danese}, {Davies}, {Davis}, {de Bernardis}, {de Rosa}, {de Zotti}, {Delabrouille}, {D{\'e}sert}, {Di Valentino}, {Dickinson}, {Diego}, {Dolag}, {Dole}, {Donzelli}, {Dor{\'e}}, {Douspis}, {Ducout}, {Dunkley}, {Dupac}, {Efstathiou}, {Elsner}, {En{\ss}lin}, {Eriksen}, {Farhang}, {Fergusson}, {Finelli}, {Forni}, {Frailis}, {Fraisse}, {Franceschi}, {Frejsel}, {Galeotta}, {Galli}, {Ganga}, {Gauthier}, {Gerbino}, {Ghosh}, {Giard},
  {Giraud-H{\'e}raud}, {Giusarma}, {Gjerl{\o}w}, {Gonz{\'a}lez-Nuevo}, {G{\'o}rski}, {Gratton}, {Gregorio}, {Gruppuso}, {Gudmundsson}, {Hamann}, {Hansen}, {Hanson}, {Harrison}, {Helou}, {Henrot-Versill{\'e}}, {Hern{\'a}ndez-Monteagudo}, {Herranz}, {Hildebrandt}, {Hivon}, {Hobson}, {Holmes}, {Hornstrup}, {Hovest}, {Huang}, {Huffenberger}, {Hurier}, {Jaffe}, {Jaffe}, {Jones}, {Juvela}, {Keih{\"a}nen}, {Keskitalo}, {Kisner}, {Kneissl}, {Knoche}, {Knox}, {Kunz}, {Kurki-Suonio}, {Lagache}, {L{\"a}hteenm{\"a}ki}, {Lamarre}, {Lasenby}, {Lattanzi}, {Lawrence}, {Leahy}, {Leonardi}, {Lesgourgues}, {Levrier}, {Lewis}, {Liguori}, {Lilje}, {Linden-V{\o}rnle}, {L{\'o}pez-Caniego}, {Lubin}, {Mac{\'\i}as-P{\'e}rez}, {Maggio}, {Maino}, {Mandolesi}, {Mangilli}, {Marchini}, {Maris}, {Martin}, {Martinelli}, {Mart{\'\i}nez-Gonz{\'a}lez}, {Masi}, {Matarrese}, {McGehee}, {Meinhold}, {Melchiorri}, {Melin}, {Mendes}, {Mennella}, {Migliaccio}, {Millea}, {Mitra}, {Miville-Desch{\^e}nes}, {Moneti}, {Montier}, {Morgante}, {Mortlock},
  {Moss}, {Munshi}, {Murphy}, {Naselsky}, {Nati}, {Natoli}, {Netterfield}, {N{\o}rgaard-Nielsen}, {Noviello}, {Novikov}, {Novikov}, {Oxborrow}, {Paci}, {Pagano}, {Pajot}, {Paladini}, {Paoletti}, {Partridge}, {Pasian}, {Patanchon}, {Pearson}, {Perdereau}, {Perotto}, {Perrotta}, {Pettorino}, {Piacentini}, {Piat}, {Pierpaoli}, {Pietrobon}, {Plaszczynski}, {Pointecouteau}, {Polenta}, {Popa}, {Pratt}, \& {Pr{\'e}zeau}}]{Planck2016}
{Planck Collaboration}, {Ade}, P.~A.~R., {Aghanim}, N., {et~al.} 2016, \aap, 594, A13, \dodoi{10.1051/0004-6361/201525830}

\bibitem[{{Poppett} {et~al.}(2024){Poppett}, {Tyas}, {Aguilar}, {Bebek}, {Bramall}, {Claybaugh}, {Edelstein}, {Fagrelius}, {Heetderks}, {Jelinsky}, {Jelinsky}, {Lafever}, {Lambert}, {Lampton}, {Levi}, {Martini}, {Rockosi}, {Schmoll}, {Sharples}, {Sirk}, {Wishnow}, {Yu}, {Ahlen}, {Bault}, {BenZvi}, {Brooks}, {Cole}, {de la Macorra}, {Dey}, {Doel}, {Fanning}, {Font-Ribera}, {Forero-Romero}, {Gazta{\~n}aga}, {Gontcho A Gontcho}, {Gonzalez-Morales}, {Hahn}, {Honscheid}, {Jimenez}, {Juneau}, {Kirkby}, {Kremin}, {Landriau}, {Le Guillou}, {Manera}, {Meisner}, {Miquel}, {Moustakas}, {Mueller}, {Mu{\~n}oz-Guti{\'e}rrez}, {Myers}, {Nie}, {Niz}, {Palanque-Delabrouille}, {Percival}, {Prada}, {Rabinowitz}, {Rezaie}, {Rossi}, {Sanchez}, {Schlafly}, {Schlegel}, {Schubnell}, {Seo}, {Sprayberry}, {Tarl{\'e}}, {Vargas-Maga{\~n}a}, {Weaver}, \& {Zhou}}]{DESI_Fiber_System_2024}
{Poppett}, C., {Tyas}, L., {Aguilar}, J., {et~al.} 2024, \aj, 168, 245, \dodoi{10.3847/1538-3881/ad76a4}

\bibitem[{{Posti} \& {Helmi}(2019)}]{Posti2019}
{Posti}, L., \& {Helmi}, A. 2019, \aap, 621, A56, \dodoi{10.1051/0004-6361/201833355}

\bibitem[{{Prudil} {et~al.}(2022){Prudil}, {Koch-Hansen}, {Lemasle}, {Grebel}, {Marchetti}, {Hansen}, {Crestani}, {Braga}, {Bono}, {Chaboyer}, {Fabrizio}, {Dall'Ora}, \& {Mart{\'\i}nez-V{\'a}zquez}}]{Prudil2022}
{Prudil}, Z., {Koch-Hansen}, A.~J., {Lemasle}, B., {et~al.} 2022, \aap, 664, A148, \dodoi{10.1051/0004-6361/202142251}

\bibitem[{{Pu} {et~al.}(2025){Pu}, {Cooper}, {Grand}, {G{\'o}mez}, \& {Monachesi}}]{Pu2025}
{Pu}, S.-Y., {Cooper}, A.~P., {Grand}, R. J.~J., {G{\'o}mez}, F.~A., \& {Monachesi}, A. 2025, \apj, 980, 63, \dodoi{10.3847/1538-4357/ada382}

\bibitem[{{Rehemtulla} {et~al.}(2022){Rehemtulla}, {Valluri}, \& {Vasiliev}}]{Rehemtulla+22}
{Rehemtulla}, N., {Valluri}, M., \& {Vasiliev}, E. 2022, \mnras, 511, 5536, \dodoi{10.1093/mnras/stac400}

\bibitem[{{Riley} {et~al.}(2025){Riley}, {Shipp}, {Simpson}, {Bieri}, {Fattahi}, {Brown}, {Oman}, {Fragkoudi}, {G{\'o}mez}, {Grand}, \& {Marinacci}}]{Riley2025}
{Riley}, A.~H., {Shipp}, N., {Simpson}, C.~M., {et~al.} 2025, \mnras, \dodoi{10.1093/mnras/staf1350}

\bibitem[{{Sakamoto} {et~al.}(2003){Sakamoto}, {Chiba}, \& {Beers}}]{Sakamoto2003}
{Sakamoto}, T., {Chiba}, M., \& {Beers}, T.~C. 2003, \aap, 397, 899, \dodoi{10.1051/0004-6361:20021499}

\bibitem[{{Schlafly} \& {Finkbeiner}(2011)}]{Schlafly2011}
{Schlafly}, E.~F., \& {Finkbeiner}, D.~P. 2011, \apj, 737, 103, \dodoi{10.1088/0004-637X/737/2/103}

\bibitem[{{Schlafly} {et~al.}(2023){Schlafly}, {Kirkby}, {Schlegel}, {Myers}, {Raichoor}, {Dawson}, {Aguilar}, {Allende Prieto}, {Bailey}, {BenZvi}, {Bermejo-Climent}, {Brooks}, {de la Macorra}, {Dey}, {Doel}, {Fanning}, {Font-Ribera}, {Forero-Romero}, {Garc{\'\i}a-Bellido}, {Gontcho A Gontcho}, {Guy}, {Hahn}, {Honscheid}, {Ishak}, {Juneau}, {Kehoe}, {Kisner}, {Kremin}, {Landriau}, {Lang}, {Lasker}, {Levi}, {Magneville}, {Manser}, {Martini}, {Meisner}, {Miquel}, {Moustakas}, {Newman}, {Nie}, {Palanque-Delabrouille}, {Percival}, {Poppett}, {Rockosi}, {Ross}, {Rossi}, {Tarl{\'e}}, {Weaver}, {Y{\`e}che}, {Zhou}, \& {DESI Collaboration}}]{DESI_Survey_Operation_2023}
{Schlafly}, E.~F., {Kirkby}, D., {Schlegel}, D.~J., {et~al.} 2023, \aj, 166, 259, \dodoi{10.3847/1538-3881/ad0832}

\bibitem[{{Schultz} \& {Wiemer}(1975)}]{Schultz1975}
{Schultz}, G.~V., \& {Wiemer}, W. 1975, \aap, 43, 133

\bibitem[{{Sesar} {et~al.}(2017){Sesar}, {Hernitschek}, {Mitrovi{\'c}}, {Ivezi{\'c}}, {Rix}, {Cohen}, {Bernard}, {Grebel}, {Martin}, {Schlafly}, {Burgett}, {Draper}, {Flewelling}, {Kaiser}, {Kudritzki}, {Magnier}, {Metcalfe}, {Tonry}, \& {Waters}}]{Sesar2017}
{Sesar}, B., {Hernitschek}, N., {Mitrovi{\'c}}, S., {et~al.} 2017, \aj, 153, 204, \dodoi{10.3847/1538-3881/aa661b}

\bibitem[{{Shen} {et~al.}(2022){Shen}, {Eadie}, {Murray}, {Zaritsky}, {Speagle}, {Ting}, {Conroy}, {Cargile}, {Johnson}, {Naidu}, \& {Han}}]{Shen2022}
{Shen}, J., {Eadie}, G.~M., {Murray}, N., {et~al.} 2022, \apj, 925, 1, \dodoi{10.3847/1538-4357/ac3a7a}

\bibitem[{{Shipp} {et~al.}(2025){Shipp}, {Riley}, {Simpson}, {Bieri}, {Necib}, {Arora}, {Fragkoudi}, {G{\'o}mez}, {Grand}, \& {Marinacci}}]{Shipp2025}
{Shipp}, N., {Riley}, A.~H., {Simpson}, C.~M., {et~al.} 2025, \mnras, 542, 1109, \dodoi{10.1093/mnras/staf1283}

\bibitem[{{Simpson} {et~al.}(2018){Simpson}, {Grand}, {G{\'o}mez}, {Marinacci}, {Pakmor}, {Springel}, {Campbell}, \& {Frenk}}]{Simpson2018}
{Simpson}, C.~M., {Grand}, R. J.~J., {G{\'o}mez}, F.~A., {et~al.} 2018, \mnras, 478, 548, \dodoi{10.1093/mnras/sty774}

\bibitem[{{Slizewski} {et~al.}(2022){Slizewski}, {Dufresne}, {Murdock}, {Eadie}, {Sanderson}, {Wetzel}, \& {Juri{\'c}}}]{Slizewski2022}
{Slizewski}, A., {Dufresne}, X., {Murdock}, K., {et~al.} 2022, \apj, 924, 131, \dodoi{10.3847/1538-4357/ac390b}

\bibitem[{{Smith} {et~al.}(2007){Smith}, {Ruchti}, {Helmi}, {Wyse}, {Fulbright}, {Freeman}, {Navarro}, {Seabroke}, {Steinmetz}, {Williams}, {Bienaym{\'e}}, {Binney}, {Bland-Hawthorn}, {Dehnen}, {Gibson}, {Gilmore}, {Grebel}, {Munari}, {Parker}, {Scholz}, {Siebert}, {Watson}, \& {Zwitter}}]{Smith2007}
{Smith}, M.~C., {Ruchti}, G.~R., {Helmi}, A., {et~al.} 2007, \mnras, 379, 755, \dodoi{10.1111/j.1365-2966.2007.11964.x}

\bibitem[{{Sohn} {et~al.}(2018){Sohn}, {Watkins}, {Fardal}, {van der Marel}, {Deason}, {Besla}, \& {Bellini}}]{Sohn2018}
{Sohn}, S.~T., {Watkins}, L.~L., {Fardal}, M.~A., {et~al.} 2018, \apj, 862, 52, \dodoi{10.3847/1538-4357/aacd0b}

\bibitem[{{Springel} {et~al.}(2008){Springel}, {Wang}, {Vogelsberger}, {Ludlow}, {Jenkins}, {Helmi}, {Navarro}, {Frenk}, \& {White}}]{Springel2008}
{Springel}, V., {Wang}, J., {Vogelsberger}, M., {et~al.} 2008, \mnras, 391, 1685, \dodoi{10.1111/j.1365-2966.2008.14066.x}

\bibitem[{{Stringer} {et~al.}(2021){Stringer}, {Drlica-Wagner}, {Macri}, {Mart{\'\i}nez-V{\'a}zquez}, {Vivas}, {Ferguson}, {Pace}, {Walker}, {Neilsen}, {Tavangar}, {Wester}, {Abbott}, {Aguena}, {Allam}, {Bacon}, {Bechtol}, {Bertin}, {Brooks}, {Burke}, {Carnero Rosell}, {Carrasco Kind}, {Carretero}, {Costanzi}, {Crocce}, {da Costa}, {Pereira}, {De Vicente}, {Desai}, {Diehl}, {Doel}, {Ferrero}, {Garc{\'\i}a-Bellido}, {Gaztanaga}, {Gerdes}, {Gruen}, {Gruendl}, {Gschwend}, {Gutierrez}, {Hinton}, {Hollowood}, {Honscheid}, {Hoyle}, {James}, {Kuehn}, {Kuropatkin}, {Li}, {Maia}, {Marshall}, {Menanteau}, {Miquel}, {Morgan}, {Ogando}, {Palmese}, {Paz-Chinch{\'o}n}, {Plazas}, {Roodman}, {Sanchez}, {Schubnell}, {Serrano}, {Sevilla-Noarbe}, {Smith}, {Soares-Santos}, {Suchyta}, {Tarle}, {Thomas}, {To}, {Varga}, {Wilkinson}, {Zhang}, \& {DES Collaboration}}]{Stringer2021}
{Stringer}, K.~M., {Drlica-Wagner}, A., {Macri}, L., {et~al.} 2021, \apj, 911, 109, \dodoi{10.3847/1538-4357/abe873}

\bibitem[{{Sun} {et~al.}(2023){Sun}, {Wang}, {Liu}, {Long}, {Chen}, \& {Gao}}]{Sun2023}
{Sun}, G., {Wang}, Y., {Liu}, C., {et~al.} 2023, Research in Astronomy and Astrophysics, 23, 015013, \dodoi{10.1088/1674-4527/ac9e91}

\bibitem[{{Sylos Labini} {et~al.}(2023){Sylos Labini}, {Chrob{\'a}kov{\'a}}, {Capuzzo-Dolcetta}, \& {L{\'o}pez-Corredoira}}]{SylosLabini2023}
{Sylos Labini}, F., {Chrob{\'a}kov{\'a}}, {\v{Z}}., {Capuzzo-Dolcetta}, R., \& {L{\'o}pez-Corredoira}, M. 2023, \apj, 945, 3, \dodoi{10.3847/1538-4357/acb92c}

\bibitem[{{Tang} {et~al.}(2014){Tang}, {Bressan}, {Rosenfield}, {Slemer}, {Marigo}, {Girardi}, \& {Bianchi}}]{Tang:2014}
{Tang}, J., {Bressan}, A., {Rosenfield}, P., {et~al.} 2014, \mnras, 445, 4287, \dodoi{10.1093/mnras/stu2029}

\bibitem[{{The pandas development Team}(2025)}]{pandas2022}
{The pandas development Team}. 2025, {pandas-dev/pandas: Pandas}, v2.3.1,  Zenodo, \dodoi{10.5281/zenodo.3509134}

\bibitem[{{Thomas} {et~al.}(2018){Thomas}, {McConnachie}, {Ibata}, {C{\^o}t{\'e}}, {Martin}, {Starkenburg}, {Carlberg}, {Chapman}, {Fabbro}, {Famaey}, {Fantin}, {Gwyn}, {H{\'e}nault-Brunet}, {Malhan}, {Navarro}, {Robin}, \& {Scott}}]{Thomas2018}
{Thomas}, G.~F., {McConnachie}, A.~W., {Ibata}, R.~A., {et~al.} 2018, \mnras, 481, 5223, \dodoi{10.1093/mnras/sty2604}

\bibitem[{{Tissera} {et~al.}(2013){Tissera}, {Scannapieco}, {Beers}, \& {Carollo}}]{Tissera2013}
{Tissera}, P.~B., {Scannapieco}, C., {Beers}, T.~C., \& {Carollo}, D. 2013, \mnras, 432, 3391, \dodoi{10.1093/mnras/stt691}

\bibitem[{{van der Walt} {et~al.}(2011){van der Walt}, {Colbert}, \& {Varoquaux}}]{numpy}
{van der Walt}, S., {Colbert}, S.~C., \& {Varoquaux}, G. 2011, Computing in Science and Engineering, 13, 22, \dodoi{10.1109/MCSE.2011.37}

\bibitem[{{Vasiliev}(2019)}]{Vasiliev2019}
{Vasiliev}, E. 2019, \mnras, 484, 2832, \dodoi{10.1093/mnras/stz171}

\bibitem[{{Vasiliev} \& {Baumgardt}(2021)}]{vasiliev_GCs}
{Vasiliev}, E., \& {Baumgardt}, H. 2021, \mnras, 505, 5978, \dodoi{10.1093/mnras/stab1475}

\bibitem[{{Vasiliev} {et~al.}(2021){Vasiliev}, {Belokurov}, \& {Erkal}}]{Vasiliev2021}
{Vasiliev}, E., {Belokurov}, V., \& {Erkal}, D. 2021, \mnras, 501, 2279, \dodoi{10.1093/mnras/staa3673}

\bibitem[{Virtanen {et~al.}(2020)Virtanen, Gommers, Oliphant, Haberland, Reddy, Cournapeau, Burovski, Peterson, Weckesser, Bright, {van der Walt}, Brett, Wilson, Millman, Mayorov, Nelson, Jones, Kern, Larson, Carey, Polat, Feng, Moore, {VanderPlas}, Laxalde, Perktold, Cimrman, Henriksen, Quintero, Harris, Archibald, Ribeiro, Pedregosa, {van Mulbregt}, \& {SciPy 1.0 Contributors}}]{2020SciPy-NMeth}
Virtanen, P., Gommers, R., Oliphant, T.~E., {et~al.} 2020, Nature Methods, 17, 261, \dodoi{10.1038/s41592-019-0686-2}

\bibitem[{{Wang} {et~al.}(2022){Wang}, {Hammer}, \& {Yang}}]{Wang2022}
{Wang}, J., {Hammer}, F., \& {Yang}, Y. 2022, \mnras, 510, 2242, \dodoi{10.1093/mnras/stab3258}

\bibitem[{{Wang} {et~al.}(2020){Wang}, {Han}, {Cautun}, {Li}, \& {Ishigaki}}]{Wang2020}
{Wang}, W., {Han}, J., {Cautun}, M., {Li}, Z., \& {Ishigaki}, M.~N. 2020, Science China Physics, Mechanics, and Astronomy, 63, 109801, \dodoi{10.1007/s11433-019-1541-6}

\bibitem[{{Wang} {et~al.}(2018){Wang}, {Han}, {Cole}, {More}, {Frenk}, \& {Schaller}}]{Wang2018}
{Wang}, W., {Han}, J., {Cole}, S., {et~al.} 2018, \mnras, 476, 5669, \dodoi{10.1093/mnras/sty706}

\bibitem[{{Wang} {et~al.}(2015){Wang}, {Han}, {Cooper}, {Cole}, {Frenk}, \& {Lowing}}]{Wang2015}
{Wang}, W., {Han}, J., {Cooper}, A.~P., {et~al.} 2015, \mnras, 453, 377, \dodoi{10.1093/mnras/stv1647}

\bibitem[{{Watkins} {et~al.}(2010){Watkins}, {Evans}, \& {An}}]{Watkins2010}
{Watkins}, L.~L., {Evans}, N.~W., \& {An}, J.~H. 2010, \mnras, 406, 264, \dodoi{10.1111/j.1365-2966.2010.16708.x}

\bibitem[{{Watkins} {et~al.}(2019){Watkins}, {van der Marel}, {Sohn}, \& {Evans}}]{Watkins2019}
{Watkins}, L.~L., {van der Marel}, R.~P., {Sohn}, S.~T., \& {Evans}, N.~W. 2019, \apj, 873, 118, \dodoi{10.3847/1538-4357/ab089f}

\bibitem[{{Wenger} {et~al.}(2000){Wenger}, {Ochsenbein}, {Egret}, {Dubois}, {Bonnarel}, {Borde}, {Genova}, {Jasniewicz}, {Lalo{\"e}}, {Lesteven}, \& {Monier}}]{Simbad}
{Wenger}, M., {Ochsenbein}, F., {Egret}, D., {et~al.} 2000, \aaps, 143, 9, \dodoi{10.1051/aas:2000332}

\bibitem[{{Wilkinson} \& {Evans}(1999)}]{Wilkinson1999}
{Wilkinson}, M.~I., \& {Evans}, N.~W. 1999, \mnras, 310, 645, \dodoi{10.1046/j.1365-8711.1999.02964.x}

\bibitem[{{Williams} {et~al.}(2017){Williams}, {Belokurov}, {Casey}, \& {Evans}}]{Williams2017}
{Williams}, A.~A., {Belokurov}, V., {Casey}, A.~R., \& {Evans}, N.~W. 2017, \mnras, 468, 2359, \dodoi{10.1093/mnras/stx508}

\bibitem[{{Williams} \& {Evans}(2015)}]{Williams2015}
{Williams}, A.~A., \& {Evans}, N.~W. 2015, \mnras, 454, 698, \dodoi{10.1093/mnras/stv1967}

\bibitem[{{Xia} {et~al.}(2025){Xia}, {Klein}, {Bullock}, {Boylan-Kolchin}, {Caudillo}, {Moreno}, {Mercado}, \& {Feldmann}}]{Xia2025}
{Xia}, L.~Y., {Klein}, C., {Bullock}, J.~S., {et~al.} 2025, arXiv e-prints, arXiv:2506.08508, \dodoi{10.48550/arXiv.2506.08508}

\bibitem[{{Xue} {et~al.}(2008){Xue}, {Rix}, {Zhao}, {Re Fiorentin}, {Naab}, {Steinmetz}, {van den Bosch}, {Beers}, {Lee}, {Bell}, {Rockosi}, {Yanny}, {Newberg}, {Wilhelm}, {Kang}, {Smith}, \& {Schneider}}]{Xue2008}
{Xue}, X.~X., {Rix}, H.~W., {Zhao}, G., {et~al.} 2008, \apj, 684, 1143, \dodoi{10.1086/589500}

\bibitem[{{Zaritsky} {et~al.}(2020){Zaritsky}, {Conroy}, {Zhang}, {Naidu}, {Bonaca}, {Caldwell}, {Cargile}, \& {Johnson}}]{Zaritsky2020}
{Zaritsky}, D., {Conroy}, C., {Zhang}, H., {et~al.} 2020, \apj, 888, 114, \dodoi{10.3847/1538-4357/ab5b93}

\bibitem[{{Zaritsky} {et~al.}(1989){Zaritsky}, {Olszewski}, {Schommer}, {Peterson}, \& {Aaronson}}]{Zaritsky1989}
{Zaritsky}, D., {Olszewski}, E.~W., {Schommer}, R.~A., {Peterson}, R.~C., \& {Aaronson}, M. 1989, \apj, 345, 759, \dodoi{10.1086/167947}

\bibitem[{{Zhou} {et~al.}(2023){Zhou}, {Li}, {Huang}, \& {Zhang}}]{Zhou2023}
{Zhou}, Y., {Li}, X., {Huang}, Y., \& {Zhang}, H. 2023, \apj, 946, 73, \dodoi{10.3847/1538-4357/acadd9}

\bibitem[{{Zinn} {et~al.}(2014){Zinn}, {Horowitz}, {Vivas}, {Baltay}, {Ellman}, {Hadjiyska}, {Rabinowitz}, \& {Miller}}]{Zinn2014}
{Zinn}, R., {Horowitz}, B., {Vivas}, A.~K., {et~al.} 2014, \apj, 781, 22, \dodoi{10.1088/0004-637X/781/1/22}

\end{thebibliography}
\bibliographystyle{aasjournal}



\appendix
\renewcommand\thefigure{\thesection\arabic{figure}}    
\setcounter{figure}{0}    

\renewcommand{\thetable}{\thesection\arabic{table}}
\setcounter{table}{0}

\section{The virial radius as a function of the model parameters}

\label{sec:appendix_rvir}

In this work, we assume that the virial radius $r_{\rm vir}$ is equivalent to  $r_{200}$, the distance at which the mean mass density $\bar{\rho}$ is equal to 200 times the critical density of the universe, $\rho_{\rm crit}$. 
The critical density $\rho_{\rm crit}$ can be expressed as a function of the present value of the Hubble constant $H_0$, in the form  

\begin{equation}
\begin{aligned}
   \rho_{\rm crit} = \frac{3H_0^2}{8\pi}.\\  
\end{aligned}
\end{equation}
\label{eq:app1}

Thus, $r_{200}$ is defined as the distance at which $\bar{\rho}(r_{200})=200\rho_{\rm crit}$.

Assuming that the mass of the MW is characterized by a spherical distribution, its mean mass density corresponds to

\begin{equation}
\begin{aligned}
   \bar{\rho}& =\frac{3}{4\pi}\frac{M(r_{200})}{{r_{200}}^3}\\
   & = \frac{3}{4\pi} \frac{\Phi_0\ \gamma\ {r_{200}}^{1-\gamma}}{ {r_{200}}^3}.\\  
\end{aligned}
\end{equation}
\label{eq:app2}

Then,

\begin{equation}
\begin{aligned}
   \frac{\Phi_0\gamma}{{r_{200}}^{2+\gamma} }& =100H_0^2,\\     
\end{aligned}
\end{equation}
\label{eq:app3}

which, solving for $r_{200}$, can be written as

\begin{equation}
\begin{aligned}
   r_{200}& = { \frac{\Phi_0\gamma}{100H_0^2} }^{ \frac{1}{2+\gamma} }. \\     
\end{aligned}
\end{equation}
\label{eq:app4}

For this work, we adopt $H_0 = 67.8$\,km\,s$^{-1}$\,Mpc$^{-1}$ \citep[][]{Planck2016}.  

\section{Cumulative mass profiles for BHBs and RRLs}
\label{sec:CMP_masses_appendix}

In Tables~\ref{tab:CMPs_appendix_BHBs} and ~\ref{tab:CMPs_appendix_RRLs}, we provide the mass enclosed at different radii computed by GME in 10\,kpc intervals for BHBs and RRLs, respectively. 
These data correspond to the values displayed in Figure~\ref{fig:CMP-rrlbhb}.

\begin{table}\scriptsize 
\caption{
Mass of the MW enclosed within different radii $r$ obtained from the DESI BHB sample. 
Here, we provide the values corresponding to the BHB data shown in Figure~\ref{fig:CMP-rrlbhb}, presented as quantiles (q2.5, q12.5, q25.0, q50.0, q75.0, q87.5, and q97.5) of the enclosed mass at each radius. 
}
\label{tab:CMPs_appendix_BHBs}
\begin{center}

\centering
\begin{tabular}{|c|c|c|c|c|c|c|c|}

\hline
  r &  q2.5 &  q12.5 &  q25.0 &  q50.0 &  q75.0 &  q87.5 &  q97.5 \\
  \multicolumn{1}{|c|}{[kpc]} &
  \multicolumn{7}{c|}{[$\times10^{12}$\,M$_\odot$]} \\  
\hline
 10 &  0.11 &   0.13 &   0.14 &   0.15 &   0.17 &   0.18 &   0.20 \\
 20 &  0.18 &   0.20 &   0.21 &   0.22 &   0.24 &   0.25 &   0.27 \\
 30 &  0.23 &   0.25 &   0.27 &   0.28 &   0.30 &   0.31 &   0.34 \\
 40 &  0.28 &   0.30 &   0.32 &   0.33 &   0.35 &   0.37 &   0.39 \\
 50 &  0.33 &   0.35 &   0.36 &   0.38 &   0.40 &   0.41 &   0.44 \\
 60 &  0.37 &   0.39 &   0.40 &   0.42 &   0.44 &   0.46 &   0.48 \\
 70 &  0.40 &   0.43 &   0.44 &   0.46 &   0.48 &   0.50 &   0.53 \\
 80 &  0.44 &   0.46 &   0.48 &   0.50 &   0.52 &   0.54 &   0.57 \\
 90 &  0.47 &   0.49 &   0.51 &   0.53 &   0.56 &   0.58 &   0.61 \\
100 &  0.50 &   0.52 &   0.54 &   0.57 &   0.59 &   0.61 &   0.65 \\
110 &  0.52 &   0.55 &   0.57 &   0.60 &   0.63 &   0.65 &   0.69 \\
120 &  0.55 &   0.58 &   0.60 &   0.63 &   0.66 &   0.68 &   0.72 \\
130 &  0.57 &   0.61 &   0.63 &   0.66 &   0.69 &   0.72 &   0.76 \\
140 &  0.59 &   0.63 &   0.65 &   0.69 &   0.72 &   0.75 &   0.80 \\
150 &  0.61 &   0.65 &   0.68 &   0.72 &   0.76 &   0.79 &   0.83 \\
160 &  0.63 &   0.68 &   0.71 &   0.75 &   0.79 &   0.82 &   0.87 \\
170 &  0.65 &   0.70 &   0.73 &   0.77 &   0.82 &   0.85 &   0.90 \\
180 &  0.67 &   0.72 &   0.75 &   0.80 &   0.85 &   0.88 &   0.94 \\
190 &  0.69 &   0.74 &   0.78 &   0.82 &   0.87 &   0.91 &   0.97 \\
200 &  0.71 &   0.76 &   0.80 &   0.85 &   0.90 &   0.94 &   1.01 \\
210 &  0.72 &   0.78 &   0.82 &   0.87 &   0.93 &   0.97 &   1.04 \\
220 &  0.74 &   0.80 &   0.84 &   0.90 &   0.95 &   1.00 &   1.07 \\
230 &  0.75 &   0.82 &   0.86 &   0.92 &   0.98 &   1.02 &   1.10 \\
240 &  0.77 &   0.84 &   0.88 &   0.94 &   1.01 &   1.05 &   1.14 \\
250 &  0.79 &   0.86 &   0.90 &   0.97 &   1.03 &   1.08 &   1.17 \\
\hline
\end{tabular}

\end{center}
\end{table}

\begin{table}\scriptsize 
\caption{
Same as Table~\ref{tab:CMPs_appendix_BHBs} but for the mass obtained with our RRL sample.  
}
\label{tab:CMPs_appendix_RRLs}
\begin{center}
\centering
\begin{tabular}{|c|c|c|c|c|c|c|c|}

\hline
  r &  q2.5 &  q12.5 &  q25.0 &  q50.0 &  q75.0 &  q87.5 &  q97.5 \\
  \multicolumn{1}{|c|}{[kpc]} &
  \multicolumn{7}{c|}{[$\times10^{12}$\,M$_\odot$]} \\  
\hline
 10 &  0.11 &   0.13 &   0.14 &   0.16 &   0.18 &   0.20 &   0.22 \\
 20 &  0.17 &   0.19 &   0.21 &   0.23 &   0.26 &   0.27 &   0.30 \\
 30 &  0.22 &   0.25 &   0.27 &   0.29 &   0.31 &   0.33 &   0.36 \\
 40 &  0.26 &   0.29 &   0.31 &   0.34 &   0.36 &   0.38 &   0.41 \\
 50 &  0.30 &   0.33 &   0.35 &   0.38 &   0.41 &   0.43 &   0.46 \\
 60 &  0.33 &   0.37 &   0.39 &   0.42 &   0.45 &   0.47 &   0.51 \\
 70 &  0.37 &   0.40 &   0.42 &   0.45 &   0.49 &   0.51 &   0.55 \\
 80 &  0.40 &   0.43 &   0.45 &   0.49 &   0.52 &   0.54 &   0.59 \\
 90 &  0.42 &   0.46 &   0.48 &   0.52 &   0.55 &   0.58 &   0.63 \\
100 &  0.45 &   0.49 &   0.51 &   0.55 &   0.59 &   0.61 &   0.67 \\
110 &  0.47 &   0.51 &   0.54 &   0.58 &   0.62 &   0.65 &   0.70 \\
120 &  0.49 &   0.54 &   0.56 &   0.60 &   0.65 &   0.68 &   0.74 \\
130 &  0.51 &   0.56 &   0.59 &   0.63 &   0.67 &   0.71 &   0.77 \\
140 &  0.53 &   0.58 &   0.61 &   0.66 &   0.70 &   0.74 &   0.80 \\
150 &  0.55 &   0.60 &   0.63 &   0.68 &   0.73 &   0.77 &   0.83 \\
160 &  0.57 &   0.62 &   0.65 &   0.70 &   0.76 &   0.80 &   0.87 \\
170 &  0.59 &   0.64 &   0.68 &   0.73 &   0.78 &   0.82 &   0.90 \\
180 &  0.61 &   0.66 &   0.70 &   0.75 &   0.81 &   0.85 &   0.93 \\
190 &  0.62 &   0.68 &   0.72 &   0.77 &   0.83 &   0.88 &   0.96 \\
200 &  0.64 &   0.70 &   0.73 &   0.79 &   0.86 &   0.90 &   0.99 \\
210 &  0.65 &   0.71 &   0.75 &   0.81 &   0.88 &   0.93 &   1.02 \\
220 &  0.67 &   0.73 &   0.77 &   0.83 &   0.90 &   0.95 &   1.05 \\
230 &  0.68 &   0.75 &   0.79 &   0.85 &   0.92 &   0.98 &   1.07 \\
240 &  0.70 &   0.76 &   0.81 &   0.87 &   0.95 &   1.00 &   1.10 \\
250 &  0.71 &   0.78 &   0.82 &   0.89 &   0.97 &   1.02 &   1.13 \\
\hline
\end{tabular}
\end{center}
\end{table}

\section{The enclosed mass of the Milky Way within 100\,kpc using the Jeans equation method}
\label{sec:nimble}

Here, we provide further details of the alternative approach used to measure the CMP of the MW described in Section~\ref{sec:comparisonLiterature}, which is based on the Jeans equation method and employs the publicly available NIMBLE (Non-parametrIc jeans Modeling with B-spLinEs) code \citep{Rehemtulla+22}.
For this comparison, we employ the latest implementation of NIMBLE.\footnote{\href{https://github.com/nabeelre/NIMBLE}{https://github.com/nabeelre/NIMBLE}.}

\begin{figure}
    \centering    
    \centering    
    \includegraphics[width=0.49\textwidth]{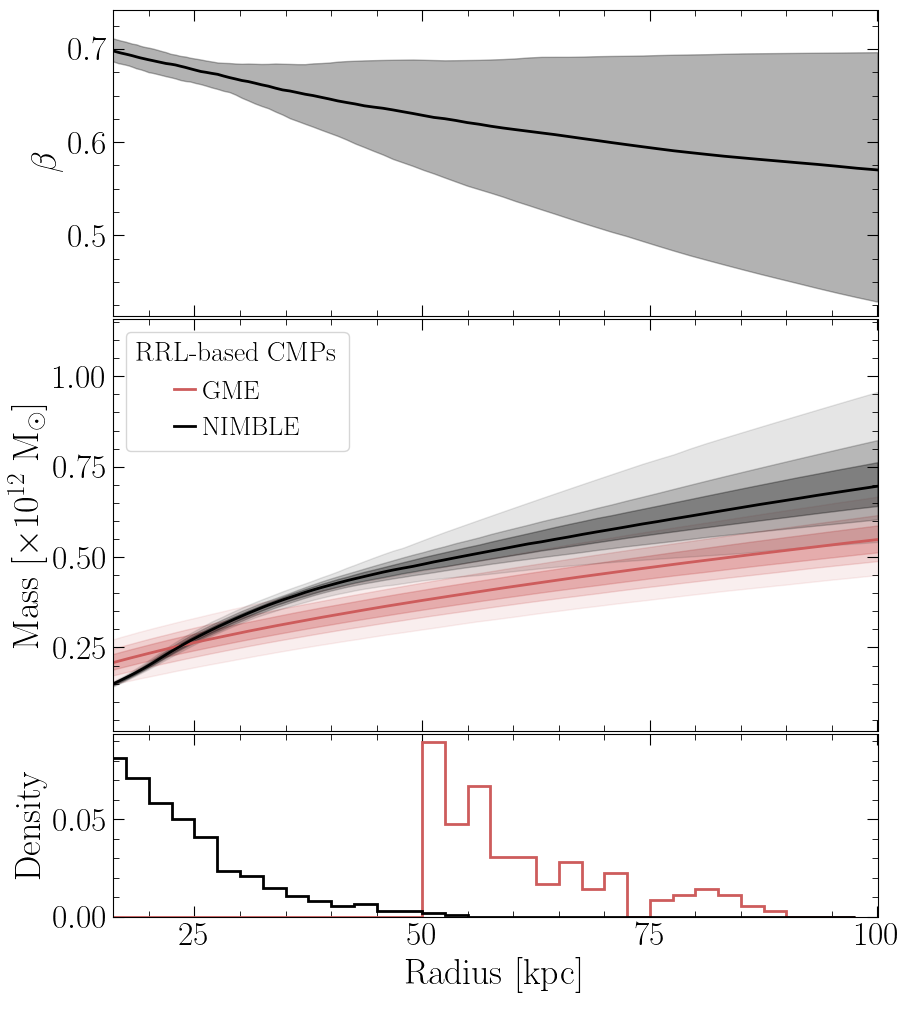}
    \caption{{\gm {\it Top}: Velocity anisotropy ($\beta$) profile derived by NIMBLE using DESI RRLs as tracers.  
    {\it Middle}: Comparison of the CMP derived by GME and NIMBLE using DESI RRLs. 
    {\it Bottom}: Normalized distance distribution of the RRLs  used as tracers for GME and NIMBLE. 
    The GME model uses 110 RRLs beyond 50\,kpc, while the NIMBLE model uses 5,240 RRLs with $R_{\rm GC}<55$\,kpc. 
    In the top panels, solid lines represent the median of the (NIMBLE or GME) MCMC chains, 
    whereas shaded regions display the 2.5, 12.5, 25.0, 75.0, 87.5, and 97.5 percentile intervals.
    The variation of $\beta$ estimated by NIMBLE displays a steady decrease from $\sim0.70$ in the inner halo to $\sim0.58$ at 100\,kpc. 
    percentile intervals that increase drastically beyond $\sim30$\,kpc.
    Although within each other's uncertainties, we find that 
    the medians of the GME and NIMBLE mass estimates differ by $\sim0.15\times10^{12}$\,M$_\odot$ at 100\,kpc.}
    }
    \label{fig:CMP-GME-nimble}
\end{figure}

\subsection{Spherical Jeans modeling with NIMBLE}
\label{sec:nimbledescription}

Spherical Jeans modeling \citep{Jeans1915} is among the oldest and most simple techniques to derive the mass of stellar systems. Deriving from the collisionless Boltzmann equation under the assumptions of spherical symmetry and dynamical equilibrium, the Jeans equations reduce to a single equation that relates the kinematics of the tracers of a system to its enclosed mass $M(<r)$. 
Most previous implementations of spherical Jeans modeling make assumptions of the velocity anisotropy profile $\beta$ \citep[e.g.,][]{Xue2008} and/or fit parametric functions (e.g., power-laws) to the velocity and density profiles \citep[e.g.,][]{Gnedin2010}. 
We make use of NIMBLE \citep{Rehemtulla+22} because it leverages the large amount of data at our disposal and does not rely on strong assumptions or the need of parametric functions. 
This method uses B-splines to non-parametrically represent the inputs of the spherical Jeans equation: the square of each spherical velocity component $v_r^2$, $v_\phi^2$, $v_\theta^2$ and the log of the tracer number density $\ln({\rho})$. 
NIMBLE also employs a Markov chain Monte Carlo routine for deconvolving effects of observational uncertainties and non-uniform spatial coverage to recover the best-fit velocity profiles, velocity anisotropy profile and intrinsic tracer density profile from the observations. These are then used to recover the enclosed mass profile from incomplete data under the assumption of spherical symmetry. 
\cite{Rehemtulla+22} extensively validated NIMBLE on a wide variety of mocks including three MW-like galaxies from the Latte suite of FIRE-2 cosmological hydrodynamic zoom-in simulations with realistic observational uncertainties (from {\it Gaia} and DESI) with the selection functions of these surveys imposed. The validation tests on the mock data from the Latte simulations indicated that the cumulative mass distribution is recovered with  $\lesssim20\%$ out to a radius of $\sim 80$\,kpc (the radius at which asymmetries and sparseness of the tracer distribution causes uncertainties to become large). Since NIMBLE does not assume a parametric form for the stellar tracer and dark matter density distributions, it is impossible to reliably extrapolate the cumulative mass distribution significantly beyond the radius where data are available.

NIMBLE was designed to run with RRLs as input under the assumption that the tracer population has a narrow range of absolute magnitudes. This assumption enables a straight forward recovery of the intrinsic 3D density distribution. Therefore, 
we apply it only to the DESI RRL catalog used in this work, as a natural extension of \citet{Rehemtulla+22}.

To enable a direct comparison with the GME-based CMPs presented in this work, we apply a similar set of selection cuts as those used for our main analysis (see Section~\ref{sec:cuts}), but include stars of brighter apparent magnitudes. 
We consider RRLs with mean {\it Gaia} magnitudes between  $14<${\sc\ int\_average\_g} $< 19$, as this magnitude range ensures a stable solution for NIMBLE while keeping a sensible completeness of the RRL catalog (see Section~\ref{sec:completeness_rrl}).
Given this restriction, the input catalog for NIMBLE covers a distance range of $R_{\rm GC}\sim5$--$55$\,kpc.

\subsection{Comparison between GME and NIMBLE mass profiles}
\label{sec:nimble_cmp}

We find the enclosed mass derived with NIMBLE to be systematically higher than the mass from GME in the distance range covered by our samples. 
More specifically, for both $R_{\rm GC}<50$\,kpc (close to the upper limit distance of NIMBLE and the lower limit of GME) and within 100\,kpc the masses inferred with NIMBLE are 25\% higher than those of our main results. 
Future versions of NIMBLE will consider other types of stars, e.g., BHBs or giant stars, which are more abundant but have a wider range of absolute magnitudes. 
Using these stars as tracers for NIMBLE will enable a careful investigation of the origin of the aforementioned discrepancy (i.e., whether it is inherent to the employed methodology or attributable to the different cuts in distance applied).

Our results are illustrated in Figure~\ref{fig:CMP-GME-nimble}. 
From the figure, two regimes are observed in the CMP inferred by NIMBLE: one characterized by a rapidly increasing enclosed mass ($\lesssim30$\,kpc) and one with a smoother increase in mass ($\gtrsim30$\,kpc), with increasingly larger confidence intervals. 
When compared with GME, we find that the enclosed mass derived by NIMBLE is larger at all radii $\gtrsim 25$\,kpc, with a difference that reaches $\sim0.15\times10^{12}$\,M$_\odot$ at 100\,kpc.
We note that the medians of the displayed mass profiles lie within each other's reported confidence intervals across the entire distance interval shown ($\sim10$--$100$\,kpc).
It should be considered, however, that no tracers are used for GME within 50\,kpc, and therefore, the enclosed mass in that range is shown only as a reference for that model.
A more detailed analysis of NIMBLE applied to halo tracers in DESI (including BHBs and RRLs), as well as a comprehensive comparison with the GME results will be presented in a separate paper, using 
data from the second data release of the DESI survey (corresponding to the first three years of survey operation).

\subsection{Velocity anisotropy of RRLs based on NIMBLE}
\label{sec:nimble_beta}

Figure~\ref{fig:CMP-GME-nimble} also shows the variation of the velocity anisotropy parameter $\beta$ as a function of galactocentric distance derived by NIMBLE. 
The figure displays a steady decrease in $\beta$ in the explored distance range, ranging from $\sim0.70$ in the inner halo to $\sim0.58$ at 100\,kpc.
In terms of the range of $\beta$ values considered by the model (from the MCMC chains of NIMBLE), and presented as shaded regions in the plot, we observe a drastic increase in the percentile intervals beyond $\sim30$\,kpc.
We note that, within that radius, the observed $\beta$ is estimated to be between 0.65 and 0.70, similar to the $\beta$ profiles presented by \citet{Medina2025a} from a sample that includes GSE stars (with more radial orbits) and stars from the metal-poor component of the halo (with more isotropic orbits). 
However, the NIMBLE estimates contrast with the values of $\beta$ derived from DESI RRLs beyond 50\,kpc in and the GME results presented in this work ($\beta\sim0.30$). 

Future avenues of exploration with NIMBLE may include, e.g., investigating the dependence of the $\beta$ profiles on metallicity \citep[similar to, e.g., ][]{Iorio2021}, and  the differences between $\beta$ profiles from different sets of tracers (e.g., BHBs, RRLs, and K giants).

\end{document}